\documentclass[pra,superscriptaddress,twocolumn,10.5pt,bibliography]{revtex4-1}

\usepackage{enumerate}    
\usepackage{amsmath}     
\usepackage{amssymb}   
\usepackage{graphicx} 
\usepackage{bm} 
\usepackage{color}  
\usepackage{hhline}
\usepackage{amsthm}
\usepackage{hyperref}
\usepackage{mathtools}
\usepackage{comment}

\newtheorem{Theorem}{Theorem}
\newtheorem{Proposition}[Theorem]{Proposition}
\newtheorem{Definition}[Theorem]{Definition}
\newtheorem{Corollary}[Theorem]{Corollary}

\theoremstyle{remark}

\newcommand{\cH}{{\mathcal H}}

\newcommand{\cT}{{\cal T}}

\newcommand{\cK}{{\cal K}}
\newcommand{\cG}{{\cal G}}

\newcommand{\cV}{\mathcal{V}}
\newcommand{\cU}{\mathcal{U}}

\newcommand{\cE}{\mathcal{E}}

\newcommand{\ket}[1]{|#1\rangle} 
\newcommand{\kett}[1]{|#1\rangle \!  \rangle} 
\newcommand{\bra}[1]{\langle#1|} 
\newcommand{\braa}[1]{\langle\! \langle #1|} 

\newcommand{\ketbra}[2]{|#1\rangle \!  \langle#2|} 
\newcommand{\kettbraa}[2]{|#1\rangle \! \rangle \! \langle \! \langle#2|}

\DeclareMathOperator{\tr}{Tr}

\newcommand{\braket}[2]{\langle #1 \vert #2 \rangle}
\newcommand{\braakett}[2]{\langle \! \langle #1 \vert #2 \rangle \! \rangle}

\begin{document}

\title{Quantum circuits for exact unitary $t$-designs and\\applications to higher-order randomized benchmarking}

\author{Yoshifumi Nakata}
\affiliation{Photon Science Center, Graduate School of Engineering,~The University of Tokyo, Bunkyo-ku,~Tokyo 113--8656, Japan}
\affiliation{JST, PRESTO, 4--1--8 Honcho, Kawaguchi, Saitama, 332--0012, Japan}

\author{Da Zhao}
\affiliation{School of Mathematical Sciences, Shanghai Jiao Tong University, Shanghai 400240, China}

\author{Takayuki Okuda}
\affiliation{Department of Mathematics, Hiroshima University, 1-3-1, Kagamiyama, Higashihiroshima 739-8562, Japan}

\author{Eiichi Bannai}
\affiliation{Faculty of Mathematics, Kyushu University (emeritus), Fukuoka 819-0385, Japan}
\affiliation{Mathematics Division, National Center for Theoretical Sciences, National Taiwan University, Taipei, 10617,
Taiwan}

\author{Yasunari Suzuki}
\affiliation{NTT Computer and Data Science Laboratories, NTT Corporation, Musashino 180-8585, Japan}
\affiliation{JST, PRESTO, 4--1--8 Honcho, Kawaguchi, Saitama, 332--0012, Japan}

\author{Shiro Tamiya}
 \affiliation{Department of Applied Physics, Graduate School of Engineering, The University of Tokyo, 7-3-1 Hongo, Bynkyo-ku, Tokyo 113-8656, Japan}

\author{Kentaro Heya}
\affiliation{Research Center for Advanced Science and Technology (RCAST), The University of Tokyo, Meguro-ku, Tokyo 153-8904, Japan}

\author{Zhiguang Yan}
\affiliation{RIKEN Center for Quantum Computing (RQC), Wako, Saitama 351-0198, Japan}

\author{Kun Zuo}
\affiliation{RIKEN Center for Quantum Computing (RQC), Wako, Saitama 351-0198, Japan}

\author{Shuhei Tamate}
\affiliation{RIKEN Center for Quantum Computing (RQC), Wako, Saitama 351-0198, Japan}

\author{Yutaka Tabuchi}
\affiliation{RIKEN Center for Quantum Computing (RQC), Wako, Saitama 351-0198, Japan}

\author{Yasunobu Nakamura}
\affiliation{Research Center for Advanced Science and Technology (RCAST), The University of Tokyo, Meguro-ku, Tokyo 153-8904, Japan}
\affiliation{RIKEN Center for Quantum Computing (RQC), Wako, Saitama 351-0198, Japan}

\begin{abstract}
A unitary $t$-design is a powerful tool in quantum information science and fundamental physics. Despite its usefulness, only approximate implementations were known for general $t$. In this paper, we provide for the first time quantum circuits that generate exact unitary $t$-designs for any $t$ on an arbitrary number of qubits. Our construction is inductive and is of practical use in small systems. We then introduce a $t$-th order generalization of randomized benchmarking ($t$-RB) as an application of exact $2t$-designs. We particularly study the $2$-RB in detail and show that it reveals \emph{self-adjointness} of quantum noise, a new metric related to the feasibility of quantum error correction (QEC). We numerically demonstrate that the $2$-RB in one- and two-qubit systems is feasible, and experimentally characterize background noise of a superconducting qubit by the $2$-RB. It is shown from the experiment that interactions with adjacent qubits induce the noise that may result in an obstacle toward a realization of QEC.
\end{abstract}


\maketitle

\section{Introduction}

Randomness in quantum systems has been driving recent progress of quantum information science~\cite{AE2007,D2005,DW2004,GPW2005,ADHW2009,DBWR2010,SDTR2013,HOW2005,HOW2007, AS2004,HLSW2004,S2005,BH2013,KRT2014,KL15,KZD2016,OAGKAL2016,B2018, G2019,BFNV2019,OSH2020,EAZ2005,KLRetc2008,MGE2011,MGE2012,B2018, PhysRevLett.112.240504, PhysRevA.93.012301, garion2020experimental, PhysRevLett.122.200502, PhysRevA.87.030301, PhysRevLett.109.240504, PhysRevLett.109.080505,OWE2019,HROWE2020} as well as fundamental physics~\cite{PSW2006,dRARDV2011,dRHRW2016, HP2007,SS2008,S2011,LSHOH2013,HQRY2016,RY2017,NWK2020,LFSLYYM2019,M2021}.
Theoretically, quantum randomness is often formulated by a unitary drawn uniformly at random, also known as a \emph{Haar} random unitary. However, the Haar randomness is physically unfeasible in large quantum systems.
From the viewpoint of applications, the unitaries that have similar properties of a Haar random unitary are of great importance since they can be used instead of the Haar one. When a random unitary has the same $t$-th order statistics as a Haar random unitary on average, it is called a \emph{unitary $t$-design}. For instance, when a protocol exploits the $t$-th power of the measurement probability after applying a Haar random unitary on any state, the protocol also works even if the Haar random unitary is replaced with a unitary $t$-design.

A unitary $t$-design can be regarded as a quantum generalization of $t$-wise independence~\cite{AE2007}, and have many applications, ranging from communication~\cite{D2005,DW2004,GPW2005,ADHW2009,DBWR2010,SDTR2013,HOW2005,HOW2007}, cryptography~\cite{AS2004,HLSW2004},  algorithms~\cite{S2005,BH2013}, sensing~\cite{KRT2014,KL15,KZD2016,OAGKAL2016}, to potentially quantum supremacy~\cite{B2018, G2019,BFNV2019}.
A unitary $t$-design is also related to another important concept in quantum information science, \emph{epsilon-net}~\cite{OSH2020}, implying more applications yet-to-be-discovered.
Furthermore, the concept of unitary designs has opened a novel research field over quantum information science, quantum thermodynamics, strongly correlated physics, and quantum gravity~\cite{PSW2006,dRARDV2011,dRHRW2016, HP2007,SS2008,S2011,LSHOH2013,HQRY2016,RY2017,NWK2020}.
Experimentally, unitary designs and related methods have been exploited for benchmarking noisy quantum devices~\cite{EAZ2005,KLRetc2008,MGE2011,MGE2012,B2018, PhysRevLett.112.240504, PhysRevA.93.012301, garion2020experimental, PhysRevLett.122.200502, PhysRevA.87.030301, PhysRevLett.109.240504, PhysRevLett.109.080505,OWE2019,HROWE2020}, realizing quantum supremacy~\cite{G2019}, demonstrating quantum chaos and quantum holography~\cite{LFSLYYM2019,M2021}.
It is also worthwhile to mention that unitary designs have been studied in combinatorial mathematics~\cite{DGS1975,DGS1977,RS2009,R2010,RS2011,BNRT2020}. Hence, developing the theory of unitary designs is of substantial interest in a wide range of science, both theoretically and experimentally.

An important question about unitary $t$-designs is how to implement them by quantum circuits. 
Many implementations of unitary $2$-designs, both approximate and exact ones, were proposed~\cite{DLT2002,BWV2008a,WBV2008,GAE2007,TGJ2007,DCEL2009,HL2009,DJ2011,BWV2008a,WBV2008,CLLW2015,NHMW2015-1}. In contrast, only \emph{approximate} implementations of unitary $t$-designs for general $t$ were known~\cite{HL2009TPE,BHH2016,NHKW2017,HM2018,HMHEGR2020}.  Explicit constructions of \emph{exact} unitary designs were left open except special cases~\cite{RS2009,BNRT2020,BNZZ2019}. 
Approximate ones typically suffice in applications, but exact designs are more preferable in certain protocols especially when they are used multiple times in a single run of the protocol. If this is the case, the error from each approximate implementation accumulates and eventually spoils the protocol. 

One of such protocols is a randomized benchmarking (RB) protocol~\cite{EAZ2005,KLRetc2008,MGE2011}, a standard method for experimentally estimating quantum noise, where unitary $2$-designs are used multiple times. Although the RB is experimentally-friendly and is widely used in various experimental systems, it reveals only the average gate fidelity. To obtain more information about the noise, a number of variants were proposed and experimentally implemented (see, e.g., Ref.\,\cite{HHFFW2019} and the references therein), which are all based on $2$-designs. It is highly expected that, by using higher-designs, much more information about the noise in quantum systems can be extracted. To this end, explicit constructions of exact unitary $t$-designs are important.

Constructing exact designs is, however, by far non-trivial. The difficulty is illustrated by a spherical $t$-design, a random real unit vector analogous to a unitary $t$-design.
The existence of exact spherical $t$-designs was proven in a non-constructive manner more than three decades ago~\cite{SZ1984}. Since then, more concise proofs and explicit constructions have been under intense investigation in combinatorial mathematics (see e.g., Refs.\,\cite{RB1991,WV1991,BB2009,BRV2013,CXX2019} and the references therein). In particular, it was only recently that constructions in general cases~\cite{X2020} and explicit constructions, in the sense that all the algorithms are given in a computable form~\cite{BNOZ2020}, were proposed.
Finding explicit constructions of exact unitary designs, since they are more complicated than spherical designs, is a rather non-trivial problem.

In this paper, we provide for the first time an explicit quantum circuit that generates an exact unitary $t$-design for any $t$ on the arbitrary number $N$ of qubits. 
More specifically, we show that an exact unitary $t$-design on $d$-dimensional Hilbert space, i.e., \emph{qudit}, can be generated from those on smaller spaces, which is obtained based on the recent mathematical results by some of the authors~\cite{BNOZ2020}.
Using this result, we provide an inductive construction of quantum circuits for exact unitary $t$-designs on $N$ qubits: we first construct a unitary $t$-design on a single qubit and then extend it to $N$ qubits. Unfortunately, the circuit fails to be efficient, but is still of practical use when the size of the system is small. 

As an application of exact unitary designs, we introduce the \emph{$t$-th order RB}, or the \emph{$t$-RB} for short, that harnesses the power of exact unitary $2t$-designs. The standard RB corresponds to the $1$-RB. The $t$-RB enables us to experimentally characterize the higher order properties of quantum noises in the manner free from state-preparation and measurement (SPAM) errors.
We especially investigate the $2$-RB in detail and show that it reveals \emph{self-adjointness} of the noise in the system. The self-adjointness is a new metric of the noise related to the feasibility of quantum error correction (QEC): small self-adjointness implies that the noise cannot be approximated by any stochastic Pauli noise. 
The noise on the system being stochastic Pauli is desirable both in theory and in practice. Stochastic Pauli noises are the commonly-used noise models in theoretical studies of QEC, since they are easy to numerically handle, and the properties of QEC, such as error thresholds and logical error rates, for  stochastic Pauli noises are well-understood. Also, there is a practical advantage if the noise on the system is stochastic Pauli since they can be corrected simply by applying Pauli operators, making the error correcting scheme easier in general.


After numerically demonstrating the feasibility of the $2$-RB, we perform the $2$-RB in a superconducting system and estimate the self-adjointness of background noise, showing that the $2$-RB experiments are feasible. From the experiment, we find that the interactions with adjacent qubits especially decrease the self-adjointness, which may lead to degradation of the performance of QEC with standard decoders. Hence, either improving the system or extending the noise model in theoretical studies of QEC, or both, is important for further experimental developments of quantum information technology.\\

This paper is organized as follows. In Sec.\,\ref{S:design}, we provide a general introduction of unitary $t$-designs. 
Our main results are summarized in Sec.\,\ref{S:main1} for the quantum-circuit construction of exact unitary $t$-designs, and in Sec.\,\ref{S:main2} for the $t$-RB protocols. 
A summary of the experiment of the $2$-RB is provided in Sec.\,\ref{S:Main3}.
After we explain the structure of the remaining paper in Sec.\,\ref{S:structure}, we provide a proof of the explicit construction in Sec.\,\ref{Sec:ConstDesign} and the theory of the $t$-RB in Sec.\,\ref{S:higher-orderRB}. The details of the experiment are provided in Sec.\,\ref{S:NumExpDetail}.
We conclude our paper with summary and discussions in Sec.\,\ref{S:SD}. Technical statements are provided in Appendices.

\section{Unitary $t$-designs} \label{S:design}

Let ${\sf U}(d)$ be the unitary group of degree $d < \infty$. The Haar measure ${\sf H}$ on ${\sf U}(d)$ is the unique unitarily invariant measure on the unitary group, i.e., it satisfies
\begin{multline}
\forall \mathcal{W} \subset {\sf U}(d), \forall V \in {\sf U}(d), \\
{\sf H}(V \mathcal{W}) = {\sf H} (\mathcal{W}V) = {\sf H}(\mathcal{W}).
\end{multline}
When it is needed to clarify the degree of the unitary group, we denote the Haar measure by ${\sf H}(d)$.

A unitary $t$-design ${\sf U}_t(d)$ is defined by a finite set of unitaries that mimics the $t$-th order statistical moment of the Haar measure ${\sf H}$.
Amongst several equivalent definitions~\cite{L2010}, we here adopt the following definition.

\begin{Definition}[Unitary $t$-design] \label{Def:des}
For $t \in \mathbb{Z}^+$, a finite set ${\sf U}_{t}(d) \subset {\sf U}(d)$ of unitaries is a unitary $t$-design if 
\begin{equation}
\mathbb{E}_{U \sim {\sf U}_{t}(d)} [U^{\otimes t} \otimes U^{\dagger \otimes t}]
=
\mathbb{E}_{U \sim {\sf H}(d)} [U^{\otimes t} \otimes U^{\dagger \otimes t}], \label{Eq:tDesign}
\end{equation}
where $\mathbb{E}_{U \sim {\sf U}_{t}(d)}$ is a uniform average over ${\sf U}_t(d)$, and $\mathbb{E}_{U \sim {\sf H}(d)}$ is the average over the Haar measure.
\end{Definition}

From an operational viewpoint, this definition implies that a unitary $t$-design cannot be distinguished from a Haar random unitary on average even when $t$ copies of the unitary are given. To clasify this, let us define a quantum operation, i.e., a completely-positive and trace-preserving (CPTP) map, $\cG^{\mu}_t$ by
\begin{equation}
\cG_{\mu}^{(t)}(\varrho) := \mathbb{E}_{U \sim \mu} \bigl[ U^{\otimes t} \varrho U^{\otimes t \dagger} \bigr], \label{Eq:gg}
\end{equation}
for any quantum state $\varrho$ on $t$ qudits, where $\mu$ is either the Haar measure ${\sf H}(d)$ on a qudit or a uniform measure over a unitary $t$-design ${\sf U}_t(d)$. Then, we can show that Definition~\ref{Def:des} is equivalent to that  (see e.g.\,\cite{L2010})
\begin{equation}
\cG^{(t)}_{\sf H} = \cG^{(t)}_{{\sf U}_t}. \label{Eq:g}
\end{equation}
This implies that, in \emph{any} experiments that use $t$ copies of a random unitary, no difference will be observed on average when a $t$-design is used instead of the Haar one.

For instance, let us consider the probability distribution $\{ p_i(U) := \tr[ P_i U \rho U^{\dagger}] \}$ when a one-qudit state is measured by a given POVM $\{P_i \}_i$ after the application of a unitary $U$. By setting the $t$-qudit state $\varrho$ in Eq.\,\eqref{Eq:gg} to $\rho^{\otimes t}$ and using Eq.\,\eqref{Eq:g}, it follows that, for any $s = 1, \dots, t$,
\begin{equation}
\mathbb{E}_{U \sim {\sf H}} \biggl[ \prod_{r=1}^s p_{i_r} (U) \biggr] = \mathbb{E}_{U \sim {\sf U}_t} \biggl[ \prod_{r=1}^s p_{i_r} (U) \biggr].
\end{equation}
Thus, the distribution of the measurement outcomes for a Haar random unitary and that for an unitary $t$-design exactly coincide up to the $t$-th order on average. Note that this is merely an example, and Eq.\,\eqref{Eq:g} implies much more: a Haar random unitary cannot be differentiated from a unitary $t$-design even by more complicated experiments over $t$ qudits.

The existence of an exact unitary $t$-design on ${\sf U}(d)$ for any $t$ and $d$ follows from the Carath\'{e}odoty's theorem and the fact that the dimension of the space, on which $U^{\otimes t} \otimes U^{\dagger \otimes t}$ is defined, is finite. Note however that the proof indicates only the existence of an exact unitary $t$-design. How to explicitly construct an exact unitary $t$-design has been a highly non-trivial problem.

In our construction, it is convenient to introduce a \emph{strong} unitary $t$-design. 

\begin{Definition}[Strong unitary $t$-design]
For $t \in \mathbb{Z}^+$, a finite set ${\sf U}_{\leq t}(d)$ of unitaries on ${\sf U}(d)$ is called a strong unitary $t$-design if
\begin{equation}
\mathbb{E}_{U \sim {\sf U}_{\leq t}(d)} [U^{\otimes r} \otimes U^{\dagger \otimes s}]
=
\mathbb{E}_{U \sim {\sf H}} [U^{\otimes r} \otimes U^{\dagger \otimes s}],
\end{equation}
for $0 \leq \forall r \leq t$ and $0 \leq \forall s \leq t$.
\end{Definition} 

Clearly, a strong unitary $t$-design is a unitary $t$-design.
Unlike standard unitary designs, strong unitary designs do not have a clear operational interpretation in quantum information processing, but we use it in the intermediate step of our construction.

\section{Main result 1 -- quantum circuits for exact unitary designs --} \label{S:main1}

In this section, we provide explicit constructions of strong unitary $t$-designs for any $t$. In particular, a quantum circuit for a strong unitary $t$-design on $N$ qubits is provided.
We start with preliminaries in Subsec.\,\ref{SS:Pre}, and provide the construction in Subsec.\,\ref{SS:InductiveConst}. We then comment on the circuit complexity of the construction in Subsec.\,\ref{SS:Efficiency}.

\subsection{Preliminaries} \label{SS:Pre}

Unitary designs have been studied in terms of representation theory~\cite{RS2009,RS2011} since the operator $U^{\otimes t} \otimes U^{\dagger \otimes t}$ in the definition can be regarded as a representation of the unitary group. Our construction is based on representation theory, where irreducible decomposition of the operator plays an important role. A brief introduction of irreducible representations (irreps) will be provided in Section~\ref{SS:unitaryrep}. Here, we mention a couple of well-known facts that are necessary to state our main result.

Any irrep of the unitary group can be indexed by a non-increasing integer sequence $\lambda :=(\lambda_1, \lambda_2, \dots, \lambda_d)$ of length $d$, i.e., $\lambda_i \in \mathbb{Z}$ for $i=1,\dots, d$, and $\lambda_1 \geq \lambda_2 \geq \dots \geq \lambda_d$~\cite{B2004}. In particular, \emph{spherical representations} of ${\sf U}(d)$ with respect to ${\sf K}:= {\sf U}(d_1) \times {\sf U}(d - d_1)$ are of great importance in the construction. Let $\Lambda_{\rm sph}(d_1,d,t)$ be a set of all non-increasing integer sequences $\lambda$ in the form of
\begin{equation}
    \lambda = (\lambda_1, \dots, \lambda_{d_1}, 0, \dots, 0, -\lambda_{d_1}, \dots, -\lambda_{1}),
\end{equation}
where $d_1 \leq d/2$ and $t \geq \lambda_1 \geq \dots \geq \lambda_{d_1} \geq 0$. The spherical representation is the irrep indexed by $\lambda \in \Lambda_{\rm sph}(d_1,d,t)$~\cite{GW2009}. For a spherical representation $\lambda \in \Lambda_{\rm sph}(d_1,d,t)$, a \emph{zonal spherical function} $Z^{(d_1)}_{\lambda}(z_1, \dots, z_{d_1})$ is defined by the unique bi-${\sf K}$-invariant function~\cite{JC1974,R2010,BNOZ2020}. The zonal spherical functions are a certain type of symmetric polynomials, and can be explicitly written down (see, e.g., Appendix A of Ref.\,\cite{BNOZ2020}).

\subsection{Inductive constructions} \label{SS:InductiveConst}

Our main technical result is to construct a strong unitary $t$-design on ${\sf U}(d)$ from those on ${\sf U}(d_1)$ and on ${\sf U}(d-d_1)$.

\begin{Theorem} \label{Thm:General}
    Let $d_1$ be a positive integer such that $d_1 \leq d/2$. Define a set of unitaries ${\sf W}_{d_1 \oplus d-d_1}$ in ${\sf U}(d)$ by 
    \begin{equation}
	    {\sf W}_{d_1 \oplus d-d_1}:= \{ U \oplus V | U \in {\sf U}_{\leq t}(d_1), V \in {\sf U}_{\leq t}(d-d_1)\},
	\end{equation}
	where ${\sf U}_{\leq t}(d_1)$ and ${\sf U}_{\leq t}(d-d_1)$ are strong unitary $t$-designs on ${\sf U}(d_1)$ and ${\sf U}(d-d_1)$, respectively. 
	Let $\boldsymbol{\theta}_{\lambda} := (\theta_{\lambda}^{(0)}, \dots, \theta_{\lambda}^{(d_1-1)})$ $(\theta_{\lambda}^{(i)} \in [0, \pi/2])$ be such that
	\begin{equation}
	    Z_{\lambda}^{(d_1)}\bigl( \cos^2 \theta_{\lambda}^{(0)}, \dots, \cos^2 \theta_{\lambda}^{(d_1-1)} \bigr) = 0,
	\end{equation}
	where $Z_{\lambda}^{(d_1)}$ is the zonal spherical function for $\lambda \in  \Lambda_{\rm sph}(d_1,d,t)$.
	Let $R_{\lambda}$ be a unitary defined by
	\begin{equation}
	R_{\lambda} = 
	    \begin{pmatrix}
		C(\boldsymbol{\theta}_{\lambda}) & i S(\boldsymbol{\theta}_{\lambda}) & 0 \\
		i S(\boldsymbol{\theta}_{\lambda}) & C(\boldsymbol{\theta}_{\lambda}) & 0 \\
		0 & 0 & I_{d-2d_1}
        \end{pmatrix},
	\end{equation}
	where $C(\boldsymbol{\theta}_{\lambda}) = {\rm diag}(\cos\theta_{\lambda}^{(0)}, \dots, \cos\theta_{\lambda}^{(d_1-1)})$ and $S(\boldsymbol{\theta}_{\lambda}) = {\rm diag}(\sin\theta_{\lambda}^{(0)}, \dots, \sin\theta_{\lambda}^{(d_1-1)})$, and $I_{d-2d_1}$ is the identity matrix of size $d- 2d_1$.
	Then, 
	\begin{equation}
	     {\sf W}_d := {\sf W}_{d_1 \oplus d-d_1} \prod_{\lambda \in \Lambda_{\rm sph}(d_1,d,t)} (R_{\lambda} {\sf W}_{d_1 \oplus d-d_1}) \label{Eq:W}
	\end{equation} 
	is a strong unitary $t$-design on ${\sf U}(d)$.  
\end{Theorem}

Theorem~\ref{Thm:General} follows from a more general result~\cite{BNOZ2020} shown by some of the authors, which works not only for the unitary group but also for a broader class of compact groups. For the sake of completeness, we provide a direct proof of Theorem~\ref{Thm:General} in Sec.\,\ref{Sec:ConstDesign}.

We then claim that 
\begin{equation}
{\sf W}_1=\{ 1, \omega, \omega^2, \dots, \omega^t \}, \label{Eq:W_1}
\end{equation}
where $\omega=\exp[\frac{2\pi}{t+1}]$ is the $(t+1)$-th root of unity, is a strong unitary $t$-design on ${\sf U}(1)$ for any $t$. This is easily checked by direct calculations:
\begin{align}
    &\mathbb{E}_{U \sim {\sf W}_1}[ U^{\otimes r} \otimes U^{\dagger \otimes s} ]
    =\frac{1}{t+1} \sum_{z \in {\sf W}_1} z^{r} \bar{z}^s = \delta_{rs},\\
    &\mathbb{E}_{U \sim {\sf H}(1)}[ U^{\otimes r} \otimes U^{\dagger \otimes s} ]
    =\int_{U(1)} z^r \bar{z}^s dz=\delta_{rs},
\end{align}
where $\delta_{rs}$ is the Kronecker delta.
Hence, we have $\mathbb{E}_{U \sim {\sf W}_1}[ U^{\otimes r} \otimes U^{\dagger \otimes s} ] = \mathbb{E}_{U \sim {\sf H}(1)}[ U^{\otimes r} \otimes U^{\dagger \otimes s} ]$ for $0 \leq  \forall s, r \leq t$, implying that ${\sf W}_1$ is a strong unitary $t$-design.

From Theorem~\ref{Thm:General} and ${\sf W}_1$, a strong unitary $t$-design on a qudit can be inductively constructed.

\begin{Corollary}\label{Cor:tDesign_qudit}
For $d \geq 1$, let ${\sf W}_{1 \oplus d-1}$ be a set of unitaries given by
\begin{equation}
{\sf W}_{1 \oplus d-1}= \{ z \oplus V | z \in {\sf W}_1, V \in {\sf U}_{\leq t}(d-1)\},
\end{equation}
where ${\sf W}_1 = \{ 1, \omega, \dots, \omega^t \}$ with $\omega$ being the $(t+1)$-th root of unity, and $\theta_{\lambda} \in [0, \pi/2]$ be such that
\begin{equation}
    Z^{(1)}_{\lambda}(\cos^2 \theta_{\lambda}) =0.
\end{equation}
Using a unitary $R_{\lambda} = e^{i \theta_{\lambda} X} \oplus I_{d-2}$, where $X$ is the Pauli-$X$ operator, we obtain that
\begin{equation}
    {\sf W}_{1 \oplus d-1} \prod_{\lambda \in \Lambda_{\rm sph}(1,d,t)} (R_{\lambda} {\sf W}_{1 \oplus d-1})
\end{equation} 
is a strong unitary $t$-design on a qudit.
\end{Corollary}

In this construction, it is important to obtain zeros for the zonal spherical functions $Z_{\lambda}^{(1)}$. This is computationally feasible since they are polynomials of one variable and are explicitly given (see Appendix A of Ref.\,\cite{BNOZ2020}). Furthermore, $\Lambda_{\rm sph}(1,d,t)$ contains only $t$ elements. Hence, we need to solve $t$ polynomials with one variable, which is tractable as far as $t$ is not too large.\\

\begin{figure}
    \centering
    \includegraphics[width=0.48\textwidth]{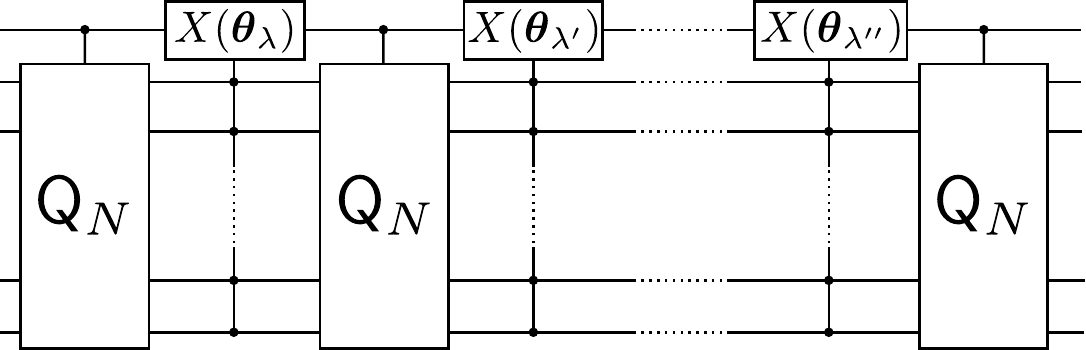}
    \caption{The quantum circuit that generates an exact unitary $t$-design on $N+1$ qubits from those on $N$ qubits. The unitary ${\sf Q}_N$ is a quantum circuit for an exact unitary $t$-design on $N$ qubits. The gate $X(\boldsymbol{\theta}_{\lambda})$ is the single-qubit $X$-rotation controlled by the other $N$ qubits, which corresponds to $R_X(\boldsymbol{\theta}_{\lambda})$ in the main text. Note that this gate can be decomposed into a sequence of two-qubit gates using a classical oracle that provides $\theta_{\lambda}^{(\boldsymbol{j})}$ from $\boldsymbol{j}$.
    The rotation angles $\boldsymbol{\theta}_{\lambda}$ are obtained by solving $Z_{\lambda}^{(D)}=0$, where $Z_{\lambda}^{(D)}$ is the zonal spherical function for the spherical representation $\lambda \in \Lambda_{\rm sph}(D, 2D, t)$ with $D = 2^N$. The number of the controlled-$X$ rotations is $|\Lambda_{\rm sph}(D, 2D, t)|$.
    By the inductive use of this quantum circuit in terms of $N$, we can decompose the circuit to that consisting only of two-qubit gates.
    }
    \label{Fig:QCforD}
\end{figure}

We now consider a strong unitary $t$-design on $N$ qubits.
Again using Theorem~\ref{Thm:General}, we obtain the quantum circuit on $(N+1)$ qubits based on that on $N$ qubits. See also Fig.\,\ref{Fig:QCforD}.

\begin{Corollary}\label{Cor:tDesign_qubit}
Let ${\sf Q}_{N}$ be a strong unitary $t$-design on $N$ qubits, and ${\rm Ctrl \mathchar`-}{\sf Q}_N$ be a set of controlled-unitaries on $N+1$ qubits, defined by
\begin{equation}
{\rm Ctrl \mathchar`-}{\sf Q}_N := \{\ketbra{0}{0} \otimes U_0 + \ketbra{1}{1} \otimes U_1 : U_0, U_1 \in {\sf Q}_N  \}.
\end{equation}
For $\lambda \in \Lambda_{\rm sph}(D,2D,t)$, where $D = 2^N$, let $\boldsymbol{\theta}_{\lambda}:=(\theta_{\lambda}^{(0)}, \dots, \theta_{\lambda}^{(D-1)})$ be such that 
\begin{equation}
Z^{(D)}_{\lambda}( \cos^2 \theta_{\lambda}^{(0)}, \dots, \cos^2 \theta_{\lambda}^{(D-1)}) = 0. \label{Eq:ZPF_qubits}
\end{equation}
Representing $\{0, \dots, D-1\}$ in binary form such as $\{\boldsymbol{j} \}_{\boldsymbol{j} \in \{ 0,1\}^N}$, we write $\theta_{\lambda}^{(j)}$ as $\theta_{\lambda}^{(\boldsymbol{j})}$. Let $R_X(\boldsymbol{\theta}_{\lambda})$ be a single-qubit $X$-rotation controlled by $N$ qubits, defined by
\begin{equation}
R_X(\boldsymbol{\theta}_{\lambda}) = \sum_{\boldsymbol{j} \in \{0, 1\}^N}  e^{i \theta_{\lambda}^{(\boldsymbol{j})} X} \otimes  \ketbra{\boldsymbol{j}}{\boldsymbol{j}}.
\end{equation}
Then,
\begin{equation}
{\rm Ctrl \mathchar`-}{\sf Q}_N \prod_{\lambda  \in \Lambda_{\rm sph}(D,2D,t)} \bigr( R_X(\boldsymbol{\theta}_{\lambda}) {\rm Ctrl \mathchar`-}{\sf Q}_N \bigl)
\end{equation} 
is a strong unitary $t$-design on $N+1$ qubits.
\end{Corollary}

Corollary~\ref{Cor:tDesign_qubit} implies that a quantum circuit for an exact unitary $t$-design can be inductively constructed from a strong unitary $t$-design on one qubit, i.e., that on ${\sf U}(2)$. 
Furthermore, a strong unitary $t$-design on ${\sf U}(2)$ can be constructed using Corollary~\ref{Cor:tDesign_qudit}. Thus, combining Corollaries~\ref{Cor:tDesign_qudit} and~\ref{Cor:tDesign_qubit}, we obtain a quantum circuit for an exact unitary $t$-design for any $t$ and on an arbitrary number of qubits.

Note that the circuit, constructed in this way, can be explicitly decomposed into two-qubit gates. The controlled unitary ${\rm Ctrl \mathchar`-}{\sf Q}_N$ part contains up to three-qubit gates, if the circuit ${\sf Q}_N$ on $N$ qubit is already decomposed into two-qubit gates. The three-qubit gates can be easily rewritten as a series of two-qubit gates.
Also, the $X$-rotation controlled by $N$ qubits, $R_X(\boldsymbol{\theta}_{\lambda})$, can be decomposed into a sequence of two-qubit gates of polynomial length using sufficiently many number of ancillary qubits, which is based on a classical oracle that computes the angle $\theta_{\lambda}^{(\boldsymbol{j})}$ from $\boldsymbol{j}$ (see Appendix~\ref{App:RX}).

In special cases, we can find a much more concise construction based on a similar technique.

\begin{Proposition}~\label{Prop:Clliford4}
Let ${\sf C}(4)$ be the Clifford group on $2$ qubits. There exists a fixed two-qubit unitary $U_c$, such that ${\sf C}(4) U_c {\sf C}(4)$ is an exact unitary $4$-design on $2$ qubits.
\end{Proposition}

Analytically, we can prove that there exist unitaries $V_1$ and $V_2$ such that ${\sf C}(4) V_1 {\sf C}(4) V_2 {\sf C}(4)$ is an exact unitary $4$-design on $2$ qubits\,\cite{BNOZ2020}. Also, an algorithm for computing the unitaries $V_1$ and $V_2$ is given. It, however, turns out from numerics that it is not necessary to apply two extra unitaries if we choose a proper unitary $U_c$, which leads to Proposition\,\ref{Prop:Clliford4}. An explicit form of the unitary $U_c$ is numerically obtained and is provided in Appendix~\ref{App:4DesClifford}. Note that the existence of $U_c$ is confirmed numerically, so the statement holds up to the numerical precision.

This construction is only for a $4$-design on $2$ qubits, but the number of unitaries in the $4$-design is much smaller than that of Corollary~\ref{Cor:tDesign_qubit}.
It is an open problem whether a similar construction works for higher-designs on a larger number of qubits.

\subsection{Efficiency and comparison with a Haar unitary} \label{SS:Efficiency}
 
To quantitatively evaluate the complexity of the quantum circuit for an exact unitary $t$-design obtained in Corollary~\ref{Cor:tDesign_qubit}, we provide an order estimate of the number $G(N)$ of two-qubit gates in the circuit. Assuming $2^N \gg t$ and using the fact that $|\Lambda_{\rm sph}(D, 2D, t)|= O(e^{\pi \sqrt{2t/3}})$ due to the Hardy and Ramanujan formula for the asymptotics of the number of partitions, we obtain 
\begin{equation}
    G(N) \approx \exp \biggl[\pi \sqrt{\frac{2t}{3}} (N-1) \biggr], \label{Eq:G}
\end{equation}
to the leading order of $N$. Hence, it is necessary to use exponentially many two-qubit gates as the number of qubits increases. This inefficiency of the quantum circuit may be intrinsic since the construction is inductive.

There is another source of inefficiency. In Corollary~\ref{Cor:tDesign_qubit}, it is necessary to find zeros of zonal spherical functions (see Eq.\,\eqref{Eq:ZPF_qubits}) for all $\lambda \in \Lambda_{\rm sph}(D, 2D, t)$. The zonal spherical function is given in terms of the summation of the (normalized) Schur polynomials (see Appendix A in Ref.\,\cite{BNOZ2020}). 
It is unlikely that the Schur polynomials have polynomial size algebraic formulas in general~\cite{SchurPoly2019}. Moreover, the number of variables for each $Z_{\lambda}^{(D)}$ is $D=2^N$. Hence, finding zeros of zonal spherical functions is computationally intractable.

In total, the construction for an exact unitary $t$-design on a large number of qubits is inefficient from both quantum-circuit and classical-computation viewpoints. We, however, think that our construction and our proof technique will form a solid basis of searching more efficient constructions of exact, as well as approximate, unitary designs. 
We also emphasize that, despite its inefficiency, our construction is of practical use on a few-qubit system, as we seek in the following sections.

Before we conclude this section, we comment on advantages of our construction of an exact unitary $t$-design over a direct implementation of a Haar random unitary. 
A naive way of implementing a Haar random unitary by a quantum circuit consists of three steps. First, we sample a Haar random unitary as a matrix by a classical computer. A classical algorithm for this is known~\cite{M2007}, but it is trivially inefficient since the size of the matrix is exponentially large. We then classically compute a decomposition of the unitary matrix into a sequence of two-qubit unitaries, providing a classical description of a quantum circuit for the unitary. This step is also inefficient, and the resulting quantum circuit is almost surely composed of the exponentially many number of two-qubit gates. Finally, we implement the circuit in practice.

This quantum circuit for a Haar random unitary is inefficient in terms of the number of qubits, and so, cannot be of practical use in a large system. Even in a small system, this naive implementation has a crucial difficulty that, every time a unitary is sampled, the above protocol outputs a quantum circuit with a rather different sequence of various two-qubit gates. This implies that, in each sampling, one needs to significantly modify the quantum circuit. This is in a sharp contrast to our quantum circuit for a unitary $t$-design based on Corollary~\ref{Cor:tDesign_qubit} since it has a \emph{fixed} structure. In each sampling, only what one needs to do is to randomly choose single-qubit gates, or more precisely elements of ${\sf U}(1)$ from ${\sf W}_1$ (see Eq.~\eqref{Eq:W_1}), and to plug them into the quantum circuit with a fixed structure. This will help practical implementations of the circuit in small systems.

It should be also noted that the single-qubit gates in our construction can be sampled from a discrete set, though sampling from a continuous gate set is necessary in the direct implementation of a Haar random unitary. This is another advantage of our construction.

\section{Main result 2 -- Higher-Order Randomized Benchmarking --} \label{S:main2}

We here introduce a higher-order generalization of the standard RB that uses exact unitary $2t$-designs. We call it the \emph{$t$-th order} RB, or simply $t$-RB. The standard RB corresponds to $1$-RB. From the higher-order RB, more information about the noise can be extracted. In particular, we show that a new characterization of the noise, which we call \emph{self-adjointness}, can be estimated from the $2$-RB.

Before we proceed, we emphasize that exact unitary designs, not approximate ones, are of crucial importance in the RB-type protocols. This is because the protocol uses unitary designs multiple times. Hence, if each unitary has an error due to the approximation, it accumulates in the whole process and results in a large error at the end. 
Since the goal of the RB-type protocol is typically very high, such as benchmarking the fidelity $> 95\%$, the error originated from the approximate designs would spoil the protocol. Hence, the use of exact unitary designs is of key importance. This point is more elaborated on in Subsec.\,\ref{SS:ExactImportant}.\\

In Subsec.\,\ref{SS:defdef}, we overview a couple of metrics of the noise, i.e., the average fidelity and unitarity, and introduce the self-adjointness. 
The importance of the self-adjointness in QEC is argued in Subsec.\,\ref{SS:impSA}. We then introduce the $t$-RB in Subsec.\,\ref{SS:tRB}.
We argue the importance of exact designs in more detail in Subsec.\,\ref{SS:ExactImportant}.
We focus on the $2$-RB in Subsec.\,\ref{SS:2RB} and show that the self-adjointness and the unitarity of the noise can be estimated from the $2$-RB at the same time.
We briefly comment on the scalability of the $t$-RB in Subsec.\,\ref{SS:Scalability}.

\subsection{Characterizing noises} \label{SS:defdef}

A noise $\cE$ acting on a $q$-qubit system is formulated by a completely-positive and trace-preserving (CPTP) map. Let $d$ be defined as $d:=2^q$.
The average fidelity and the unitarity are defined by
\begin{align}
&F(\cE) := \int d \varphi \bra{\varphi} \cE \bigl( \ketbra{\varphi}{\varphi} \bigr) \ket{\varphi},\\
&u(\cE) := \frac{d}{d-1} \int d \varphi \bigl|\! \bigr| \cE' ( \ketbra{\varphi}{\varphi} ) \bigl|\! \bigr|_2^2,
\end{align}
respectively, where $\cE'(\rho) := \cE(\rho - I/d)$, and $|\!| A |\!|_2 = (\tr[A^{\dagger} A])^{1/2}$ is the Schatten 2-norm.
The average fidelity satisfies $1/(d+1) \leq F(\cE) \leq 1$, and $F(\cE)=1$ if and only if the system is noiseless, i.e., $\cE$ is the identity channel, while the unitarity satisfies $0 < u(\cE) \leq 1$, and $u(\cE) = 1$ if and only if the noise is coherent, i.e., $\cE$ is a unitary channel. 
The unitarity is an important metric in the context of QEC since coherent noise is known to be hard to correct in general~\cite{KLDF2016,SWS2015,SFK2017}.

In the RB-type protocols, it is more natural to use a \emph{fidelity parameter} $f(\cE)$ rather than the average fidelity itself. It is defined by
\begin{equation}
f(\cE) = \frac{d F(\cE) - 1}{d-1},
\end{equation}
and satisfies $-1/(d^2-1) \leq f(\cE) \leq 1$.

We next introduce a self-adjointness of the noise. For any linear map $\cE$, an \emph{adjoint} map $\cE^{\dagger}$ is defined by $\tr[ A \cE(B)] =\tr[ \cE^{\dagger}(A) B]$. A noise $\cE$ is called self-adjoint if $\cE = \cE^{\dagger}$, which is equivalent to that all the Kraus operators of $\cE$ are self-adjoint.

The \emph{self-adjointness} $H(\cE)$ of the noise $\cE$ is defined by
\begin{equation}
H(\cE) := 1- \frac{d+1}{2d} \int d \varphi \bigl|\! \bigr| \cE( \ketbra{\varphi}{\varphi}) - \cE^{\dagger} ( \ketbra{\varphi}{\varphi}) \bigl|\! \bigr|_2^2.
\end{equation}
The normalization constant is chosen such that $0 \leq H(\cE) \leq 1$. Obviously, $H(\cE) = 1 $ if and only if $\cE$ is self-adjoint, i.e. $\cE=\cE^{\dagger}$.
Note that the self-adjointness has two contributions from the noisy map $\cE$, one is from the unital part and the other from the non-unital part. The non-unital part of the noise makes the self-adjointness less than one since, if $\cE$ is not unital, then $\cE^{\dagger}$ is not trace-preserving, which implies that $\cE \neq \cE^{\dagger}$.

To clearly separate the two contributions, we introduce a \emph{self-adjointness parameter} $h(\cE)$. Using $\cE'(\rho) = \cE(\rho- I/d)$, we defined it by
\begin{equation}
h(\cE) := \frac{d}{d-1} \int d \varphi \tr[\cE'(\varphi)   \cE'^{\dagger}(\varphi)].
\end{equation}
The self-adjointness parameter $h(\cE)$ is related to the self-adjointness $H(\cE)$ and the unitarity $u(\cE)$ by
\begin{align}
&H(\cE) = 1 - \frac{d^2-1}{d^2}\bigl( u(\cE) - h(\cE) \bigr) - \frac{d+1}{2d^2} | \alpha_{\cE} |^2, \label{Eq:66}
\end{align}
where $| \alpha_{\cE} |$ is a measure of the non-unital part of the noise (see Subsec.\,\ref{SS:L} for the definition). We can clearly observe that $H(\cE)$ consists of two factors, the unital part $h(\cE)$ and the non-unital part $|\alpha_{\cE}|$.

The three metrics of noises, namely, fidelity, unitarity, and self-adjointness, all capture different properties of the noises. The fidelity reveals the first order property of the noises, while the unitarity and the self-adjointness, which are independent to each other, reveal the second-order.
In order to improve noisy quantum devices, it is of crucial importance to obtain the information of noise as much as possible. Hence, it is certainly of practical use to introduce the self-adjointness as a new metric of noise. In addition, we argue in the next subsection that the self-adjointness has important implications for QEC.

\subsection{Importance of self-adjointness in QEC} \label{SS:impSA}

The most important family of self-adjoint noises is stochastic Pauli noises, whose Kraus operators are all proportional to Pauli matrices. In QEC, Pauli noises are the standard yet most important class of noises both in theory and in practice. From a theoretical perspective, Pauli noises are easy to numerically handle. Hence, most numerical calculations have been carried out by assuming Pauli noises, and it has been confirmed that QEC has preferable features, such as exponential decreases and threshold behaviors of logical error rates, if the noise is Pauli.

The noise being Pauli is also practically preferable in experimental realizations of QEC since it typically simplifies the decoding tasks.
This is especially the case for stabilizer codes, such as surface and color codes, whose standard decoders are to estimate what types of Pauli operators should be applied on which physical qubits during recovery operations.  
For stochastic Pauli noises, if the estimation goes well, the state is fully retrieved with high probability by applying Pauli operators to the suitable physical qubits. In contrast, it is not possible to fully correct non-Pauli noises by applying Pauli operators since they generate undesired coherence between different code spaces. Thus, QEC of non-Pauli noises generally suffers from degradation of logical error rates when the standard decoders are used~\cite{SFK2017,BEKP2018} or requires more complicated algorithms for retrieving the performance of QEC. Neither of them is preferable in practice since it induces additional experimental difficulties.

For these reasons, it is desirable to check that the noise on an experimental system is stochastic Pauli. To this end, the self-adjointness provides useful information since, if $H(\cE) \ll 1$, then the noise is far from self-adjoint and cannot be approximated by Pauli noises. 
This implies that the practical situation differs from the standard assumption in theoretical studies of QEC and incurs additional difficulties on decoding procedure. Thus, the self-adjointness provides practical information about the feasibility of QEC using Pauli-based decoders.

Note that the difficulty of QEC for non-Pauli noises, captured by the self-adjointness, highly depends on the assumptions in quantum error correction schemes. When \emph{any} decoding procedure is available, it would not be so important whether the noise is Pauli or non-Pauli. When this is the case, the unitarity will be a more suitable metric of noise relevant to the feasibility of QEC~\cite{KLDF2016,SWS2015,SFK2017}. Note also that non-Pauli noises can be always transformed to a Pauli noise by Pauli-twirling. However, Pauli-twirling induces additional noise onto the system and, as a result, the performance of QEC will degrade. Thus, it is practically desirable to manufacture the system so that the noise is stochastic Pauli.

We also provide a pedagogical example of noise, where performance of QEC can be directly captured by the self-adjointness but not by fidelity nor unitarity.
Consider a $\theta$-rotation error around the $X$-axis on one qubit, i.e., $\exp[i \theta X/2]$, where $X$ is the Pauli-$X$ operator. The average fidelity $F_{\theta}$ and the self-adjointness $H_{\theta}$ can be obtained as 
\begin{align}
1/3 = F_{\pi} &< F_{\pi/2}=2/3,\\
0 = H_{\pi/2} &< H_{\pi}=1.
\end{align}
The unitarity is $1$ for any $\theta$.

One may expect that the $\pi/2$-rotation error is easier to correct than the $\pi$-rotation since the former has higher fidelity than the latter.
However, this is not the case since $\pi$-rotation is simply a perfect bit-flip that can be trivially corrected, while the $\pi/2$-rotation error is known to be particularly hard to correct~\cite{DP2017}. 
Thus, neither the average fidelity nor the unitarity, which is $1$ for both errors, is a good metric of the error correctability. In contrast, the self-adjointness clearly captures whether the error can be corrected, at least in this case, since $H_{\pi/2}$ and $H_{\pi}$ are the minimum and the maximum values of the self-adjointness, respectively.


\subsection{General description of the $t$-th order RB} \label{SS:tRB}

We now introduce the $t$-RB using an exact unitary $2t$-design ${\sf U}_{2t}:= \{ U_i \}_i$. As is the case for the standard RB, we assume that the noise is gate- and time-independent, so that the noisy implementation of ${\sf U}_{2t}$ is given by $\{ \cG_i := \cE \circ \cU_i\}_i$, where $\cE$ is the CPTP map that represents the noise, and we used the notation that $\cU(\rho) := U \rho U^{\dagger}$.

Let $O_{\rm ini}$ and $O_{\rm meas}$ be the initial and measurement operators, respectively, which we assume to be Hermitian. 
We first apply a sequence of unitaries $U_{\boldsymbol{i}} = U_{i_m} \dots U_{i_1}$ onto the initial operator $O_{\rm ini}$. Each $U_{i_n}$ is chosen uniformly at random from ${\sf U}_{2t}$, which we denote by $U_{\boldsymbol{i}} \sim {\sf U}_{2t}^{\times m}$. We then apply its inverse $U_{i_{m+1}} := U_{\boldsymbol{i}}^{-1}$, and measure $O_{\rm meas}$. 

If the system is noiseless, $\cE = {\rm id}$, this protocol results in a trivial expectation value that
\begin{equation}
\tr \bigl[ O_{\rm meas} \cU_{i_{m+1}} \circ \cU_{i_m} \circ \dots \circ \cU_{i_1}(O_{\rm ini}) \bigr] = \tr[O_{\rm meas} O_{\rm ini}]
\end{equation}
due to the inverse unitary $U_{i_{m+1}}$. 
However, when the system is noisy, the expectation value becomes 
\begin{equation}
\langle O_{\rm meas} \rangle_{O_{\rm ini}, \boldsymbol{i}}
:=
\tr \bigl[ O_{\rm meas} \cG_{i_{m+1}} \circ \cG_{i_m} \circ \dots \circ \cG_{i_1}(O_{\rm ini}) \bigr],
\end{equation}
which in general differs from $\tr[O_{\rm meas} O_{\rm ini}]$. The basic idea of the RB-type protocol is to extract some information about the noise $\cE$ from the difference. 

In the $t$-RB, we especially focus on the average of the $t$-th power of the expectation value over all choices of the unitary sequence. That is,
\begin{equation}
V^{(t)}(m, \cE| O_{\rm ini}, O_{\rm meas}) :=
\mathbb{E}_{U_{\boldsymbol{i}} \sim {\sf U}_{2t}^{\times m}} \bigl[
\bigl( \langle O_{\rm meas} \rangle_{O_{\rm ini}, \boldsymbol{i}} \bigr)^t
\bigr].
\end{equation}
Using the representation-theoretic technique, it can be shown that $V^{(t)}(m, \cE| O_{\rm ini}, O_{\rm meas})$ is generally given in the following form:
\begin{equation}
V^{(t)}(m, \cE| O_{\rm ini}, O_{\rm meas}) 
= \sum_{\lambda} \tr \bigl[ \hat{A}^{(t)}_{\lambda}  \bigl( \hat{C}^{(t)}_{\lambda}(\cE) \bigr)^m \bigr], \label{Eq:general}
\end{equation}
where $\lambda$ labels the irreps of the unitary group, $\hat{A}^{(t)}_{\lambda}$ and $\hat{C}^{(t)}_{\lambda}(\cE)$ are $m_{\lambda} \times m_{\lambda}$ matrices  with $m_{\lambda}$ being the multiplicity of the irrep $\lambda$. 
This is well-known in the literature of RB-type protocols, but we provide a proof in Sec.~\ref{SS:generaltRB} for completeness.

Despite its abstract expression, Eq.\,\eqref{Eq:general} has an important implication that the matrix $\hat{C}^{(t)}_{\lambda}(\cE)^m$ depends only on $\cE$ and $m$, but not on $O_{\rm ini}$ and $O_{\rm meas}$. Hence, from the experimental data of $V^{(t)}(m, \cE| O_{\rm ini}, O_{\rm meas})$ for various $m$, it is in principle possible to estimate the matrix $\hat{C}^{(t)}_{\lambda}(\cE)$, which contains certain information of the noise $\cE$, in the way independent of $O_{\rm ini}$ and $O_{\rm meas}$. 

In practice, the most important situation is when the representation is multiplicity-free, i.e., $m_{\lambda} = 1$ for any $\lambda$. In this case, $V^{(t)}$ reduces to a much simpler form:
\begin{equation}
V^{(t)}(m, \cE| O_{\rm ini}, O_{\rm meas}) = \sum_{\lambda} A^{(t)}_{\lambda} \bigl(C^{(t)}_{\lambda}(\cE)\bigr)^m,  \label{Eq:tRBExpDecay}
\end{equation}
where $A^{(t)}, C^{(t)}_{\lambda}(\cE)  \in \mathbb{R}$. Note that $|C^{(t)}_{\lambda}(\cE)| \leq 1$ since $V^{(t)}$ is a bounded function. Hence, in this case, $V^{(t)}$ becomes a sum of some exponentially decreasing functions with respect to $m$.

To be more concrete, let us recall the standard RB, corresponding to the $1$-RB. As shown in Ref.\,\cite{EAZ2005}, $V^{(1)}$ is given by
\begin{equation}
V^{(1)}(m, \cE| O_{\rm ini}, O_{\rm meas}) = A_0^{(1)} + A_1^{(1)} f(\cE)^m, \label{Eq:korekore}
\end{equation}
where $A_0^{(1)}$ and $A_1^{(1)}$ depend only on $O_{\rm ini}$ and $\cE(O_{\rm meas})$, and $f(\cE)$ is the fidelity parameter of the noise $\cE$. 
Thus, by fitting experimentally obtained data of $V^{(1)}$ for different $m$ with the fitting function $F(m)=A + B \alpha^m$, we can estimate the fidelity parameter $f(\cE)$.

\subsection{Importance of exact designs in RB} \label{SS:ExactImportant}
In the RB protocol, it is important to use \emph{exact} unitary designs because designs are used many times, sometimes a few hundreds to a thousand, in a single run of the protocol.
To illustrate this, let us consider the $1$-RB when the unitary $2$-design in the protocol is $\epsilon$-approximate. 

Let $m$ be the length of the unitary sequence as above. It is straightforward to show that, 
\begin{multline}
    V^{(1)}(m, \cE| O_{\rm ini}, O_{\rm meas}) \\
    \approx
    A_0' + A_1 f^m + \epsilon (m-2) (E_2 f^2 + E_1 f + E_0) + O(m^2 \epsilon^2), \label{Eq:korekoredame}
\end{multline}
where $E_i$'s are some constants that depend on $O_{\rm ini}$, $\cE(O_{\rm meas})$, $f$, and how the design differs from the exact one. 
See Subsec.\,\ref{SS:1RBdetail} for the derivation.
Compared to the $1$-RB with exact ones, i.e., Eq.\,\eqref{Eq:korekore}, fitting this function with respect to $m$ is much harder since it is not a simple exponential decay.

The fitting may go well if $\epsilon \ll f^m/m$. This requires a very high precision of the design since $m$ can be a few hundreds in actual experiments. For instance, when $f=0.95$, the degree $\epsilon$ of approximation of the unitary design should be order $10^{-5}$ or so. Although it is possible to achieve this degree of approximation by a sufficiently long quantum circuit~\cite{BHH2016,NHKW2017,HMHEGR2020}, the RB becomes unpractical if we use such a long circuit at every use of a unitary design in the protocol and repeat it a few hundreds times.

There might be a possibility to improve Eq.\,\eqref{Eq:korekoredame} by using different constructions of approximate unitary designs at every step, by which the differences from the exact design may become random so that they cancel out in total. This will be an interesting question, but at this point, it is not clear if such a technique works. Also, even if it works, we need to assume additional structures of approximate constructions. 

The higher-order RB with approximate designs will incur more difficulty in practice. Since it uses higher moment of the outcomes, the fitting function becomes more complicated than Eq.\,\eqref{Eq:korekoredame} when one uses approximate designs. Similarly to the $1$-RB with approximate $2$-designs, much better degree of approximation, that is, longer quantum circuits, will be needed, which is not practical.
Thus, we conclude that exact unitary designs are of crucial importance in a practical implementation of the $t$-RB.

\subsection{Second-order RB} \label{SS:2RB}
We next focus on the $2$-RB using exact unitary $4$-designs, and show that the $2$-RB reveals the self-adjointness of the noise.
To this end, we set the initial operator $O_{\rm ini}$ to a traceless one, i.e., $\tr[O_{\rm ini}] = 0$. This setting, together with the fact that the noise is trace-preserving, makes the representation multiplicity-free (see Appendix~\ref{App:Irreps}). Hence, the expectation value $V^{(2)}(m, \cE| O_{\rm ini}, O_{\rm meas})$ for the $2$-RB is given by a sum of exponentially decaying functions as shown in Eq.\,\eqref{Eq:tRBExpDecay}.  

Note that the expectation value for a traceless initial operator can be obtained by performing the same experiment for two different quantum states $\rho$ and $\rho'$, and by taking the difference of the expectation values before they are squared. That is,
\begin{multline}
V^{(2)}(m, \cE| \Delta, O_{\rm meas})\\
=
\mathbb{E}_{U_{\boldsymbol{i}}} \bigl[
\bigl( \langle O_{\rm meas} \rangle_{\rho, \boldsymbol{i}}
-
\langle O_{\rm meas} \rangle_{\rho', \boldsymbol{i}}
\bigr)^2
\bigr]
\end{multline}
where $\Delta = \rho - \rho'$ is a traceless operator.

Our second main result in this paper is about $V^{(2)}(m, \cE| \Delta, O_{\rm meas})$ as summarized in Theorem~\ref{Thm:2RB}.

\begin{Theorem} \label{Thm:2RB}
In the above setting, $V^{(2)}(m, \cE| \Delta, O_{\rm meas})$ is given as follows.
For single-qubit systems,
\begin{multline}
V^{(2)}(m, \cE| \Delta, O_{\rm meas}) \\= A_0 u(\cE)^m + A_1 \biggl( \frac{9}{10}f(\cE)^2 - \frac{1}{5} u(\cE) + \frac{3}{10} h(\cE) \biggr)^m, \label{Eq:15}
\end{multline}
where $f(\cE), u(\cE)$, and $h(\cE)$ are the fidelity parameter, the unitarity, and the self-adjointness parameter of the noise $\cE$, respectively.
For multi-qubit systems,
\begin{equation}
V^{(2)}(m, \cE| \Delta, O_{\rm meas}) = A_0 u(\cE)^m + \sum_{\lambda={\rm I}, {\rm II}, {\rm III}} A_{\lambda} C_{\lambda}(\cE)^m, \label{Eq:16}
\end{equation}
where $0 \leq C_{\lambda}(\cE) \leq 1$ depend only on the noise $\cE$.
Moreover, they satisfy
\begin{multline}
\sum_{\lambda={\rm I}, {\rm II}, {\rm III}} D_{\lambda} C_{\lambda}(\cE) \\
=
\frac{(d^2-1)^2}{2} f(\cE)^2 - u(\cE) + \frac{d^2-1}{2} h(\cE), \label{Eq:19p}
\end{multline}
where
\begin{align}
&D_{\rm I} = \frac{d^2(d-1)(d+3)}{4},\\
&D_{\rm II} = \frac{d^2(d+1)(d-3)}{4},\\
&D_{\rm III} = d^2-1.
\end{align}
\end{Theorem}
See Subsec.\,\ref{SS:2RBdetail} for the proof.\\

In the single-qubit case, $V^{(2)}$ is a sum of two exponentially decaying functions with respect to $m$. Hence, from the double-exponential fitting of the experimental data of $V^{(2)}$, we can simultaneously estimate $u(\cE)$ and $\frac{9}{10}f(\cE)^2 - \frac{1}{5} u(\cE) + \frac{3}{10} h(\cE)$.
Since it can be shown that the former is not less than the latter, we can estimate which of the two decaying rates corresponds to which quantity without any ambiguity.
It is also possible to estimate the fidelity parameter $f(\cE)$ from the same data set by computing $V^{(1)}(m, \cE| \Delta, O_{\rm meas})$ because a unitary $2$-design is also a unitary $1$-design. Thus, from the experiment of the $2$-RB on a single qubit, all of $f(\cE),$ $u(\cE)$, and $h(\cE)$ can be estimated simultaneously.

In multi-qubit systems, $V^{(2)}$ has a little more complicated form and consists of four exponentially decaying functions. Also, the decaying rates do not directly correspond to neither the unitarity nor the self-adjointness parameter. We observe from Eq.\,\eqref{Eq:19p} that $h(\cE)$ can be obtained from a linear combination of the decaying rates $C_{\lambda}(\cE)$, the value of $u(\cE)$, and $f(\cE)$.

One may think that, in the case of multiple qubits, it is practically intractable to accurately fit four exponentially decaying functions from experimental data because each data point has an error.
This difficulty can be circumvented by choosing appropriate initial and measurement operators. By doing so, we can set some of $A_{\lambda}$ zero in the ideal situation (see Tab.\,\ref{Tab:coeff}). This allows us to estimate the decaying rates one by one. 
Note that the initial and measurement operators in Tab.\,\ref{Tab:coeff} are all diagonal in the computational basis. Hence, it suffices to perform the experiments for the four initial operators $\ket{00}, \ket{01}, \ket{10}$, and $\ket{11}$, with the measurement in the computational basis. From the data of these experiments, it is possible to reproduce all cases listed in Tab.\,\ref{Tab:coeff} by post-processing.

In the multi-qubit case, the ambiguity remains to decide which of the decaying rates corresponds to which quantity. This is the case even when we use the above step-by-step estimation of the rates since, for instance, it is not clear if the unitarity $u(\cE)$ is larger or smaller than $C_{\rm I}$. In this case, we need to additionally perform the unitarity benchmarking~\cite{WGHF2015,HHFFW2019} to separately estimate $u(\cE)$. 
If we have an estimated value of $u(\cE)$, the step-by-step estimation allows us to decide all decaying rates without any ambiguity.

See Sec.\,\ref{SS:NumericalEval} and Sec.\,\ref{SS:NumericalEvalDetail} for the performance of 2-RB in concrete cases.

\begin{table}[t!]
\centering
\begin{tabular}{c||c|c|c}
$(\Delta, O_{\rm meas})$ & $\bigl(ZZ, \ketbra{00}{00}\bigr)$ & $(ZZ, ZZ)$ & $\bigl(\rho_-, \rho_- \bigr) $\\ \hline \hline
$A_0$ & $1/5$ & 16/15 & 4/15 \\
$A_{\rm I}$ & $4/5$ & 48/5 & 41/15\\
$A_{\rm II}$ & $0$ & 16/3 & 1/3 \\
$A_{\rm III}$ & $0$& 0 & 2/3 \\
\end{tabular}
\caption{A table of coefficients $A_0, A_{\rm I}, A_{\rm II}$, and $A_{\rm III}$ appeared in Eq.\,\eqref{Eq:16}, for the $2$-qubit case. The first row provides a pair of the initial and measurement operators $(\Delta, O_{\rm meas})$. We have assumed that the average fidelity of the noise $\cE$ is close to $1$, so that the inverse unitary $U_{i_{m+1}}$ can be applied nearly noiseless (see Subsec.\,\ref{SS:2RBdetail} for the detail).
The operator $ZZ$ is $Z \otimes Z$, and $\rho_- := \ketbra{00}{00}-\ketbra{11}{11}$. By choosing proper operators, we can set some coefficients zero, so that experimental estimations of $C_{\lambda}(\cE)$ become easy.}\label{Tab:coeff}
\end{table}

\begin{figure*}[tb!]
    \centering
    \includegraphics[width=\textwidth]{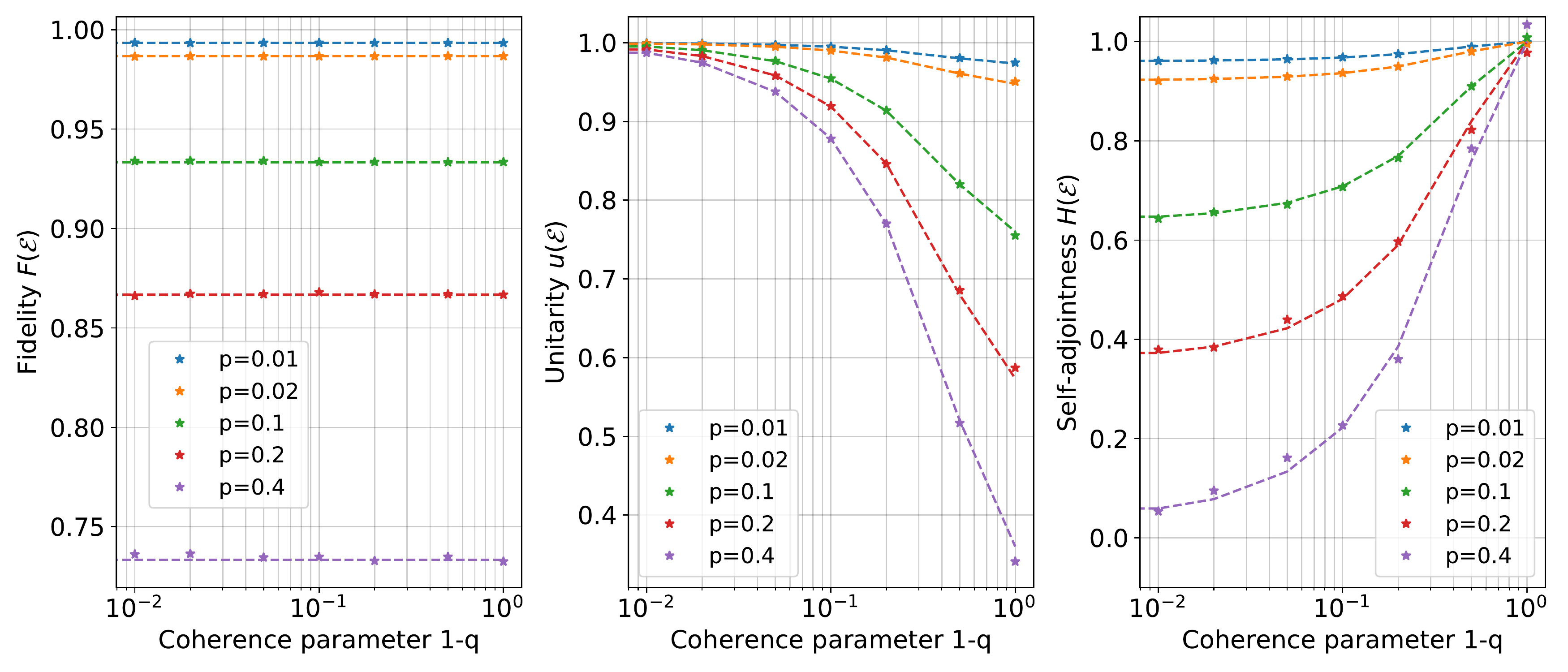}
    \caption{The estimated values of $F(\mathcal{E}_1), u(\mathcal{E}_1),$ and $H(\mathcal{E}_1)$ obtained by 2-RB on one-qubit system for various parameters $p$ and $q$, where we have taken $5000$ samplings both for measurement and for unitary sequences. The dots show the fitting results, and the dashed lines represent theoretical values.}
    \label{Fig:numerical_overview_1q}
\end{figure*}

\subsection{Scalability} \label{SS:Scalability}

The $t$-RB for $t \geq 2$ inherits most of the desired properties of the RB-type protocols. For instance, it is experimentally-friendly since, apart from using higher-designs, the difference of the $t$-RB from the standard RB ($1$-RB) is only taking the $t$-th power of the expectation value before the average. It is also true that the $t$-RB is free from SPAM errors (see Eqs.\,\eqref{Eq:general} and~\eqref{Eq:tRBExpDecay}).

The property that the standard RB does have and the $t$-RB does not in general is the scarability. This is for two reasons.
First, no efficient construction of exact unitary $2t$-designs is known for $t \geq 2$ so far.
Second, in the $t$-RB protocol, it is necessary to apply the inverse unitary at the end of the unitary sequence. Hence, we need to beforehand compute the inverse of each sequence. When the system is large, the task is intractable in general. This difficulty is avoided in the standard RB by using the Clifford group, which is an exact unitary $2$-design. Since the inverse is contained in the group, we can find the inverse relatively easily.
One may expect that the difficulty of finding the inverse could be also avoided in the $t$-RB by using the $2t$-design that is also a group, which is called a unitary $2t$-group~\cite{BNRT2020}. However, it is known that unitary $2t$-groups do not exist for $t \geq 2$ if the number of qubits $ \geq 3$. Thus, in the $t$-RB for $t \geq 2$, the hardness of finding the inverse  in a large system is inevitable.

Nonetheless, we emphasize that, in the current experimental situations, the RB-type protocols for more than three qubits are practically intractable due to the limitation of the coherent time. Thus, the experimental use of the RB-type protocols is currently aiming to characterize the noise on one- or two-qubit systems in a concise manner. Considering this fact, even if the $t$-RB is not scalable, it is practically useful and beneficial: it is as concise as the standard RB and provides more information about the noise, such as self-adjointness.

\section{Main result 3 -- $2$-RB in a superconducting system --} \label{S:Main3}
We finally implement the $2$-RB in a superconducting system and estimate the self-adjointness of background noise.
Unlike the analytical studies, the expectation values and the average over a unitary $4$-design cannot be taken with arbitrary precision in experiments since the number of repetitions of experiment is practically limited. To check that this limitation does not cause any problem in the evaluation of the self-adjointness, we start with numerically investigating the feasibility of the $2$-RB in Subsec.\,\ref{SS:NumericalEval}. We then provide a summary of experimental results in Subsec.\,\ref{SS:Experiments}.

In recent years, a number of experiments have been performed to characterize various noises on superconducting quantum systems in detail\,\cite{Wilen2021, HAN202110, McEwen2021, mcewen2021resolving}. From our experiments, we show that
the interactions with the adjacent qubits particularly decrease the self-adjoinenss and may cause problems toward realizations of QEC. In particular, our result implies that there exists a gap between the superconducting system and the common noise model used in theoretical studies of QEC, and also that the standard decoders of stabilizer codes may suffer from degradation of logical errors. Hence, toward the realization of QEC, it is desired to further improve the system or to develop the theory of QEC.

\subsection{Numerical evaluation} \label{SS:NumericalEval}

When the $2$-RB is practically implemented, there are two additional concerns. One is originated from the fact that the expectation value $\langle O_{\rm meas} \rangle_{O_{\rm ini}, \boldsymbol{i}}$ is obtained from a limited number of measurements in the basis of $O_{\rm meas}$, resulting in an error due to a finite number of measurements.
The other originates from the evaluation of the average $\mathbb{E}_{U_{\boldsymbol{i}} \sim {\sf U}_{4}^{\times m}}$ over the sequence of unitaries in the $4$-design. Ideally, all sequences in ${\sf U}_{4}^{\times m}$ should be taken, but practically, the average is often evaluated from a small subset in ${\sf U}_{4}^{\times m}$ of randomly chosen sequences, leading to an additional error of estimation.

Taking sufficiently many measurements and samplings of unitary sequences will reproduce the analytical results with high accuracy. However, it is complicated to analytically derive the numbers sufficient for achieving a desired accuracy. We hence perform numerical experiments and show that experimentally-tractable numbers of samplings are sufficient for a reliable $2$-RB.

\subsubsection{One-qubit cases}

In the case of single-qubit systems, we consider a specific noisy map given by 
\begin{equation}
\mathcal{E}_1(\rho) = q e^{i\theta X}  \rho e^{- i\theta X} + (1-q)((1-p) \rho + p X \rho X), \label{Eq:Noise1}
\end{equation}
which is characterized by three parameters $p, q, \theta$. 
The first term of the right-hand side represents a unitary part and the second term represents a stochastic part of the noise. A parameter $q$ determines a ratio between them. Hence, we can consider $q$ as a coherent parameter of noise, e.g., noise is unitary when $q=1$ and is a probabilistic Pauli noise when $q=0$. 
The parameters $\theta$ and $p$ represent the rotation angle of the unitary part and the error probability of the stochastic part, respectively. For simplicity, we choose $\theta$ such that the fidelity parameters of unitary and stochastic parts are equal, that is, $p = \sin^2 \theta$. Then, the fidelity parameter $f(\mathcal{E}_1)$ becomes independent of the coherent parameter $q$. 

To perform the $2$-RB for this noise, we may use the exact $4$-design constructed in Corollary~\ref{Cor:tDesign_qubit}. However, it is known that the icosahedral group, which we denote by ${\sf I}$, forms an exact $4$-design on one qubit~\cite{RS2009}. Since the icosahedral group has less cardinality than our inductive construction, we use it in the following analysis.

The numerical results for the $2$-RB on a single qubit are shown in Fig.\,\ref{Fig:numerical_overview_1q}. For each sequence length $m$, we have taken $5000$ random unitary sequences from ${\sf I}^{\times m}$ and have had $5000$ measurements to obtain a single data point of $V^{(1)}$. A detailed fitting procedure is provided in Subsec.\,\ref{SS:NumericalEvalDetail}. 

To check the accuracy of the $2$-RB, we consider the relative errors $|y-\tilde{y}|/(1-y)$, where $y$ and $\tilde{y}$ are the theoretical value and the fitting value, respectively. Note that $1-y \sim 0$ for all the fitting values when a fidelity close to unity is achieved.
For almost all data points of $F(\mathcal{E}), u(\mathcal{E}),$ and $H(\mathcal{E})$, we find that the relative errors are less than $5.0\%$, except the case when $p$ is large, or equivalently, when the fidelity is small. The relative error becomes moderately large, such as $35\%$, when $p=0.4$ and $q\geq 0.1$, corresponding to $F(\cE_1)=0.7$.
This is because the decaying rate of the second term in Eq.\,\eqref{Eq:15} is rather small, making the fitting difficult. However, such a case is not practically relevant since the fidelity is typically $>90\%$. Thus, we conclude that the 2-RB on 1-qubit systems works well in practice.

To analyze the dependence of the accuracy of the $2$-RB on the number of measurements and samplings of random unitary sequences, we additionally perform the 2-RB on one qubit with the various numbers of measurements and samplings. The results are summarized in Tab.\,\ref{Tab:SR_shift_for_1q}, where we set the noise parameters to $p=0.02$ and $q=0.02$. From these results, it appears that setting the numbers of measurements and samplings of random sequences to a few hundreds is sufficient for a good estimate. These results further indicate that increasing the number of random sequences rather than the number of measurements is preferable to improve the accuracy. See Subsec.\,\ref{SS:NumericalEvalDetail} for the details.

\begin{table}[t!]
\centering
\caption{Numerically estimated values of the fidelity, unitarity, and self-adjointness from the single-qubit $2$-RB with the finite numbers of measurements and samplings of unitary sequences. We set $p=q=0.02$. The theoretical values are shown at the bottom of the table.}
\begin{tabular}{c||c|r|r|r}
\# of meas. & \# of sequences & $F$ & $u$ & $H$ \\ \hline \hline
10 & 100 & $0.986(6)$ & $0.9985(5)$ & $0.92(6)$\\ 
10 & 500 & 0.986(1) & 0.9984(1) & 0.93(1)\\
10 & 1000 & 0.986(6) & 0.9984(3) & 0.92(3) \\
100 & 100 & 0.986(3) & 0.9980(8) & 0.92(2) \\
100 & 500 & 0.986(5)& 0.9980(4)& 0.92(3)\\
100 & 1000 & 0.986(6) & 0.9979(4)& 0.92(1)  \\
1000 & 100 & 0.986(3) & 0.9980(1)& 0.92(3)\\
1000 & 500 & 0.986(4) & 0.9979(5)& 0.92(3)\\
1000 & 1000 & 0.986(6) & 0.9978(9)& 0.92(2) \\ \hline
 & & 0.9866 & 0.99793 & 0.9247
\end{tabular}
\label{Tab:SR_shift_for_1q}
\end{table}

\subsubsection{Two-qubit cases}

\begin{figure*}[tbh!]
    \centering
    \includegraphics[width=\textwidth]{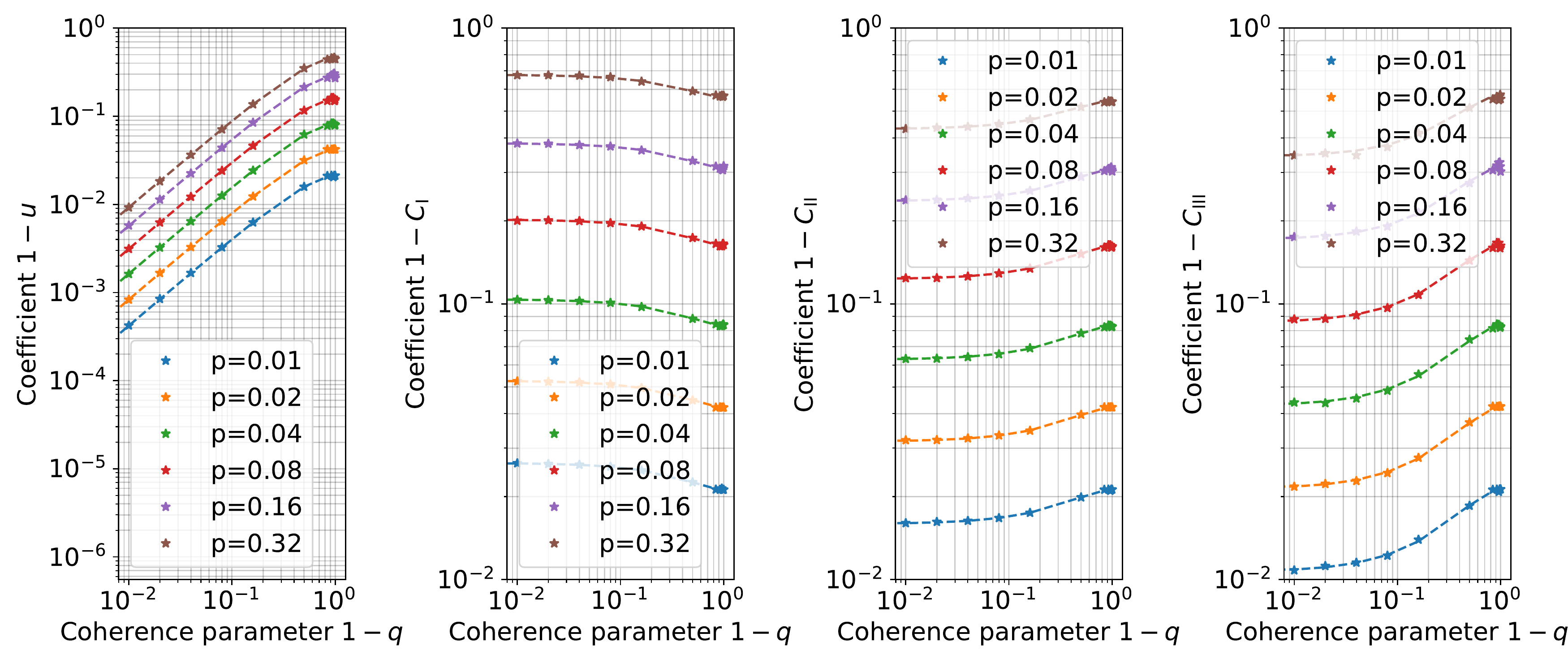}
    \caption{Four coefficients for exponentially decaying functions in the 2-RB are plotted according to coherence parameter $q$, where we had $10^4$ measurement and $10^4$ samplings of unitary sequences. Dashed lines are theoretical values. Each color corresponds to each value of the parameter $p$.}
    \label{Fig:numerical_overview_2q}
\end{figure*}

For two-qubit systems, we consider the noise given by
\begin{align}
\mathcal{E}_2(\rho) &= q e^{i \theta (X\otimes X)}  \rho e^{- i\theta (X \otimes X)} \nonumber \\ 
&+ (1-q)((1-p) \rho + p (X \otimes X) \rho (X\otimes X) ),
\end{align}
which is similar to the one-qubit case. We choose $\theta$ as $p = \sin^2 \theta$, so that $f(\mathcal{E}_2)$ is independent of the coherent parameter $q$. In this case, we use the construction of exact unitary $4$-designs given in Proposition~\ref{Prop:Clliford4}.

In the two-qubit case, it is needed to fit the experimental data by a sum of four exponentially decaying functions, which is in general not easy especially when each data point has errors caused by the finite number of measurements and samplings of unitary sequences. To avoid this difficulty, we use the method explained in Subsec.\,\ref{SS:2RB}, and determine the four decaying rates, i.e., $u(\cE_2), C_{\rm I}(\cE_2), C_{\rm II}(\cE_2)$, and $C_{\rm III}(\cE_2)$ in Eq.\,\eqref{Eq:16}, one by one.

The results are shown in Fig.\,\ref{Fig:numerical_overview_2q}. 
We have taken $10^4$ random unitary sequences for each sequence length $m$ and the parameters $p,q$. A detailed process of fittings are explained step by step in Sec.\,\ref{SS:NumericalEvalDetail}. In the figure, fitted values are shown as data points. Dashed lines are drawn with theoretically calculated values. 

Similarly to the case of the single-qubit $2$-RB, we have checked the relative errors of the fitting results to the theoretical values. The errors are all below $3\%$ for all the points except the case when the theoretical value is exactly zero. 
As in the case of single-qubit 2-RB, when we calculate $F(\mathcal{E}), u(\mathcal{E}),$ and $H(\mathcal{E})$ from the fitting values, the relative values of almost all the data points are less than $4.0\%$. While the relative errors become large when $p=0.4$, such a case is not a problem in typical calibration scenario. Thus, the $2$-RB works in actual situations also in the case of two-qubit systems.


\subsection{Experimental implementations of the 2-RB} \label{SS:Experiments}
We demonstrate the $2$-RB in a superconducting-qubit system.
We first explain the setup of our experiments, and then verify the feasibility of the $2$-RB experiment by comparing the unitarity obtained from the $2$-RB with that from the unitarity benchmarking (UB)~\cite{WGHF2015,HHFFW2019}. 
We finally characterize background noise of the system. As the background noise is gate- and time-independent, it satisfies the assumptions of the $2$-RB (see Subsec.\,\ref{SS:ExperimentDetail} for the detail).

\subsubsection{Experimental setup}

We use two superconducting qubits ($Q_1$ and $Q_2$) coupled with each other via an electric dipole interaction, which are a part of our $16$-qubit device~\cite{tamate2021scalable}. In all the experiments below, we use the qubit $Q_1$ as a target qubit of the single-qubit $2$-RB and, in some experiments, $Q_2$ as an environmental qubit that induces additional error onto $Q_1$.

The simplified system Hamiltonian $H$ is formulated as follows,
\begin{align}
\frac{H}{\hbar} = \frac{\omega_1}{2} Z\otimes I + \frac{\omega_2}{2} I\otimes Z + \frac{\chi_{ge}}{2} Z\otimes Z,
\label{Eq:ExpHam}
\end{align}
where $\omega_i/2\pi$ is the eigenfrequency of the $i$-th qubit and $\chi_{ge}/2\pi=-0.760~{\rm MHz}$ is an effective interaction strength between the qubits~\cite{gambetta2006qubit}.
It can be interpreted that the eigenfrequency of $Q_1$ switches depending on the quantum state of $Q_2$.
When $Q_2$ is in the $\ket{0}$ ($\ket{1}$) state, $Q_1$ has the eigenfrequency $(\omega_1 + \chi_{ge})/2\pi$  ($(\omega_1 - \chi_{ge})/2\pi$).
In the Bloch sphere representation, the state vector of the qubit rotates around the $Z$-axis with its eigenfrequency as the angular velocity.

We use a local oscillator synchronized with the eigenfrequency of the qubit for observation.
The state vector is stationary in a rotating frame of the local oscillator since the $Z$-axis rotation speed of the Bloch vector matches with that of the measurement basis.
The rotation frame picture also holds when the qubit $Q_1$ couples to the adjacent qubit $Q_2$ when the qubit $Q_2$ is in the $\ket{0}$ or $\ket{1}$ state.
For instance, when the qubit $Q_2$ is always in the $\ket{0}$ state, the eigenfrequency of $Q_1$ is $(\omega_1 + \chi_{ge})/2\pi$. We can detune the frequency of the local oscillator from the qubit frequency $\omega_1$ by $\chi_{ge}$ to make the state vector of $Q_1$ stationary.

It is, however, impossible to keep track of the eigenfrequency of the qubit
when the state of the adjacent qubit varies. This results in an inevitable $Z$-rotation occurring in the quantum state.
In an actual experiment involving multiple qubits, the frequency of the local oscillator is usually set to $\omega_1/2\pi$ to minimize the average $Z$-rotation angle.
See Subsec.\,\ref{SS:ExperimentDetail} for the detail.

\subsubsection{Comparison with the UB}

\begin{table}[t!]
\centering
\caption{The estimated values of the fidelity, unitarity, and self-adjointness from the experiment of the $2$-RB and that of the UB in the superconducting qubit system.}
\begin{tabular}{c||c|c|c|c|c}
Experiment  & Group         & $F$       & $u$      & $H$ \\ \hline \hline
$2$-RB      & Icosahedral   & 0.926(6)  & 0.970(1) & 0.6(1) \\
UB          & Clifford      &    --       & 0.977(1) &  --  \\ \hline
Theoretical  &  --   &    0.936  & 1 &  0.655
\end{tabular}
\label{Tab:Experiment_Sanity_Check}
\end{table}

In the experiment aiming to compare the $2$-RB and the UB on a single qubit, we use only $Q_1$ and add an artificial noise after applying each gate. 
The isolation of the qubit $Q_1$ from the qubit $Q_2$ can be done by keeping the qubit $Q_2$ in the state $\ket{0}$ and by setting the frequency of the local oscillator to $(\omega_1 + \chi_{ge})/2\pi$, which effectively cancel the coupling between $Q_1$ and $Q_2$. About the noise, we especially choose a single-qubit $Z$-rotation by angle $0.2 \pi$, denoted by $R_Z(0.2\pi)$.

Both in the case of the $2$-RB and the UB, we use the icosahedral group ${\sf I}$ and the Clifford group on a single qubits, respectively. Note that the former is an exact $4$-design on a single qubit, and the latter is an exact $2$-design.

We have taken $100$ and $1000$ random sequences for the $2$-RB and the UB, respectively. This is because the UB with the Clifford group converges slower than the $2$-RB with the icosahedral group, which is likely due to the fact that the former and the latter are based on unitary $2$- and $4$-designs, respectively. A higher-design typically leads to a quick convergence since it is more concentrating around the average~\cite{L2009LDB}. 
A faster convergence of the UB with $4$-design is expected, which highlights the potential use of a higher-design also for the UB.
We have taken $10^4$ measurements for each random sequence to obtain a data point of $V^{(2)}$.
The results are summarized in Tab.\,\ref{Tab:Experiment_Sanity_Check}.

From the results, we observe that the unitarity characterized by the $2$-RB matches with that by the UB. This indicates that the $2$-RB on our single-qubit system works to characterize the gate performance. 

Note that the difference between the unitarity from the $2$-RB and that from the UB is slightly beyond the standard deviation. This is likely because the noise property varies in the UB experiment. As mentioned, we have taken $1000$ random sequences in the UB to ensure the convergence of the statistical average, which has taken more than 10 hours in total. Since the noise in the experimental system drifts in such a long timescale, the situation of the experiment deviates from the ideal situation, where time-independence of the noise is assumed. Indeed, unlike the theoretical prediction of the UB, the data is slightly different from a single-exponential decay. This deviation is expected to be the origin a less precise value of the unitarity estimated from the UB.


\begin{figure*}[tbh!]
    \centering
    \includegraphics[width=\textwidth]{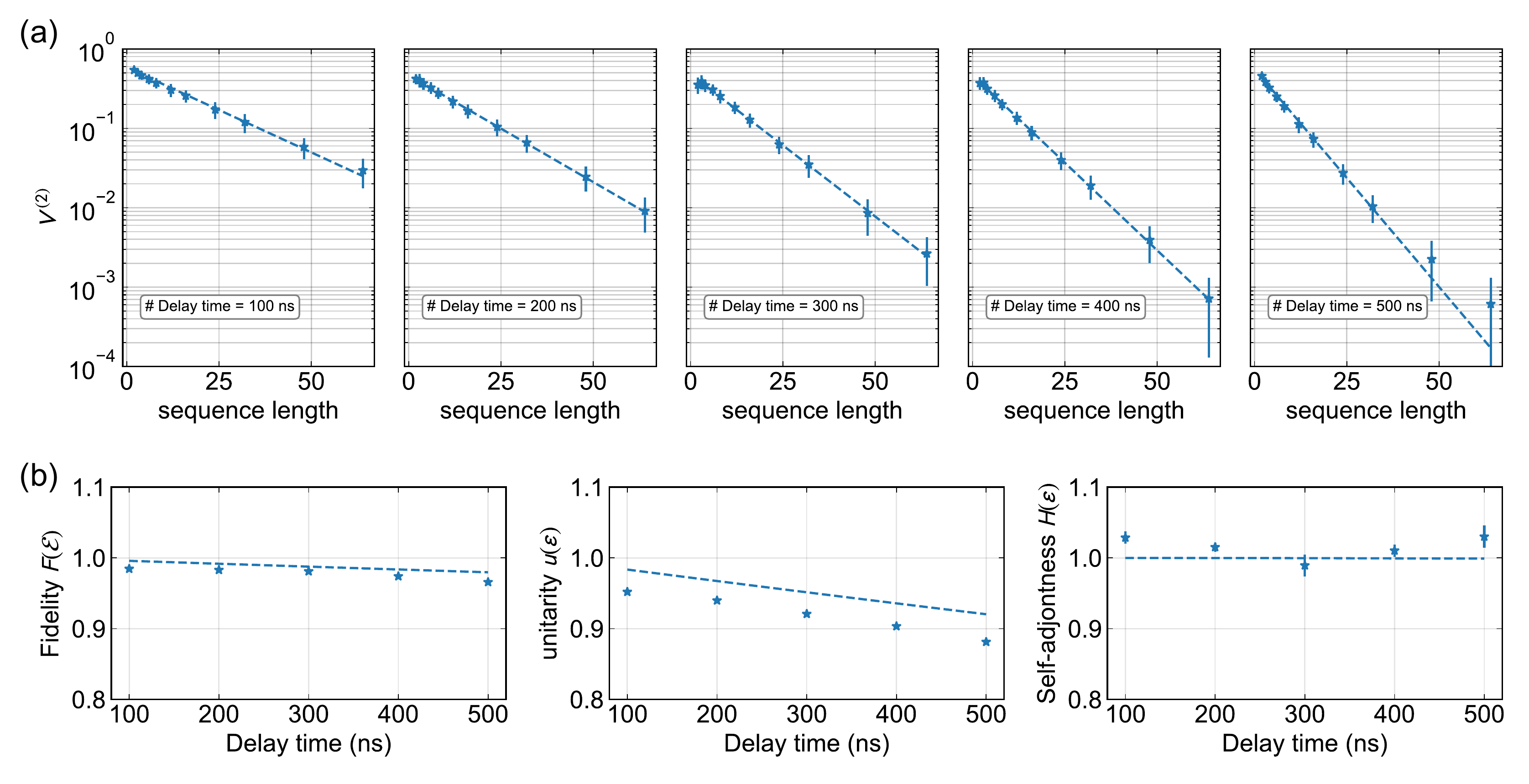}
    \caption{(a) Experimental results for the single-qubit $2$-RB on an isolated qubit when the inverleaved delay time $t$ are swept.
    We have taken $100$ random unitary sequences from ${\sf I}^{\times m}$ for each sequence length $m$ and have had $10^4$ measurements for each sequence to obtain a data point of $V^{(2)}$.
    The stars represent the values of $V^{(2)}$ and the dashed lines are fitting results.
    The error bars represent the standard deviation of $V^{(2)}$.
    Since the vertical axis of the figure is logarithmic notation, the error bars at the bottom of the figure are displayed larger.
    (b) Estimated values of $F(\mathcal{E}_{t}), u(\mathcal{E}_{t}),$ and $H(\mathcal{E}_{t})$ obtained by the $2$-RB.
    The stars show the fitting results, and the dashed lines are the predicted values from the phenomenological model.}
    \label{Fig:experimental_1q_sweep_delay}
\end{figure*}

\subsubsection{Characterizing background noise}
We next perform the single-qubit $2$-RB, aiming to characterize background noise of the qubit $Q_1$ in the experimental system. 
We intentionally insert a delay time $t$ after each application of a gate to extract the information of the background noise.


In the following experiments, we have taken $100$ random unitary sequences from ${\sf I}^{\times m}$, where ${\sf I}$ is the icosahedral group, for each sequence length $m$ and have had $10^4$ measurements for each random sequence to get a data point of $V^{(2)}$.\\


\begin{figure*}[tbh!]
    \centering
    \includegraphics[width=\textwidth]{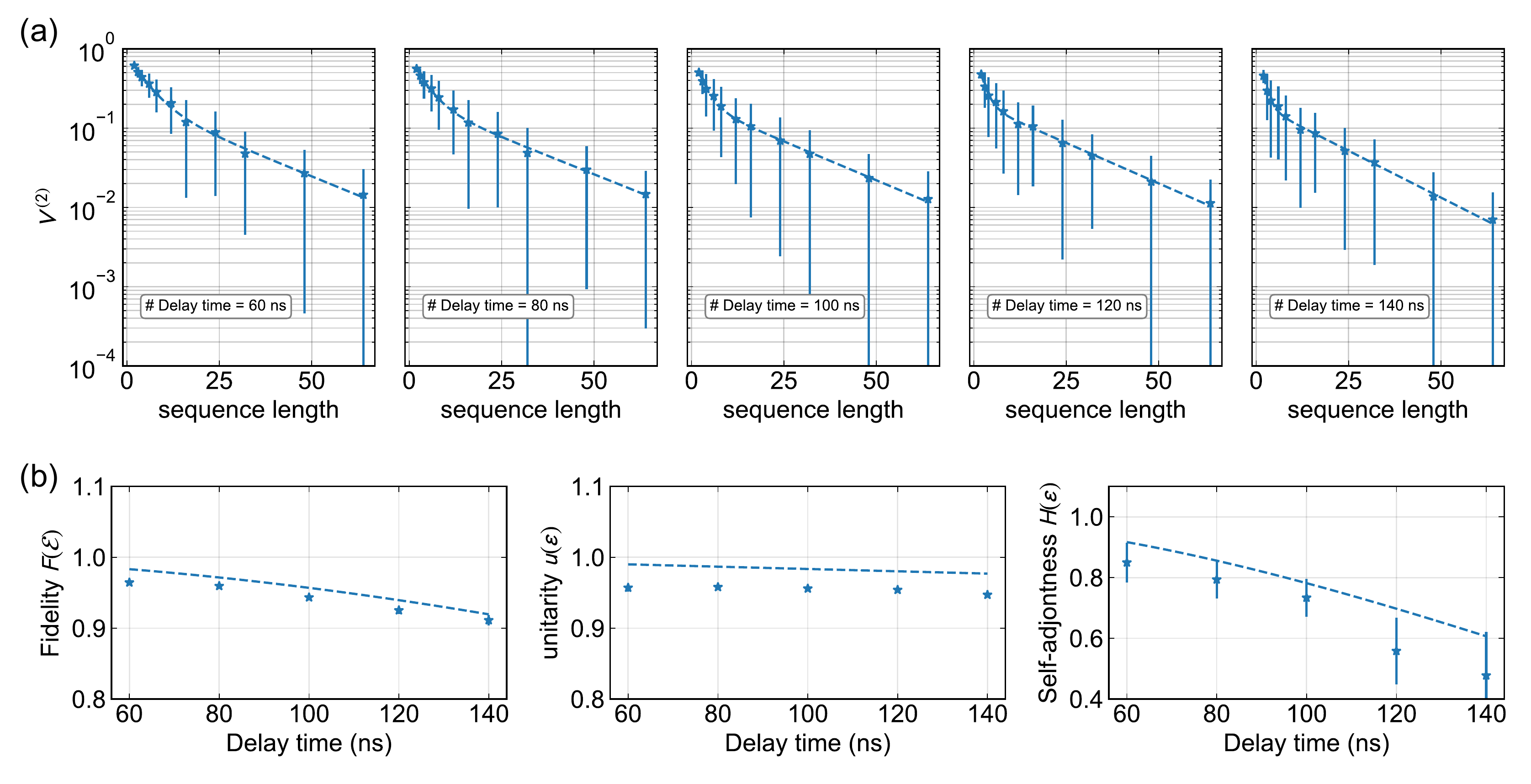}
    \caption{(a) Experimental results for the single-qubit $2$-RB on the qubit coupled to another qubit, where we sweep the interleaved delay time $t$.
    We have taken $100$ random unitary sequences from ${\sf I}^{\times m}$ for each sequence length $m$ and have had $10^4$ measurements for each random sequence to take a data point of $V^{(2)}$.
    The stars correspond to the values of $V^{(2)}$ and the dashed lines are fitting results.
    (b) Estimated values of $F(\mathcal{E}_{t}), u(\mathcal{E}_{t}),$ and $H(\mathcal{E}_{t})$ obtained by the $2$-RB.
    The stars show the fitting results, and the dashed lines are the predicted values from the phenomenological model.}
    \label{Fig:experimental_1q_sweep_delay_zz}
\end{figure*}

In the first experiment, we set the frequency of the local oscillator to $(\omega_1 + \chi_{ge})/2\pi$ and treat the qubit $Q_1$ as a target qubit isolated from the qubit $Q_2$.
The background noise of the isolated qubit is often phenomenologically modeled by the Lindblad Master equation given by
\begin{align}
\frac{d\rho}{dt} &=
\sum_{k\in [1,2]}
L_k^\dagger \rho L_k - \frac{1}{2}\left\{L_k^\dagger L_k, \rho\right\}, \label{Eq:NoiseTheoretical1}
\end{align}
where
$L_1 = \hat{a}/\sqrt{T_1}$ represents the energy dissipation with the relaxation time $T_1$, $\hat{a}=(X+iY)/2$ is an annihilation operator of the qubit, and $L_2 = Z/\sqrt{2T_{\phi}}$ represents the phase dissipation with the relaxation time $T_\phi=1/(1/T_2 - 1/2T_1)$.
By solving the Eq.\,\eqref{Eq:NoiseTheoretical1}, we can obtain phenomenological predictions about the background noise $\mathcal{E}_{t}$ corresponding to the delay time $t$. 

We sweep the delay time $t$ from $100$ to $500~{\rm ns}$. 
The value $V^{(2)}$ obtained from the experiments is shown in Fig.\,\ref{Fig:experimental_1q_sweep_delay}~(a). We estimate the unitarity and the self-adjointness from $V^{(2)}$ through a fitting based on a sum of \emph{two} exponentially-decaying curves given in Eq.\,\eqref{Eq:15}. 
However, we observe single-exponential decays from the results.
This indicates two possibilities. One is that the two decaying rates are nearly the same. The other is that one of the two decaying rates is much smaller than the other, so that one exponentially-decaying curve becomes quickly negligible as $m$ increases. 

In our experiment, the former is the case because the average fidelity is high, which is confirmed from the $1$-RB.
We can analytically show that the two decaying rates typically coincide when the fidelity is sufficiently high.
More specifically, we have (see Eq.\,\eqref{Eq:97779} in Subsec.\ref{SS:PropSA})
\begin{equation}
   1-4 \epsilon \lesssim \frac{h(\cE) + u(\cE)}{2}, \label{Eq:eerrwwww}
\end{equation}
to the first order of $\epsilon$, where $\epsilon$ is the infidelity $1-F(\cE)$.
This implies that the two decaying rates in Theorem~\ref{Thm:2RB} are approximately greater than $1-4 \epsilon$ and $1- 6 \epsilon$, respectively.
Thus, if $\epsilon \ll 1$, which is indeed the case in our system, the two decaying rates are hard to distinguish, making the curve of $V^{(2)}$ a single-exponential decay.

We, hence, estimate the single-exponential decay rate from the experimental data of the $2$-RB and derive $u(\mathcal{E}_{t})$ and $h(\mathcal{E}_{t})$ from
\begin{equation}
    u(\cE_t) = \frac{9}{10}f(\cE_t)^2 - \frac{1}{5}u(\cE_t) + \frac{3}{10}h(\cE_t).
\end{equation}
Here, $u(\cE_t)$ is obtained from the estimated decaying rate, and $f(\cE_t)$ from the $1$-RB (See Eq.\,\eqref{Eq:15}).

The obtained fidelity, unitarity, and self-adjointness are summarized in Fig.\,\ref{Fig:experimental_1q_sweep_delay}~(b).
In calculating self-adjointness, we solved Eq.\,\eqref{Eq:66}, where we substituted $\alpha_{\mathcal{E}}$ of the phenomenological prediction.
They reveal that the background noise of the isolated qubit has the unitarity $u(\mathcal{E}_{t})$ that slowly decreases as the delay increases, while its self-adjoingness $H(\mathcal{E}_{t})$ is nearly independent of the delay.
As we have explained in Subsec.\,\ref{SS:defdef}, a problem may occur when the unitarity is high and the  self-adjointness is low, which is not observed in this experiment. Hence, we conclude that the background noise in this case would not cause any problem toward the realization of QEC. \\

In the second experiment, we set the frequency of the local oscillator to $\omega_1/2\pi$ and treated $Q_1$ as a target qubit exposed to the noise induced by the adjacent qubit $Q_2$. In this experiment, no control pulses are applied to $Q_2$, so that $Q_2$ is expected to remain in the $\ket{0}$ state. This leads to a continuous rotation of the state vector of $Q_1$ by the interaction Hamiltonian term of $\chi_{ge} Z/2$.

In this case, the background noise with the interaction Hamiltonian is modeled by the Lindblad Master equation written as follows,
\begin{align}
\frac{d\rho}{dt} =\left[\frac{\chi_{ge}}{2}Z,\rho\right] + 
\sum_{k\in [1,2]}
L_k^\dagger \rho L_k - \frac{1}{2}\left\{L_k^\dagger L_k, \rho\right\},
\label{Eq:NoiseTheoretical2}
\end{align}
providing a phenomenological model.

Similarly to the first experiment, we sweep the delay time $t$ from $60~{\rm ns}$ to $180~{\rm ns}$.
The delay time is set to a shorter time than the first experiment because the fidelity deteriorates due to the Z-rotation error.
Since the Z-rotation error does not affect the unitarity, we conclude that the decay rate, which is less sensitive to the delay time than the other, corresponds to the unitarity.
The results of the experiment are shown in Fig.\,\ref{Fig:experimental_1q_sweep_delay_zz}~(a).
As seen from the results, the curves $V^{(2)}$ obey double-exponential decay. From the two decaying rates, we obtain the unitarity $u(\mathcal{E}_{t})$ and the self-adjointness $H(\mathcal{E}_{t})$ as a function of the delay time as depicted in Fig.\,\ref{Fig:experimental_1q_sweep_delay_zz}~(b).
Note that, although the unitarity may seem different from the former experiment, it is merely due to the different time scale of the horizontal axis. The unitarities in the two experiments indeed coincide within the standard deviation (see, e.g, the delay time $100$ (ns)).

The experimental results qualitatively coincide with the phenomenological predictions obtained from Eq.\,\eqref{Eq:NoiseTheoretical2} (see Fig.\,\ref{Fig:experimental_1q_sweep_delay_zz}~(b)). However, the experimental values tend to be smaller. This indicates that there exist noise sources not included in the phenomenological model.
The candidates of the additional noise sources are calibration errors in the $R_X(\pi/2)$ gates, the initial thermal excitation rate of $Q_2$ ($7.2~\%$), and the interaction of $Q_1$ with adjacent qubits other than $Q_2$.
Note that the initial thermal excitation of $Q_2$ makes the noise time-dependent due to the relaxation, and hence, makes the result different from the theoretical prediction of the $2$-RB.

Compared to the first experiment (Fig.\,\ref{Fig:experimental_1q_sweep_delay}), we observe from Fig.\,\ref{Fig:experimental_1q_sweep_delay_zz} that the fidelity $F(\cE_t)$ and the self-adjointness $H(\cE_t)$ quickly decrease as the delay time $t$ increases. The latter decreases especially quickly: $H(\cE_t) \approx 0.48$ at $t=140$ (ns). 
This implies that, even if the fidelity is moderately high ($F(\cE_t) \approx 0.91$ at $t=140$ (ns)), the extra $Z$-rotation induced by the interaction with another qubit radically changes the property of the noise and makes the noise far from self-adjoint. Consequently, as the delay increases, the noise quickly becomes the one that cannot be approximated by any stochastic Pauli noise.

This result has an important implication toward a realization of QEC.
As mentioned in Subsec.\,\ref{SS:defdef}, theoretical studies of QEC commonly assume stochastic Pauli noises to numerically compute error thresholds and error rates.
Our result implies that, when the  interaction with another qubit is non-negligible, we cannot directly apply the theoretical predictions based on Pauli noises.
This problem will be more prominent when the system size grows since, in a large system, a qubit interacts with more qubits in an uncontrolled manner, making the noise much less self-adjoint and much far from Pauli noises.
To circumvent this, effective cancellation of the dipole interaction is of great importance in the further improvement since the dominant interaction between qubits should be originated from the electric dipole interaction.

This feature of the noise, i.e., interactions with other qubits induce small self-adjointness and the difficulty of approximating the noise by a Pauli noise, is expected to be common in any experimental systems. The $2$-RB experiment and the self-adjointness offer a useful method and measure, respectively, to experimentally evaluate the noise in the system from this perspective.

\section{Structure of the remaining paper} \label{S:structure}

The remaining of this paper is organized as follows.
In Sec.\,\ref{Sec:ConstDesign}, a proof of Theorem~\ref{Thm:General} is provided. A brief introduction of representations of the unitary group is also provided before the proof. We then explain the higher-order RB in Sec.\,\ref{S:higher-orderRB}, including the proof of Theorem~\ref{Thm:2RB}. The methods used in the numerical analysis, and the experimental demonstrations are provided in Sec.\,\ref{S:NumExpDetail}.
After we summarize the paper in Sec.\,\ref{S:SD}, we prove technical statements in Appendices.

\section{Constructions of exact designs} \label{Sec:ConstDesign}
In this section, we provide a proof of Theorem~\ref{Thm:General}. We start with a brief introduction of representations of the unitary group in Subsec.\,\ref{SS:unitaryrep}, and prove Theorem~\ref{Thm:General} in Subsec.\,\ref{SS:ProofThm1}.

\subsection{Unitary $t$-designs and representation theory} \label{SS:unitaryrep}

Unitary $t$-designs are closely related to representations of the unitary group since the operator $U^{\otimes t} \otimes U^{\dagger \otimes t}$ in the definition can be regarded as a representation $\rho$ of $U \in {\sf U}(d)$ on $\cH_d^{\otimes 2t}$ with $\cH_d$ being the Hilbert space with dimension $d$, i.e., $\rho(U) = U^{\otimes t} \otimes U^{\dagger \otimes t}$.
It is natural to consider irreps of the unitary group.

A well-known fact is that each irrep can be indexed by a non-increasing integer sequence $\lambda :=(\lambda_1, \lambda_2, \dots, \lambda_d)$, i.e. $\lambda_1 \geq \lambda_2 \geq \dots \geq \lambda_d$, of length $d$. 
In particular, each irrep in $U^{\otimes t} \otimes U^{\dagger \otimes t}$ can be indexed by an element of a set $\Lambda(d,t)$ defined by
\begin{multline}
    \Lambda(d, t) := \\
    \{\lambda= (\lambda_1, \lambda_2, \dots, \lambda_d)| \lambda_1 \geq \dots \geq \lambda_d, \lambda^+ = \lambda^- \leq t\},
\end{multline}
where $\lambda^+$ and $\lambda^-$ are the absolute value of sum of positive and negative $\lambda_i$'s, respectively.
Using this notation, the representation space $\cH_d^{\otimes 2t}$ is irreducibly decomposed into
\begin{equation}
    \cH_d^{\otimes 2t} = \bigoplus_{\lambda \in \Lambda(d, t)} V_{\lambda}^{\oplus m_{\lambda}}, \label{Eq:IrrepsVectorSpace}
\end{equation}
where $m_{\lambda}$ is the multiplicity of the irrep $\lambda$.
Accordingly, the map $\rho$ is also decomposed into the irreducible ones $\rho_{\lambda}$. 

Based on the irrep $(\rho_{\lambda}, V_{\lambda})$ of the unitary group, a unitary $t$-design ${\sf U}_t(d)$ can be characterized in a representation-theoretic manner: for any $\lambda \in \Lambda(d,t)$,
\begin{equation}
    \mathbb{E}_{U \sim {\sf U}_t(d)} [\rho_{\lambda}(U)] 
    = 
    \mathbb{E}_{U \sim {\sf H}(d)} [\rho_{\lambda}(U)]. \label{Eq:RepDesign}
\end{equation}
The strong unitary $t$-designs are similarly characterized in terms of irreps~\cite{RS2009}. To this end, let $\Lambda_{\leq}(d,t)$ be
\begin{equation}
    \Lambda_{\leq}(d, t) := \{\lambda= (\lambda_1, \lambda_2, \dots, \lambda_d)| \lambda_1 \geq \dots \geq \lambda_d, \lambda^{\pm} \leq t\},
\end{equation}
where $\lambda^+$ is not necessarily equal to $\lambda^-$. Then, a strong unitary $t$-design ${\sf U}_{\leq t}(d)$ satisfies
\begin{equation}
    \mathbb{E}_{U \sim {\sf U}_{\leq t}} [\rho_{\lambda}(U)] 
    = 
    \mathbb{E}_{U \sim {\sf H}} [\rho_{\lambda}(U)], \label{Eq:RepHaarAv}
\end{equation}
for any $\lambda \in \Lambda_{\leq}(d,t)$. 

One of the merits in this characterization is that the right-hand-sides of  Eqs.\,\eqref{Eq:RepDesign} and~\eqref{Eq:RepHaarAv} are zero for all non-trivial irreps due to the Schur's orthogonality relation, which states that, for any unitarily inequivalent irreps $\lambda$ and $\lambda'$,  
\begin{equation}
\mathbb{E}_{U \sim {\sf H}} \bigl[(\rho_{\lambda}(U))_{ij} (\rho_{\lambda'}(U))_{i'j'} \bigr] = 0,
\end{equation}
for any $i,j, i', j'$, where $(\rho_{\lambda}(U))_{ij}$ is the $(i,j)$ element of the matrix.
By setting the irrep $\rho_{\lambda'}$ to a trivial irrep, i.e., $\rho_{\lambda'}(U) = 1$ for any $U \in {\sf U}(d)$, we have
\begin{equation}
\mathbb{E}_{U \sim {\sf H}} \bigl[\rho_{\lambda}(U) \bigr] = 0,
\end{equation}
for any non-trivial irrep $\lambda$.
On the other hand, for any trivial irrep $\lambda$, it is trivial that 
\begin{equation}
\mathbb{E}_{U \sim {\sf H}} \bigl[\rho_{\lambda}(U) \bigr] = 1.
\end{equation}
From these facts, (strong) unitary $t$-designs can be defined in terms of representation as follows:

\begin{Definition}[Unitary designs in representation theory] \label{Def:zero}
An ensemble ${\sf U}_{t}(d)$ of unitaries is an exact unitary $t$-design if it holds for any irrep $\rho_{\lambda}$ with $\lambda \in \Lambda(d,t)$ that
\begin{equation}
    \mathbb{E}_{U \sim {\sf U}_t(d)} [\rho_{\lambda}(U)] = 
    \begin{cases}
1 & \text{if the irrep is trivial,}\\
0 & \text{otherwise.}
\end{cases} \label{Eq:aaartcct}
\end{equation}
An ensemble ${\sf U}_{\leq t}(d)$ is a strong unitary $t$-design if Eq.\,\eqref{Eq:aaartcct} holds for any irrep $\rho_{\lambda}$ with $\lambda \in \Lambda_{\leq}(d,t)$.
\end{Definition}

\subsection{Proof of Theorem~\ref{Thm:General}} \label{SS:ProofThm1}

We now prove Theorem~\ref{Thm:General}, which states that ${\sf W}_d$ defined by
\begin{equation}
     {\sf W}_d := {\sf W}_{d_1 \oplus d-d_1} \prod_{\lambda \in \Lambda_{\rm sph}(d_1,d,t)} (R_{\lambda} {\sf W}_{d_1 \oplus d-d_1}),
\end{equation}
is a strong unitary $t$-design on ${\sf U}(d)$. Here, 
 \begin{align}
    &{\sf W}_{d_1 \oplus d-d_1}= \{ U \oplus V | U \in {\sf U}_{\leq t}(d_1), V \in {\sf U}_{\leq t}(d-d_1)\},
\end{align}
where ${\sf U}_{\leq t}(d)$ and ${\sf U}_{\leq t}(d-d_1)$ are strong unitary $t$-designs on ${\sf U}(d)$ and ${\sf U}(d-d_1)$, respectively, and $R_{\lambda}$ is constructed by solving the zonal spherical function $Z_{\lambda}$.

It suffices to show 
\begin{equation}
    \mathbb{E}_{U \sim {\sf W}_d} [\rho_{\lambda}(U)] = 0, \label{Eq:Goal}
\end{equation}
for all non-trivial irreps indexed by $\lambda \in \Lambda_{\leq}(d, t)$.
Note that the average over $U \sim {\sf W}_d$ consists of the independent averages over all ${\sf W}_{d_1 \oplus d-d_1}$, further consisting of those over the strong unitary $t$-designs ${\sf U}_{\leq t}(d_1)$ and ${\sf U}_{\leq t}(d-d_1)$. 

Let us first fix a non-trivial irrep $\lambda \in \Lambda_{\leq}(d,t)$ and consider $W_{\lambda}$ defined by
\begin{equation}
W_{\lambda} := \mathbb{E}_{U \sim {\sf W}_{d_1 \oplus d-d_1}} [\rho_{\lambda}(U)].
\end{equation}
Since we consider only irreps $\lambda \in \Lambda_{\leq}(d, t)$, this average can be replaced with the averages over the product ${\sf H}(d_1) \times {\sf H}(d-d_1)$ of the Haar measures on ${\sf K}:= {\sf U}(d_1) \times {\sf U}(d - d_1)$. That is,
\begin{align}
    W_{\lambda} &=\mathbb{E}_{U \sim {\sf H}(d_1) \times {\sf H}(d-d_1)} [\rho_{\lambda}(U)].
\end{align}

To investigate $W_{\lambda}$, we consider the irreps of ${\sf K}$.
Since ${\sf K}$ is a subgroup of ${\sf U}(d)$, each irreducible space $V_{\lambda}$ of ${\sf U}(d)$ is decomposed into a direct sum of those of irreps of ${\sf K}$.
For the same reason as in Definition~\ref{Def:zero}, every non-trivial irrep of ${\sf K}$ becomes zero by taking the average over ${\sf H}(d_1) \times {\sf H}(d-d_1)$. 
Hence, if the non-trivial irreducible representation space $V_{\lambda}$ of ${\sf U}(d)$ does not contain trivial irreps of ${\sf K}$, $W_{\lambda} = 0$.
In contrast, if a non-trivial irrep $\lambda$ of ${\sf U}(d)$ contains trivial irreps of ${\sf K}$, then the matrix elements of $W_{\lambda}$ corresponding to the trivial irreps of ${\sf K}$ are one, and the others are zero.

Trivial irreps of ${\sf K}$ in a non-trivial irrep $\lambda$ of ${\sf U}(d)$ were studied in a great detail since $({\sf K}, {\sf U}(d))$ is an example of a Gelfand pair~\cite{T1994,W2007}. It is known that the irreps of ${\sf U}(d)$ indexed by $\lambda \in \Lambda_{\rm sph}(d_1,d,t)$ contains only one trivial irreps of ${\sf K}$, and that other irreps of ${\sf U}(d)$ contain no trivial irrep of ${\sf K}$~\cite{GW2009}.
Since trivial irreps are one-dimensional, we denote by $\ket{w_{\lambda}} \in V_{\lambda}$ a unit vector that spans the trivial irrep of ${\sf K}$ in the spherical representation $\lambda \in \Lambda_{\rm sph}(d_1,d,t)$ of ${\sf U}(d)$.
Then, we have 
\begin{equation}
    W_{\lambda} = 
    \begin{cases}
    0 & \text{\ if\ } \lambda \notin \Lambda_{\rm sph}(d_1,d,t),\\
    \ketbra{w_{\lambda}}{w_{\lambda}} & \text{\ if\ } \lambda \in \Lambda_{\rm sph}(d_1,d,t).
    \end{cases} \label{Eq:AvW}
\end{equation}

If $\lambda \notin \Lambda_{\rm sph}(d_1,d,t)$, we immediately obtain $\mathbb{E}_{U \sim {\sf W}_{d_1 \oplus d-d_1}} [\rho_{\lambda}(U)]=0$ from the definition of $W_{\lambda}$, which implies Eq.\,\eqref{Eq:Goal}.

If $\lambda \in \Lambda_{\rm sph}(d_1,d,t)$, we define a matrix $M_{\lambda}(U)$ on $V_{\lambda}$ ($U \in {\sf U}(d)$) by
\begin{align}
    M_{\lambda}(U) &:= \mathbb{E}_{V_1, V_2 \sim {\sf H}(d_1) \times {\sf H}(d-d_1)} [\rho_{\lambda}(V_1 U V_2)].
\end{align}
Importantly, for any $\lambda \in \Lambda_{\rm sph}(d_1,d,t)$, there exists at least one $R_{\lambda} \in {\sf U}(d)$ such that $\bra{w_{\lambda}} M_{\lambda}(R_{\lambda}) \ket{w_{\lambda}}=0$. This follows from the fact that
\begin{align}
     \mathbb{E}_{U \sim {\sf H}(d)}[ \bra{w_{\lambda}} M_{\lambda}(U)  \ket{w_{\lambda}}]
     &= \bra{w_{\lambda}}  \mathbb{E}_{U \sim {\sf H}(d)}[ \rho_{\lambda}(U) ] \ket{w_{\lambda}}\\
     &=0,
\end{align}
where we have used the unitary invariance of ${\sf H}(d)$ and that the irrep $\rho_{\lambda}$ is non-trivial, so that $\mathbb{E}_{U \sim {\sf H}(d)}[ \rho_{\lambda}(U) ]=0$.
Due to the intermediate value theorem, there always exists at least one unitary $R_{\lambda} \in {\sf U}(d)$ such that $\bra{w_{\lambda}} M_{\lambda}(R_{\lambda}) \ket{w_{\lambda}}=0$.

Using such $R_{\lambda} \in {\sf U}(d)$ and Eq.\,\eqref{Eq:AvW}, it is straightforward to observe that $W_{\lambda} M_{\lambda} (R_{\lambda}) W_{\lambda} = 0$.
Furthermore, it follows that
\begin{align}
    W_{\lambda} M_{\lambda} (R_{\lambda}) W_{\lambda}
    &=
    \mathbb{E}_{U, U', V_1, V_2 \sim {\sf W}_{d_1 \oplus d-d_1}} [\rho_{\lambda}(U V_1  R_{\lambda} V_2 U')]\\
    &=
    \mathbb{E}_{W_1, W_2 \sim {\sf W}_{d_1 \oplus d-d_1}} [\rho_{\lambda }(W_1  R_{\lambda} W_2)]\\
    &=
    \mathbb{E}_{U \sim {\sf W}_{d_1 \oplus d-d_1} R_{\lambda} {\sf W}_{d_1 \oplus d-d_1}} [\rho_{\lambda}(U)].
\end{align}
We, hence, obtain
\begin{equation}
    \mathbb{E}_{U \sim {\sf W}_{d_1 \oplus d-d_1} R_{\lambda} {\sf W}_{d_1 \oplus d-d_1}} [\rho_{\lambda}(U)] = 0. \label{Eq:eerrcr}
\end{equation}

Thus, the finite set of unitaries ${\sf W}_{d_1 \oplus d-d_1} R_{\lambda} {\sf W}_{d_1 \oplus d-d_1}$ satisfies the condition for the design, i.e., Eq.\,\eqref{Eq:Goal} for any non-trivial irrep $\lambda \in \Lambda_{\leq}(d,t)$, which leads to the statement that the set of unitaries defined by
\begin{equation}
     {\sf W}_d = {\sf W}_{d_1 \oplus d-d_1} \prod_{\lambda \in \Lambda_{\rm sph}(d_1,d,t)} (R_{\lambda} {\sf W}_{d_1 \oplus d-d_1}),
\end{equation}
is a strong $t$-design on ${\sf U}(d)$.

Finally, let us clarify the relation between $R_{\lambda}$ and the zero of the zonal spherical function. To this end, we first observe that the matrix element $\bra{w_{\lambda}} M_{\lambda}(U) \ket{w_{\lambda}}$ of $M_{\lambda}$ is the zonal spherical function $Z_{\lambda}(U)$.
This can be checked by a simple calculation: for any $W_1, W_2 \in K$ and $U \in {\sf U}(d)$, we have
\begin{align}
    Z_{\lambda}(W_1 U W_2)&= \bra{w_{\lambda}} \mathbb{E}[\rho_{\lambda} (V_1 W_1 U W_2V_2)] \ket{w_{\lambda}}\\
    &= \bra{w_{\lambda}} \mathbb{E}[\rho_{\lambda}(V_1 U V_2)] \ket{w_{\lambda}} \label{Eq:ZSF}\\
    &=Z_{\lambda}(U),
\end{align}
where the averages are all taken over $V_1, V_2 \sim {\sf H}(d_1) \times {\sf H}(d-d_1)$. 
Thus, $Z_{\lambda}(U)$ is bi-${\sf K}$-invariant, and so, is the zonal spherical function. This implies that $R_{\lambda}$ is indeed a zero of the zonal spherical function.

Based on this fact, we can provide a matrix form of $R_{\lambda}$ in the fixed basis in which a unitary in ${\sf W}_{d_1 \oplus d-d_1}$ is represented as $U \oplus V$.
To this end, it is important to notice that the bi-{\sf K}-invariance of the zonal spherical function implies that it is characterized by the cosets of ${\sf K} = {\sf U}(d_1) \times {\sf U}(d-d_1)$ in ${\sf U}(d)$. The cosets can be further identified with $d_1$-dimensional subspaces corresponding to the support on which ${\sf U}(d_1)$ acts. For instance, the identity element in the coset of ${\sf K}$ corresponds to the subspace $V_0$ spanned by the first $d_1$ vectors of the fixed basis.
The matrix form of $R_{\lambda}$ is obtained by specifying the relation between $V_0$ and the subspace corresponding to another representative of the coset. 

To characterize the relation between two subspaces, we use the \emph{principal} angles. For two subspaces $X$ and $Y$, let us refer to $\theta = \min {\rm argcos} |\braket{x}{y}|$, where the minimum is taken over all unit vectors $\ket{x} \in X, \ket{y} \in Y$, as the minimum angle between $X$ and $Y$.
The principal angles $(\theta_0, \dots, \theta_{m-1})$ between two $m$-dimensional subspaces $X$ and $Y$ are then defined as follows: $\theta_0$ is the minimum angle between $X$ and $Y$, and $\theta_{i+1}$ is the minimum angle between $X \cap {\rm span}\{\ket{x_0}, \dots, \ket{x_{i}}\}$ and $Y \cap {\rm span}\{\ket{y_0}, \dots, \ket{y_{i}}\}$, where $(\ket{x_j}, \ket{y_j})$ is a pair of the unit vectors that leads to $\theta_j$.

The cosine of the principal angles between $V_0$ and the subspace corresponding to another representative in the coset determines the value of the zonal spherical function $Z_{\lambda}$, and so, $Z_{\lambda}$ can be written as $Z_{\lambda}(\cos^2 \theta_0, \dots, \cos^2 \theta_{d_1-1})$~\cite{R2010,BNOZ2020}. See, e.g., Refs.\,\cite{JC1974,R2010,BNOZ2020} for the explicit form of $Z_{\lambda}$ as a polynomial of $(\cos^2 \theta_1, \dots, \cos^2 \theta_{d_1})$. 

By solving the polynomial, we obtain the principal angles $(\theta^{(0)}_{\lambda}, \dots, \theta^{(d_1-1)}_{\lambda})$ between $V_0$ and the subspace corresponding to the zero of $Z_{\lambda}$.
Recalling the definition of the principal angles and using the left- and right-invariance of the coset by any unitary in ${\sf K}$, we can take a matrix form of $R_{\lambda}$ as follows:
\begin{equation}
	R_{\lambda} = 
	    \begin{pmatrix}
		C(\boldsymbol{\theta}_{\lambda}) & i S(\boldsymbol{\theta}_{\lambda}) & 0 \\
		i S(\boldsymbol{\theta}_{\lambda}) & C(\boldsymbol{\theta}_{\lambda}) & 0 \\
		0 & 0 & I_{d-2d_1}
        \end{pmatrix},
\end{equation}
where $C(\boldsymbol{\theta}_{\lambda}) = {\rm diag}(\cos\theta_{\lambda}^{(0)}, \dots, \cos\theta_{\lambda}^{(d_1-1)})$ and $S(\boldsymbol{\theta}_{\lambda}) = {\rm diag}(\sin\theta_{\lambda}^{(0)}, \dots, \sin\theta_{\lambda}^{(d_1-1)})$, and $I_{d-2d_1}$ is the identity matrix of size $d- 2d_1$. Note that $R_{\lambda}$ is not necessarily in this form since the coset is invariant under the action of ${\sf K}$.
$\hfill \blacksquare$

\section{Higher-order RB} \label{S:higher-orderRB}

In this section, we investigate the higher-order RB in detail. 
We begin with a preliminary in Subsec.\,\ref{SS:L} and explain several basic properties of the self-adjointness in Subsec.\,\ref{SS:PropSA}. We consider the $t$-RB for general $t$ and the $2$-RB in Subsecs.\,\ref{SS:generaltRB} and~\ref{SS:2RBdetail}, respectively. 

\subsection{Liouville representation} \label{SS:L}

Let $\sigma_0 = I/\sqrt{2}$, $\sigma_1 = X/\sqrt{2}$, $\sigma_2 = Y/\sqrt{2}$, and $\sigma_3 = Z/\sqrt{2}$ be normalized Pauli operators on one qubit, where normalization is in terms of the Hilbert-Schmidt inner product. 
For $q$ qubits, we introduce a vector $\vec{n} = (n_1, n_2, \dots, n_{q})$ ($n_i \in \{0,1,2,3\}$) and use the notation that
\begin{equation}
\sigma_{\vec{n}}:= \sigma_{n_1} \otimes \dots \otimes \sigma_{n_{q}}.
\end{equation}
We also denote $2^q$ by $d$ in this section.

The Liouville representation is a matrix representation of quantum channels, also known as the Pauli transfer matrix. See, e.g., Refs.\,\cite{WGHF2015,KLDF2016,DHW2019}.
Let $\kett{\cdot}$ be a linear map from a set of all linear operators on a $d$-dimensional Hilbert space to a $d^2$-dimensional vector space that specifically maps $\sigma_{\vec{n}}$ to a canonical orthonormal basis vector $e_{\vec{n}}$.
Since the map is linear, we have
\begin{equation}
\kett{A} := \sum_{\vec{n}} \tr[\sigma_{\vec{n}} A] \kett{\sigma_{\vec{n}}},
\end{equation}
for any linear operator $A$. Note that $\braakett{A}{B} = \tr[A^{\dagger} B]$.

Based on this vector representation of linear operators, a linear supermap $\cE$ can be represented by a matrix.
The Liouville representation of a linear supermap $\cE$ is defined by
\begin{equation}
L_{\cE} := \sum_{\vec{n}} \kettbraa{\cE(\sigma_{\vec{n}})}{\sigma_{\vec{n}}},
\end{equation}
which is a regular matrix of size $d^2$.
The matrix element in the canonical basis of $\{\kett{\sigma_{\vec{n}}}\}_{\vec{n}}$ is given by
\begin{equation}
\bigl( L_{\cE} \bigr)_{\vec{n} \vec{m}} = \braakett{\sigma_{\vec{n}}}{\cE(\sigma_{\vec{m}})} = \tr[\sigma_{\vec{n}} \cE(\sigma_{\vec{m}})].
\end{equation}


The vector and Liouville representations satisfy the following properties:
\begin{enumerate}
\item $L_{\cE} \kett{\rho} = \kett{\cE(\rho)}$, 
\item $L_{\cE_2 \circ \cE_1} = L_{\cE_2}L_{\cE_1}$,
\item $L_{\alpha \cE_1 + \beta \cE_2} = \alpha L_{\cE_1} + \beta L_{\cE_2}$ ($\alpha, \beta \in \mathbb{C}$),
\item $L_{\cE_1 \otimes \cE_2} = L_{\cE_1} \otimes L_{\cE_2}$,
\item $L_{\cE^{\dagger}} = L_{\cE}^{\dagger}$.
\end{enumerate}

Properties of a linear supermap $\cE$ can be also expressed in terms of the Liouville representation. For instance, the linear map $\cE$ is TP if and only if $(L_{\cE})_{\vec{0}\vec{0}}=1$ and $(L_{\cE})_{\vec{0}\vec{n}} = 0$ for any $\vec{n} \neq \vec{0}$.
Since we are interested in the CPTP map $\cE$ that represents a noise, its Liouville representation is always in the form of
\begin{equation}
L_{\cE} = 
\begin{pmatrix}
        1 & 0 \\
        \alpha_{\cE} & \tilde{L}_{\cE}
\end{pmatrix}, \label{Eq:Li}
\end{equation}
where $0$ is a row vector of length $d^2-1$ with all elements being zero, $\alpha_{\cE}$ is a column vector of length $d^2-1$, called a \emph{non-unital part} of the noise, and $\tilde{L}_{\cE}$ is a $(d^2-1) \times (d^2-1)$ matrix.
The non-unital part $\alpha_{\cE}$ of the noise is the zero vector if and only if the map $\cE$ is unital, i.e., $\cE(I)=I$ with $I$ being the identity operator. 

In the Liouville representation, the fidelity parameter $f(\cE)$ and the unitarity $u(\cE)$ of a noisy CPTP map $\cE$ are given by
\begin{align}
f(\cE) &= \frac{1}{d^2-1} \sum_{\vec{n} \neq \vec{0}} \braa{\sigma_{\vec{n}}} L_{\cE} \kett{\sigma_{\vec{n}}}\\
&= \frac{1}{d^2-1} \tr[\tilde{L}_{\cE}],\\
u(\cE) &= \frac{1}{d^2-1} \sum_{\vec{n} \neq \vec{0}} \braa{\sigma_{\vec{n}}} L_{\cE}^{\dagger} L_{\cE} \kett{\sigma_{\vec{n}}},\\
&= \frac{1}{d^2-1} \tr[\tilde{L}_{\cE}^{\dagger} \tilde{L}_{\cE}],
\end{align}
respectively.

\subsection{Properties of the self-adjointness} \label{SS:PropSA}

For a CPTP map $\cE$, the self-adjointness $H(\cE)$ and the self-adjointness parameter $h(\cE)$ are defined by
\begin{align}
H(\cE) &:= 1- \frac{d+1}{2d} \int d \varphi \bigl|\! \bigr| \cE( \ketbra{\varphi}{\varphi}) - \cE^{\dagger} ( \ketbra{\varphi}{\varphi}) \bigl|\! \bigr|_2^2,\\
h(\cE) &:= \frac{d}{d-1} \int d \varphi \tr[\cE'(\varphi)   \cE'^{\dagger}(\varphi)],\\
&= \frac{1}{d^2-1} \tr[ \tilde{L}_{\cE}^2], \label{Eq:B1rr} 
\end{align}
where $\cE'(\rho) = \cE(\rho- I/d)$, and the last line is shown in Appendix~\ref{App:CharaNoise}.

We first show the relation between $H(\cE)$ and $h(\cE)$, i.e., Eq.\,\eqref{Eq:66} in Subsec.\,\ref{SS:defdef}:
\begin{align}
H(\cE) &= 1 - \frac{d^2-1}{d^2}\bigl( u(\cE) - h(\cE) \bigr) - \frac{d+1}{2d^2} | \alpha_{\cE} |^2. \label{Eq:ice}
\end{align}
From the definition of $H(\cE)$, we have
\begin{multline}
\frac{2d}{d+1} \bigl(1- H(\cE)\bigr)\\
= \int d \varphi \biggl[   \tr[ \cE( \ketbra{\varphi}{\varphi})^2] + \tr[ \cE^{\dagger}( \ketbra{\varphi}{\varphi})^2] \\-2 \tr[\cE( \ketbra{\varphi}{\varphi})\cE^{\dagger}( \ketbra{\varphi}{\varphi})] \biggr]
\end{multline}
By rewriting $\cE$ with $\cE'$, the first term in the right-hand side is expressed in terms of the unitarity $u(\cE)$, such as
\begin{align}
&\int d \varphi  \tr[ \cE( \ketbra{\varphi}{\varphi})^2] = \frac{d-1}{d} u(\cE) + \tr\bigl[ \cE (I/d)^2 \bigr].
\end{align}
By using the swap operator $\mathbb{F}:= \sum_{\vec{n}} \sigma_{\vec{n}} \otimes \sigma_{\vec{n}}$, and the property that $\tr[MN]=\tr[\mathbb{F} (M \otimes N)]$ for any matrices $M$ and $N$, which is called a \emph{swap trick}, it follows that 
\begin{align}
\tr\bigl[ \cE (I/d)^2 \bigr] &= \tr\bigl[ \mathbb{F} \cE(I/d)^{\otimes 2} \bigr]\\
&= \frac{1}{d}\sum_{\vec{n}} \braa{\sigma_{\vec{n}}} L_{\cE} \kett{\sigma_{\vec{0}}}^2\\
&= \frac{1}{d} | \alpha_{\cE} |^2  + \frac{1}{d},
\end{align}
which leads to
\begin{align}
&\int d \varphi  \tr[ \cE( \ketbra{\varphi}{\varphi})^2] = \frac{d-1}{d} u(\cE) + \frac{1}{d} | \alpha_{\cE} |^2  + \frac{1}{d}.
\end{align}
Similarly, we obtain
\begin{align}
&\int d \varphi  \tr[ \cE^{\dagger}( \ketbra{\varphi}{\varphi})^2] = \frac{d-1}{d} u(\cE) + \frac{1}{d},
\end{align}
from the facts that $L_{\cE^{\dagger}} = L_{\cE}^{\dagger}$ and that $| \alpha_{\cE^{\dagger}} | = 0$ for any TP map $\cE$.

From the definition of $\cE'$, it is straightforward to show that the self-adjointness parameter $h(\cE)$ is given by
\begin{align}
h(\cE) &=\frac{1}{d-1}\biggl[ d \int \tr \bigl[ \cE(\varphi) \cE^{\dagger}(\varphi) \bigr] d\varphi  - 1 \biggr]. \label{Eq:B12rr}
\end{align}
Combining these altogether, we arrive at
\begin{equation}
\frac{2d}{d+1} \bigl(1- H(\cE)\bigr)
=
\frac{2(d-1)}{d} \bigl( u(\cE) - h(\cE) \bigr) + \frac{1}{d}| \alpha_{\cE} |^2,
\end{equation}
implying Eq.\,\eqref{Eq:ice}.\\

The self-adjointness parameter also satisfies the following properties. They are all shown in Appendix~\ref{App:CharaNoise}.
\begin{enumerate}
\item $-\frac{1}{d^2-1} \leq h(\cE) \leq u(\cE)$.
\item $h(\cE) = u(\cE)$ if and only if $\tilde{L}_{\cE} = \tilde{L}_{\cE}^{\dagger}$. For a unital noise $\cE$, $h(\cE) = u(\cE)$ if and only if the noise is self-adjoint ($\cE = \cE^{\dagger}$).
\item $h(\cE) =-\frac{1}{d^2-1}$ if and only if $\tr[K_iK_j]  = 0$ for any $i, j$, where $\{ K_i \}$ are the Kraus operators of $\cE$.
\item the average gate fidelity $F(\cE)$ is bounded from above by $u(\cE)$ and $h(\cE)$:
 \begin{equation}F(\cE) \leq \frac{d-1}{d} \sqrt{\frac{h(\cE) + u(\cE)}{2}} + \frac{1}{d}. \label{Eq:97779}\end{equation}
\end{enumerate}

\subsection{A general expression for the $t$-RB} \label{SS:generaltRB}
We here show that the expectation value $V^{(t)}(m, \cE| O_{\rm ini}, O_{\rm meas})$ in the $t$-RB has a general form of
\begin{equation}
V^{(t)}(m, \cE| O_{\rm ini}, O_{\rm meas}) 
= \sum_{\lambda} \tr \bigl[ \hat{A}_{\lambda} (\hat{C}_{\lambda}(\cE))^m\bigr], \label{Eq:general2}
\end{equation}
where $\hat{A}_{\lambda}$ is a regular matrix depending on $O_{\rm ini}$ and $\cE(O_{\rm meas})$, and $\hat{C}_{\lambda}(\cE)$ is a regular matrix depending only on $\cE$. As we will see below, $\lambda$ labels the irreps of a $t$-copy representation of the unitary group, and the size of the matrices is equal to the multiplicity of each irrep. 

The expectation value is defined by
\begin{multline}
V^{(t)}(m, \cE| O_{\rm ini}, O_{\rm meas}) :=\\ 
\mathbb{E}_{U_{\boldsymbol{i}}} \bigl[ \bigl(\tr \bigl[ O_{\rm meas} \cG_{i_{m+1}} \circ \cG_{i_m} \circ \dots \circ \cG_{i_1}(O_{\rm ini}) \bigr] \bigr)^t \bigr],
\end{multline}
where $\mathbb{E}_{U_{\boldsymbol{i}}}$ is the average over all unitary sequences $U_{\boldsymbol{i}} \sim {\sf U}_{2t}^{\times m}$. Note that $\cG_i = \cE \circ \cU_i$ and that $\cU_i$ is the unitary channel defined by $\cU_i(\rho) = U_i \rho U_i^{\dagger}$.
In terms of the Liouville representation, we have
\begin{multline}
\tr \bigl[ O_{\rm meas} \cG_{i_{m+1}} \circ \cG_{i_m} \circ \dots \circ \cG_{i_1}(O_{\rm ini}) \bigr]\\ = \braa{O'_{\rm meas}} L_{\cU'_m \circ \cE \circ \cU^{'\dagger}_m} \dots L_{\cU'_1 \circ \cE \circ \cU^{'\dagger}_1} \kett{O_{\rm ini}},
\end{multline}
where we have used that  $\cG_i = \cE \circ \cU_i$, $\cU'_n = \cU_n \circ \cU_{n-1} \circ \dots \circ \cU_2 \circ \cU_1$, and $O'_{\rm meas} = \cE(O_{\rm meas})$.

Noticing the $t$-th power and the fact that each unitary is independently chosen from a unitary $2t$-design ${\sf U}_{2t}$, we obtain
\begin{align}
&V^{(t)}(m, \cE| O_{\rm ini}, O_{\rm meas})= \bigl(\braa{O'_{\rm meas}}^{\otimes t} \bigr) (L_{\rm av})^m  \bigl( \kett{O_{\rm ini}}^{\otimes t} \bigr), \label{Eq:kore}
\end{align}
where $L_{\rm av}$ is defined by
\begin{align}
L_{\rm av} : &= \mathbb{E}_{U \sim {\sf U}_{2t}}[ (L_{\cU \circ \cE \circ \cU^{\dagger}})^{\otimes t}],\\
&=\mathbb{E}_{U \sim {\sf U}_{2t}}[ (L_{\cU} L_{\cE} L_{\cU^{\dagger}})^{\otimes t}]\\
&= \mathbb{E}_{U \sim {\sf H}}[  (L_{\cU} L_{\cE} L_{\cU^{\dagger}})^{\otimes t}].
\end{align}
The last line follows since ${\sf U}_{2t}$ is an exact unitary $2t$-design.

To write down $L_{\rm av}$ explicitly, let us consider the tensor-$t$ Liouville representation given by 
\begin{equation}
{\sf U}(d) \ni U \rightarrow L_{\cU^{\otimes t}} \in GL(\cK),
\end{equation}
where $GL(\cK)$ is the general linear group acting on the $d^q$-dimensional vector space $\cK$ defined by
\begin{equation}
\cK := {\rm span}\{ \bigotimes_{s = 1}^t \kett{\sigma_{\vec{n}_s}} : \vec{n}_s \in \{0,1,2,3 \}^{q}, s \in [1, t] \}. \label{Eq:repspace}
\end{equation}
We denote the irreducible decomposition by 
\begin{equation}
\cK = \bigoplus_{\lambda} \cK_{\lambda}^{\oplus m_{\lambda}},
\end{equation}
where $\lambda$ labels the irreps, and $m_{\lambda}$ is the multiplicity of the irrep labeled by $\lambda$.

The key observation is that 
\begin{equation}
\forall V \in {\sf U}(d), \quad [L_{\rm av}, L_{\cV^{\otimes t}}] = 0,
\end{equation}
which simply follows from the unitary invariance of the Haar measure. This implies that $L_{\rm av} \in {\rm End}_{\sf U}(\cK)$, where ${\rm End}_{\sf U}(\cK)$ is a set of all endomorphisms of $\cK$ that commute with the tensor-$t$ Liouville action of ${\sf U}(d)$.
It is well-known that ${\rm End}_{\sf U}(\cK)$ is isomorphic to the direct sum of matrix algebras:
\begin{equation}
{\rm End}_{\sf U}(\cK) \simeq \bigoplus_{\lambda} M(m_{\lambda}, \mathbb{C}), \label{Eq:End}
\end{equation}
where $M(m_{\lambda}, \mathbb{C})$ is a set of all $m_{\lambda} \times m_{\lambda}$ matrices over $\mathbb{C}$.
Thus, the operator $L_{\rm av} \in {\rm End}_{\sf U}(\cK)$ can be represented by a direct sum of matrices.

To obtain the explicit form of $L_{\rm av}$, let $\cK_{\lambda}^{(1)} \oplus \dots \oplus \cK_{\lambda}^{(m_{\lambda})}$ be a fixed decomposition of $\cK_{\lambda}^{\oplus m_{\lambda}}$, and denote $\eta_{\lambda}^{p \rightarrow q}$ be the isomorphism from $\cK_{\lambda}^{(p)}$ to $\cK_{\lambda}^{(q)}$. We also denote by $\Pi_{\lambda}^{(p)}$ the projection onto $\cK_{\lambda}^{(p)}$. Then, from the explicit form of the isomorphism, we have
\begin{equation}
L_{\rm av}
=
\sum_{\lambda} \sum_{p, q =1}^{m_{\lambda}} (\hat{C}_{\lambda}(\cE))_{pq} \eta_{\lambda}^{p \rightarrow q} \Pi_{\lambda}^{(p)},\label{Eq:EndMat}
\end{equation}
where $\hat{C}_{\lambda}(\cE) \in M(m_{\lambda}, \mathbb{C})$. Each element $\hat{C}_{\lambda}(\cE)$ is given by
\begin{align}
( \hat{C}_{\lambda} (\cE) )_{pq} &= \tr[ L_{\rm av} \eta_{\lambda}^{q \rightarrow p} \Pi_{\lambda}^{(q)}],\\
&= \tr \bigl[  \mathbb{E}_{U \sim {\sf H}}[  (L_{\cU} L_{\cE} L_{\cU^{\dagger}})^{\otimes t}]\eta_{\lambda}^{q \rightarrow p} \Pi_{\lambda}^{(q)} \bigr],\\
&=   \mathbb{E}_{U \sim {\sf H}} \tr \bigl[  L_{\cE}^{\otimes t} L_{\cU^{\dagger}}^{\otimes t}\eta_{\lambda}^{q \rightarrow p} \Pi_{\lambda}^{(q)} L_{\cU}^{\otimes t} \bigr],\\
&=   \mathbb{E}_{U \sim {\sf H}} \tr \bigl[  L_{\cE}^{\otimes t} \eta_{\lambda}^{q \rightarrow p} \Pi_{\lambda}^{(q)} \bigr],\\
&= \tr \bigl[  L_{\cE}^{\otimes t} \eta_{\lambda}^{q \rightarrow p} \Pi_{\lambda}^{(q)} \bigr],
\end{align}
where we have used the irreducibility in the fourth line.

Consequently, it follows that
\begin{equation}
(L_{\rm av})^m
=
\sum_{\lambda} \sum_{p, q =1}^{m_{\lambda}} \bigl(\hat{C}_{\lambda}(\cE)^m)_{pq} \eta_{\lambda}^{p \rightarrow q} \Pi_{\lambda}^{(p)}.
\end{equation}
Substituting this into Eq.\,\eqref{Eq:kore}, we obtain
\begin{equation}
V^{(t)}(m, \cE| O_{\rm ini}, O_{\rm meas})\\
=
\sum_{\lambda} \tr \bigl[ \hat{A}_{\lambda}( \hat{C}_{\lambda}(\cE) )^m\bigr],
\end{equation}
where the $m_{\lambda} \times m_{\lambda}$ matrices $\hat{A}_{\lambda}$ are given by
\begin{equation}
( \hat{A}_{\lambda} )_{pq}
=
\braa{\cE(O_{\rm meas})^{\otimes t}}
\eta_{\lambda}^{q \rightarrow p} \Pi_{\lambda}^{(q)}
\kett{O_{\rm ini}^{\otimes t}}.
\end{equation}
This completes the proof. \hfill $\blacksquare$

\subsection{The first-order RB} \label{SS:1RBdetail}
Let us briefly overview the $1$-RB using an exact unitary $2$-design, namely, the standard RB. We also explain how the result changes when the $2$-design is an approximate one rather than the exact one.

In the $1$-RB, the representation space is given by
\begin{equation}
    \cK = {\rm span} \bigl\{ \kett{\sigma_{\vec{n}}}: \vec{n} \in \{ 0,1,2,3\}^q \bigr \}.
\end{equation}
We need to find a irreducible decomposition of $\cK$ under the action of a unitary group ${\sf U}(d)$ as $U \rightarrow L_{\cU}$. The Liouville representation $L_{\cU}$ is defined by $L_{\cU}\kett{\rho} =  \kett{U \rho U^{\dagger}}$.
Hence, $\cK$ is irreducibly decomposed to
\begin{equation}
    \cK = \cK_0 \oplus \cK_1,
\end{equation}
where
\begin{align}
    &\cK_0={\rm span} \{ \kett{\sigma_{\vec{0}}} \},\\
    &\cK_1 = {\rm span}\{ \kett{\sigma_{\vec{n}}}: \vec{n} \in \{ 0,1,2,3\}^q, \vec{n} \neq \vec{0} \}.
\end{align}

Denoting by $\Pi_0$ and $\Pi_1$ projectors onto $\cK_0$ and $\cK_1$, respectively, we have
\begin{align}
    L_{\rm av} &:= \mathbb{E}_{U \sim {\sf U}_2}[(L_{\cU \circ \cE \circ \cU^{\dagger}})], \\
    &= \Pi_{0} + f(\cE) \Pi_{1}, \label{Eq:exactapp}
\end{align}
where $f(\cE)$ is the fidelity parameter. Note that $ {\sf U}_2$ is an exact unitary $2$-design.
We thus obtain that
\begin{equation}
V^{(1)}(m, \cE| \Delta, O_{\rm meas}) = A_0 + A_1 f(\cE)^m, \label{Eq:korekoredetail}
\end{equation}
where $A_i = \braa{\cE(O_{\rm meas})}\Pi_i \kett{O_{\rm ini}}$ for $i=0,1$.\\

When the $2$-design is an approximate one ${\sf U}^{(\epsilon)}_2$, Eq.\,\eqref{Eq:exactapp} holds only approximately. The degree of approximation depends on how we measure it, but we here assume that the design is $\epsilon$-approximate when Eq.\,\eqref{Eq:exactapp} holds up to $\epsilon$-approximation.
That is, we assume that
\begin{align}
    L_{\rm av}^{(\epsilon)} &:= \mathbb{E}_{U \sim {\sf U}_2^{(\epsilon)}}[(L_{\cU \circ \cE \circ \cU^{\dagger}})],\\
    &= L_{\rm av} + \epsilon \Delta,
\end{align}
where $\Delta$ is some operator of $O(1)$.
Note that standard definitions of approximate designs require harder criteria (see, e.g., Ref.\,\cite{L2010}).
In this case, instead of Eq.\, \eqref{Eq:korekoredetail}, we have
\begin{multline}
    V^{(1)}(m, \cE| \Delta, O_{\rm meas}) 
    =
    A_0 + E+ A_1 f^m \\
+ \epsilon (m-2) (E_2 f^2 + E_1 f + E_0) +  O(m^2 \epsilon^2), \label{Eq:koredamedetail}
\end{multline}
where
\begin{align}
    &E_2 = \braa{\cE(O_{\rm meas})}\Pi_0 \Delta \Pi_0 \kett{O_{\rm ini}},\\
    &E_1 =\braa{\cE(O_{\rm meas})}( \Pi_0 \Delta \Pi_1 + \Pi_1 \Delta \Pi_0) \kett{O_{\rm ini}},\\
    &E_0 = \braa{\cE(O_{\rm meas})}\Pi_1 \Delta \Pi_1 \kett{O_{\rm ini}},\\
    &E = \braa{\cE(O_{\rm meas})} \{\Pi_0, \Delta\} + f \{\Pi_1, \Delta\} \kett{O_{\rm ini}}.
\end{align}

Comparing Eqs.\,\eqref{Eq:korekoredetail} and~\eqref{Eq:koredamedetail}, we observe that using approximate unitary $2$-designs result in more complicated form or the fitting function.

\subsection{The second-order RB} \label{SS:2RBdetail}
We now focus on the $2$-RB. Although the representation space in this case is
\begin{equation}
\cK = {\rm span} \bigl\{ \kett{\sigma_{\vec{n}_1  \otimes \vec{n}_2}} : \vec{n}_1, \vec{n}_2 \in \{0,1,2,3 \}^{q} \bigr\},\label{Eq:erccc}
\end{equation}
where we have used the notation that $\sigma_{\vec{n}_1  \otimes \vec{n}_2} = \sigma_{\vec{n}_1}  \otimes \sigma_{\vec{n}_2}$,
it is not necessary to consider the whole space because we assume that the initial operator $\Delta$ is traceless. This, together with the fact that the noise map is trace-preserving, implies that the operator remains traceless during the whole process. We also observe that the whole process is symmetric under the exchange of the first and the second spaces, each labeled by $\vec{n}_1$ and $\vec{n}_2$ in Eq.\,\eqref{Eq:erccc}. Hence, in the analysis of the $2$-RB, the relevant space is only the \emph{traceless symmetric} subspace defined by 
\begin{multline}
\cK_{TS} := {\rm span}\{ \kett{\sigma_{\vec{n}_1  \otimes \vec{n}_2} + \sigma_{\vec{n}_2  \otimes \vec{n}_1}}\\
 : \vec{n}_1, \vec{n}_2 \in \{0,1,2,3 \}^{q}, (\vec{n}_1, \vec{n}_2) \neq (\vec{0}, \vec{0}) \},
\end{multline}
where $\vec{0} = (0, \dots, 0)$.
The irreducible decomposition of $\cK_{TS}$ can be obtained by an extensive use of the result in Ref.\,\cite{HWW2018} (see Appendix~\ref{App:Irreps}), based on which we explicitly compute $V^{(2)}(m, \cE| \Delta, O_{\rm meas})$.

It turns out that the situation differs depending on whether $q=1$ or $q \geq 2$. We, hence, deal with the two cases separately.

\subsubsection{$2$-RB in a single-qubit system}
When $q = 1$, the irreducible decomposition of $\cK_{TS}$ is given by
\begin{equation}
\cK_{TS} = \cK_0 \oplus \cK_1,
\end{equation}
which is multiplicity-free. Here, $\cK_0$ and $\cK_1$ are 
\begin{align}
&\cK_0 := {\rm span} \{ \kett{\sigma_{1}^{\otimes 2} + \sigma_{2}^{\otimes 2} + \sigma_{3}^{\otimes 2}} \}, \label{Eq:1qv0}\\
&\cK_1 := {\rm span} \{ \kett{S_{1,2}}, \kett{S_{1,3}}, \kett{S_{2,3}}, \notag \\
&\hspace{15mm} \kett{\sigma_1^{\otimes 2} - 2 \sigma_2^{\otimes 2} + \sigma_3^{\otimes 2}}, \kett{\sigma_1^{\otimes 2} - \sigma_3^{\otimes 2}} \}, \label{Eq:1qv1}
\end{align}
respectively, with $S_{n, m} := (\sigma_n \otimes \sigma_m  + \sigma_m \otimes \sigma_n)/\sqrt{2}$. It is obvious that $d_0 := \dim \cK_0 = 1$ and $d_1 := \dim \cK_1 = 5$.

This decomposition implies that the expectation $V^{(2)}(m, \cE| \Delta, O_{\rm meas})$ is in the form of
\begin{equation}
V^{(2)}(m, \cE| \Delta, O_{\rm meas}) = A_0 C_0(\cE)^m + A_1 C_1(\cE)^m,
\end{equation}
where both $A_{\lambda}$ and $C_{\lambda}$ are given by
\begin{align}
&A_{\lambda} = \braa{\cE(O_{\rm meas})^{\otimes 2}} \Pi_{\lambda} \kett{\Delta^{\otimes 2}} \label{Eq:Ai1q},\\
&C_{\lambda} (\cE) = \frac{\tr[ \Pi_{\lambda} L_{\cE}^{\otimes 2} ]}{\tr[\Pi_{\lambda}]},
\end{align}
with $\Pi_{\lambda}$ being the projections onto $\cK_{\lambda}$. Since the projections can be explicitly constructed from Eqs.\,\eqref{Eq:1qv0} and~\eqref{Eq:1qv1}, we can compute $C_{\lambda}(\cE)$.

First, we have
\begin{align}
C_0(\cE) &= \frac{1}{3} \sum_{n, m=1}^3 \bigl(\braa{\sigma_n} L_{\cE} \kett{\sigma_m}\bigr)^2,\\
&=\frac{1}{3} \sum_{n=1}^3 \braa{\sigma_n} L_{\cE}^{\dagger} L_{\cE} \kett{\sigma_n},\\
&= u(\cE). \label{Eq:unitarity}
\end{align}

For $C_1 (\cE)$, we start from the relation that 
\begin{equation}
\Pi_1 = \Pi_{\rm sym} - \Pi_0 - \kettbraa{\sigma_0^{\otimes 2}}{\sigma_0^{\otimes 2}} - \sum_{n=1}^3\kettbraa{S_{0,n}}{S_{0,n}}, \label{Eq:1qubitRBstart}
\end{equation}
where $\Pi_{\rm sym}$ is the projection onto the symmetric subspace of $\cK^{\otimes 2}$. 
The projection $\Pi_{\rm sym}$ is also expressed by $(\mathbb{I} + \mathbb{F} )/2$. Here, $\mathbb{I}$ is the identity operator on $\cK^{\otimes 2}$ and $\mathbb{F}$ is the swap operator on $\cK^{\otimes 2}$ defined by $\sum_{n,m=0}^3 \kettbraa{\sigma_n}{\sigma_m} \otimes \kettbraa{\sigma_m}{\sigma_n}$. Using the swap trick, we have
\begin{align}
\tr[ \Pi_{\rm sym} L_{\cE}^{\otimes 2}] =\frac{1}{2}\bigl( \tr[L_{\cE}]^2 +\tr[L_{\cE}^2] \bigr).
\end{align}
Moreover, from the direct calculations, we obtain
\begin{align}
&\tr[ \kettbraa{\sigma_0^{\otimes 2}}{\sigma_0^{\otimes 2}} L_{\cE}^{\otimes 2}] = L_{00}^2,\\
&\tr[ \kettbraa{S_{0,n}}{S_{0,n}} L_{\cE}^{\otimes 2}] = L_{00}L_{nn} + L_{0n}L_{n0},
\end{align}
where $L_{nm} = \braa{\sigma_n} L_{\cE} \kett{\sigma_m}$.
Further using the relations 
\begin{align}
&\tr[L_{\cE}] = L_{00} + \tr[\tilde{L}_{\cE}],\\
&\tr[L_{\cE}^2] = L_{00}^2 + 2 \sum_{n=1}^3 L_{0n}L_{n0} + \tr[\tilde{L}_{\cE}^2],
\end{align}
we obtain from Eq.\,\eqref{Eq:1qubitRBstart} that
\begin{align}
\tr[ \Pi_1 L_{\cE}^{\otimes 2}]
&=
\frac{1}{2} \tr[\tilde{L}_{\cE}^2] - u(\cE)+\frac{1}{2} \tr[\tilde{L}_{\cE}]^2,\\
&=
\frac{3}{2} h(\cE) - u(\cE)+\frac{9}{2} f(\cE)^2.
\end{align}

Altogether, we obtain
\begin{multline}
V(m, \cE| \Delta, O_{\rm meas}) \\= A_0 u(\cE)^m +
A_1  \biggl(\frac{9}{10} f(\cE)^2 - \frac{1}{5}u(\cE) +   \frac{3}{10} h(\cE)\biggr)^m.
\end{multline}

\subsubsection{$2$-RB in a multi-qubit system}
For a multi-qubit system ($q \geq 2$), the traceless symmetric subspace is decomposed into four irreducible subspaces:
\begin{equation}
\cK_{TS} = \cK_0 \oplus \cK_{\rm I} \oplus \cK_{\rm II} \oplus \cK_{\rm III},
\end{equation}
where $\cK_0 = {\rm span} \{ \sum_{\vec{n} \neq \vec{0}} \kett{ \sigma_{\vec{n} \otimes \vec{n}}} \}$ and the others are given in Appendix~\ref{App:Irreps}. Each irrep is multiplicity-free.
We denote by $D_{\lambda}$ the dimension of each subspace, which are 
\begin{align}
&D_0 = 1,\\
&D_{\rm I} = \frac{d^2(d-1)(d+3)}{4},\\
&D_{\rm II}  = \frac{d^2(d + 1)(d-3)}{4},\\
&D_{\rm III}  = d^2-1.
\end{align}

Since the decomposition is multiplicity-free, $V^{(2)}(m, \cE| \Delta, O_{\rm meas})$ is a sum of four exponentially decaying functions. Furthermore, from the fact that $\cK_0 ={\rm span}\{ \sum_{\vec{n} \neq \vec{0}} \kett{ \sigma_{\vec{n} \otimes \vec{n}}} \}$, we obtain that $C_0(\cE) = u(\cE)$. Hence, we have
\begin{equation}
V^{(2)}(m, \cE| \Delta, O_{\rm meas}) = A_0  u(\cE)^m
+
\sum_{\lambda={\rm I}, {\rm II}, {\rm III}} A_{\lambda} C_{\lambda}(\cE)^m,
\end{equation}
where 
\begin{align}
&A_{\lambda} = \braa{\cE(O_{\rm meas})^{\otimes 2}} \Pi_{\lambda} \kett{O_{\rm ini}^{\otimes 2}} \label{Eq:Ai1q},\\
&C_{\lambda} (\cE) = \frac{\tr[ \Pi_{\lambda} L_{\cE}^{\otimes 2} ]}{\tr[\Pi_{\lambda}]},
\end{align}
with $\Pi_{\lambda}$ being the projections onto the irrep $\cK_{\lambda}$. 

It is not clear whether each $C_{\lambda}(\cE)$ ($\lambda = {\rm I}, {\rm II}, {\rm III}$) has a clear physical meaning, such as $C_0(\cE) = u(\cE)$ being the unitarity. However, a linear combination of them  does. To see this, we use the relation that 
\begin{multline}
\Pi_{\rm I} + \Pi_{\rm II} +  \Pi_{\rm III} =\\
\Pi_{\rm sym} -\Pi_0 
- \kettbraa{\sigma_{\vec{0}\otimes \vec{0}}}{\sigma_{\vec{0} \otimes \vec{0}}} -  \sum_{\vec{n} \neq \vec{0}} \kettbraa{S_{\vec{0} \vec{n}}}{S_{\vec{0} \vec{n}}},
\end{multline}
where $S_{\vec{0} \vec{n}} := (\sigma_{\vec{0} \otimes \vec{n}} + \sigma_{\vec{n} \otimes\vec{0}})/\sqrt{2}$.
From this relation, we can show, by a calculation similar to the one-qubit case, that
\begin{multline}
\tr\bigl[ (\Pi_{\rm I} + \Pi_{\rm II} +  \Pi_{\rm III} ) L_{\cE}^{\otimes 2} \bigr]\\
=
\frac{(d^2 - 1)^2}{2} f(\cE)^2  - u(\cE) + \frac{d^2-1}{2} h(\cE).
\end{multline}
Since $\tr[\Pi_{\lambda} L_{\cE}^{\otimes 2}] = \tr[\Pi_{\lambda}] C_{\lambda}(\cE) = D_{\lambda} C_{\lambda}(\cE)$, we obtain
\begin{equation}
\sum_{\lambda={\rm I}, {\rm II}, {\rm III}} D_{\lambda} C_{\lambda}(\cE)
=
\frac{(d^2 - 1)^2}{2} f(\cE)^2  - u(\cE) + \frac{d^2-1}{2} h(\cE). \label{Eq:ervrcr}
\end{equation}

We finally note that Tab.\,\ref{Tab:coeff} in Subsec.\,\ref{SS:2RB} is obtained by constructing the orthonormal basis in each subspace $\cK_{\rm I}$, $\cK_{\rm II}$, and $\cK_{\rm III}$ (see Appendix~\ref{App:Irreps}). We also assume that the noise $\cE$ is weak, so that $\cE(O_{\rm meas}) \approx O_{\rm meas}$. Based on this assumption, we have
\begin{equation}
A_{\lambda} \approx \braa{O_{\rm meas}^{\otimes 2}} \Pi_{\lambda} \kett{O_{\rm ini}^{\otimes 2}},
\end{equation}
enabling us to compute $A_{\lambda}$ for given initial and measurement operators.

\begin{figure*}
    \centering
    \includegraphics[width=\textwidth]{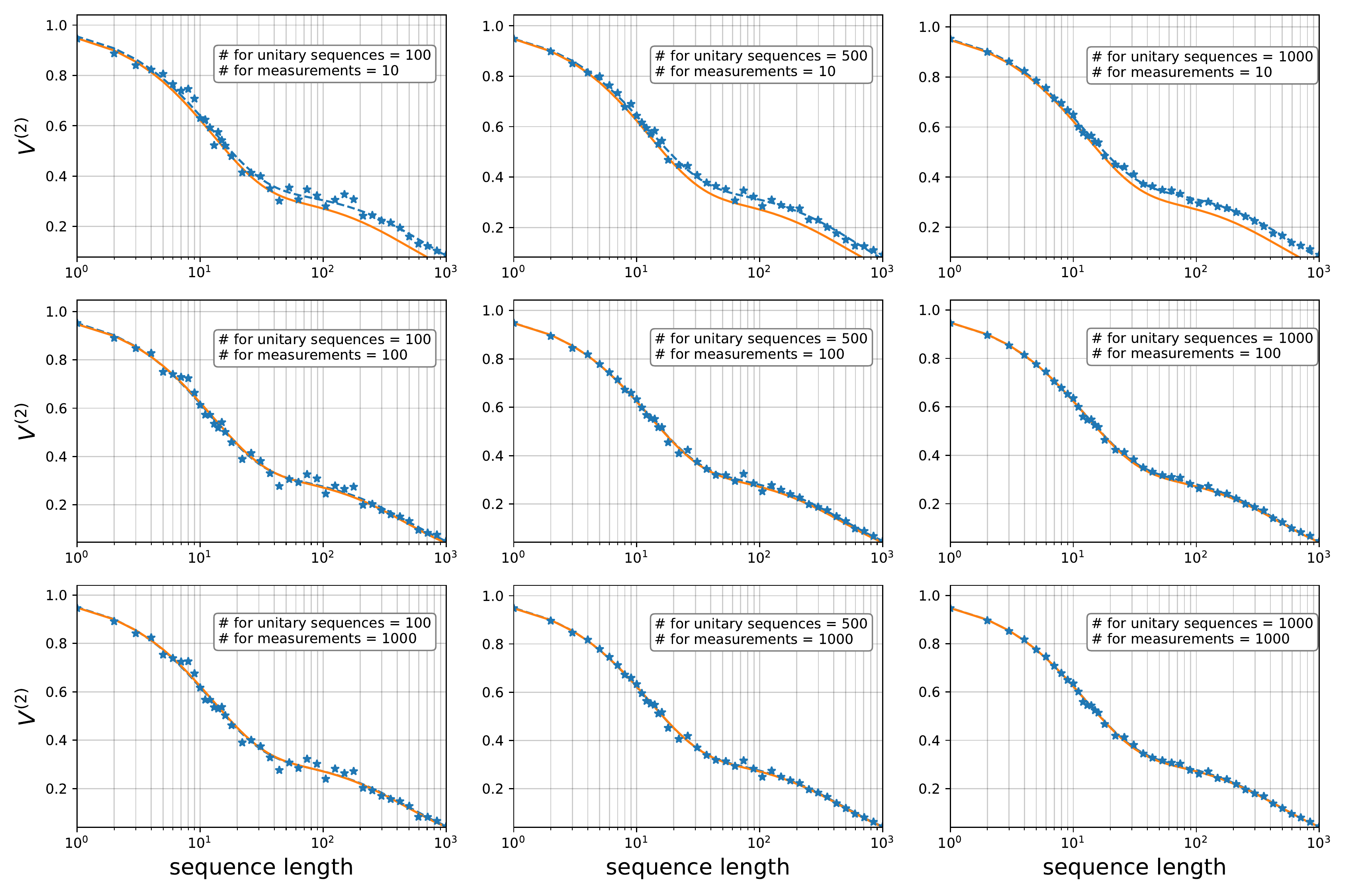}
    \caption{Numerical results for the single-qubit $2$-RB with various numbers of samplings for measurement and those for unitary sequences when the noise parameters are set to $p=q=0.02$. The dots correspond to the values of $V^{(2)}$ obtained from the given numbers of samplings for measurement and for unitary sequences, and the dashed lines are fitting results. Theoretical values are shown in the solid lines.}
        \label{fig:one_qubit_decay}
\end{figure*}

\section{Experimental realization of $2$-RB} \label{S:NumExpDetail}
Based on the former sections, we explain in detail how we have experimentally implemented the $2$-RB and estimated the self-adjointness of the noise in the system.
In Subsec.\,\ref{SS:NumericalEvalDetail}, we provide the details of the numerical evaluation of the one- and two-qubit $2$-RB discussed in Sec.\,\ref{SS:NumericalEval}.
The details of experiment is given in Subsec.\,\ref{SS:ExperimentDetail}.

\subsection{Numerical analysis} \label{SS:NumericalEvalDetail}

\subsubsection{Single-qubit systems}

We explain the fitting procedure of the 2-RB on one-qubit systems in detail. 
The noise we consider is given by the following CPTP map: 
\begin{equation}
\mathcal{E}_1(\rho) = q e^{i\theta X}  \rho e^{- i\theta X} + (1-q)((1-p) \rho + p X \rho X), \label{Eq:1qubitnoise}
\end{equation}
which is characterized by three parameters $p, q$, and $\theta$. We particularly choose $\theta$ as $p = \sin^2 \theta$ for the fidelity parameter $f(\mathcal{E}_1)$ to be independent of the coherence parameter $q$. 
Using the Liouville representation, it is straightforward to compute the fidelity parameter, the unitarity, and the self-adjointness parameter of this noise. They are, respectively, given by
\begin{align}
    &f(\cE) = 1 - \frac{4}{3}p,\\
    &u(\cE) = 1 - \frac{8}{3}p(1-p)(1-q^2),\\
    &h(\cE) = 1- \frac{8}{3}p(1-p)(1+q^2).
\end{align}
As shown in Theorem~\ref{Thm:2RB}, $V^{(2)}$ for one qubit is
\begin{multline}
V^{(2)}(m, \cE_1| \Delta, O_{\rm meas}) \\= A_0 u(\cE_1)^m + A_1 \biggl( \frac{9}{10}f(\cE_1)^2 - \frac{1}{5} u(\cE_1) + \frac{3}{10} h(\cE_1) \biggr)^m. \label{Eq:162}
\end{multline}

To obtain Fig.\,\ref{Fig:numerical_overview_1q} in Subsec.\,\ref{SS:NumericalEval}, we first estimate the fidelity parameter $f(\mathcal{E}_1)$ from the $1$-RB, i.e., by fitting $V^{(1)}(m, \mathcal{E}_1| \ket{0}\bra{0}, \ket{0}\bra{0})$ and then obtain $u(\mathcal{E}_1)$ and $h(\mathcal{E}_1)$ from the fitting results of $V^{(2)}(m, \mathcal{E}_1|Z, \ket{0}\bra{0})$ and the estimated value of $f$. 
The fitting of $V^{(2)}$ is first performed based on Eq.\,\eqref{Eq:162} by regarding the coefficients $A_0, A_1$ and the two exponential decaying rates as free parameters. 
Then, we subtract the first exponential curve of Eq.\,\eqref{Eq:162} from the data and carry out the fitting of the second exponential curve again where we consider $A_1$ and the base of the second exponential curve as free parameters. This procedure is redundant, but turns out to improve the accuracy of the fitting because, in most cases, the first exponential decaying rate is larger than the second one, and thus the second exponential curve is clearly visible in the region of long sequence length. 

We also estimate how many measurements and unitary sequences suffice to obtain a good estimate of the noise parameters. In Fig.\,\ref{fig:one_qubit_decay}, we provide $V^{(2)}$ for various numbers of measurement and samplings of unitary sequences. We observe that $1000$ for both suffice to obtain a good estimate when $p = q = 0.02$, which corresponds to the gate fidelity $97.3\%$.

When the number of measurements is small, experimental values are positively biased from theoretical values with the infinite number of samplings (see the figures in the top line of Fig.\,\ref{fig:one_qubit_decay}).
This difference can be understood as follows. Let $\langle O_{\rm meas} \rangle_{O_{\rm ini}, \bm{i}, n}$ be the expectation value for a random sequence described by $\bm{i}$ with the finite number $n$ of measurements. We describe the expectation value and variance of this random variable averaged over all choices of the unitary sequence as $\mu$ and $\sigma^2/n$, respectively. Note that this mean value is independent of $n$, and this variance is inverse proportional to $n$ since $\langle O_{\rm meas} \rangle_{O_{\rm ini}, \bm{i}, n}$ is a linear combination of binomial distribution. $V^{(2)}$ is the expectation value of squared random variable $\langle O_{\rm meas} \rangle^2_{O_{\rm ini}, \bm{i}}$, and its expectation value over random sequences is derived as 
\begin{align}
\mathbb{E}[\langle O_{\rm meas} \rangle^2_{O_{\rm ini}, \bm{i}}] &= \mathbb{E}[(\langle O_{\rm meas} \rangle_{O_{\rm ini}, \bm{i}} - \mu)^2] + \mu^2 \\
&= \sigma^2/n + \mu^2.
\end{align}
Therefore, an experimentally obtained value with a finite number of measurements is positively biased by $\sigma^2/n$. In practice, we can remove the effect of this bias by increasing the number of sampling $n$ and using the region satisfying $\mu^2 \ll \sigma^2/n$ for fitting.

\begin{figure}[tb!]
    \centering
    \includegraphics[width=.4\textwidth]{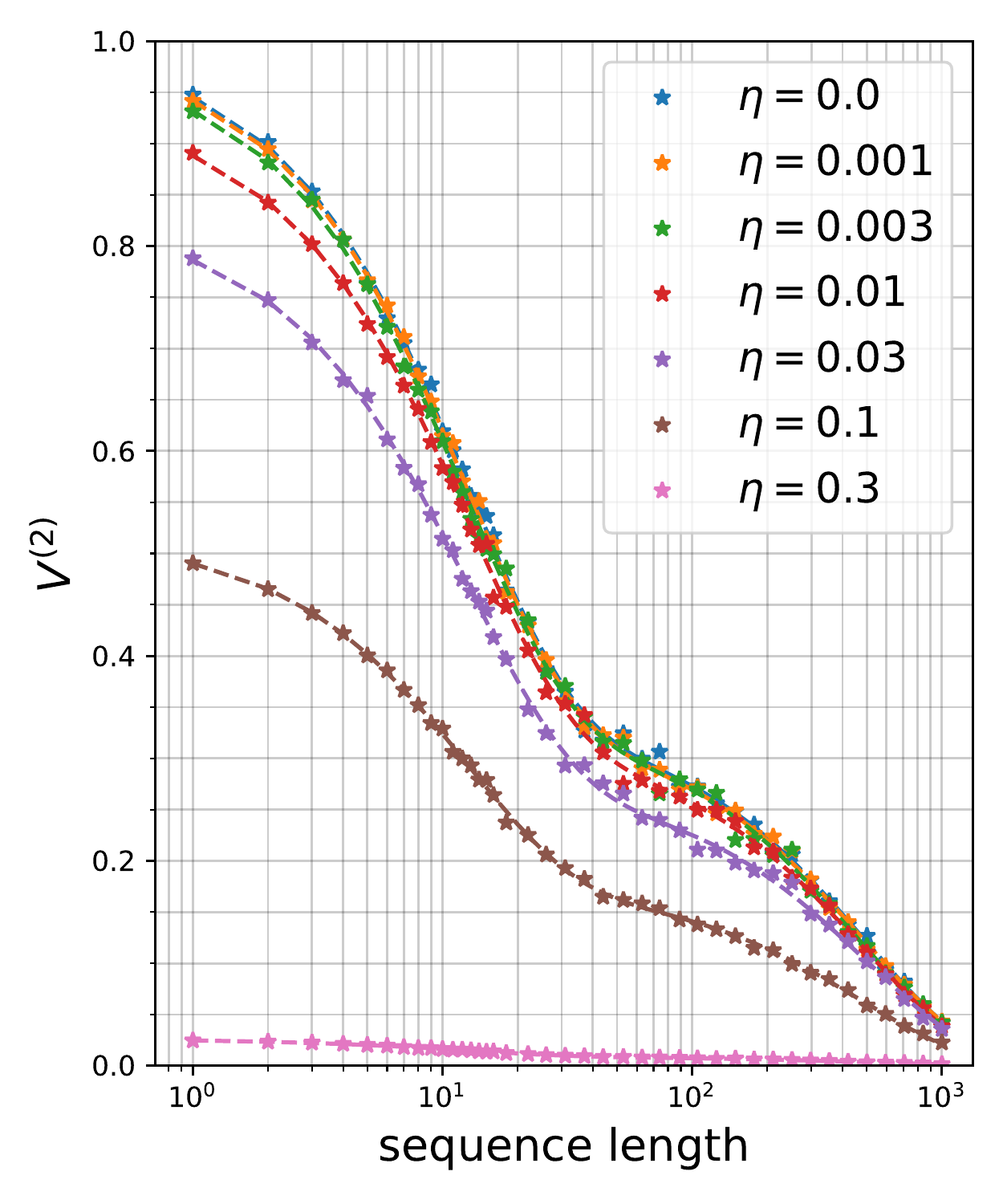}
    \caption{Numerical results for the single-qubit $2$-RB with SPAM errors for $p=q=0.02$. We have taken $1000$ measurements and $1000$ random unitary sequences. The dots represent the numerically obtained data, and the dashed lines are the fitting curves.}
    \label{Fig:numerical_SPAM_1q}
\end{figure}

We finally check the robustness of the $2$-RB on single-qubit systems against SPAM errors. Although the $2$-RB is ideally SPAM-error free, the fitting may become harder with the existence of SPAM errors. Our analysis, however, reveals that this is unlikely the case.
Here, we model the state-preparation error $\eta_{\mathrm{prep}}$ as $\rho=\eta_{\mathrm{prep}}\ketbra{0}{0}+(1-\eta_{\mathrm{prep}})\ketbra{1}{1}$ and $\rho'=(1-\eta_{\mathrm{prep}})\ketbra{0}{0}+\eta_{\mathrm{prep}}\ketbra{1}{1}$, and measurement error $\eta_{\mathrm{meas}}$ as readout bit-flip error, i.e., the POVM is $\{\Pi_{x'}|x' \in \{0,1\} \}$ where $\Pi_{x'}=\sum_{x \in \{0,1\}}\eta_{\mathrm{meas}}(x'|x) \Pi_x$ and $\eta_{\mathrm{meas}}(x'|x)$ is conditional probability.
In this numerical experiments, we assume that $\eta=\eta_{\mathrm{prep}}=\eta_{\mathrm{meas}}(0|1)=\eta_{\mathrm{meas}}(1|0)$ and get 1000 samples and 1000 random sequences. We set the parameters $p = q= 0.02$. 

The results are provided in Fig.\,\ref{Fig:numerical_SPAM_1q} and Tab.\,\ref{Tab:numerical_SPAM_1q}. The relative errors of estimates for $F, u$, and $H$ are within $5\%$ except for the estimate for $u$ when $\eta=0.3$, thus the 2-RB is likely to work well even in realistic situations with SPAM errors as expected from the analytical studies.

\begin{table}[t!]
\centering
\caption{The estimates of $F,u$,and $H$ from the numerics shown in Fig.\,\ref{Fig:numerical_SPAM_1q}. The parameter $\eta$ is for SPAM errors such as $\eta = \eta_{\mathrm{prep}}=\eta_{\mathrm{meas}}(0|1)=\eta_{\mathrm{meas}}(1|0)$. The theoretical values are shown at the bottom of the table.}
\begin{tabular}{c||r|r|r}
SPAM error $\eta$ & $F$ & $u$ & $H$ \\ \hline \hline
0.0  & $0.986(6)$ & $0.9979(5)$ & $0.92(5)$\\ 
0.001 & 0.986(5) & 0.9979(6) & 0.92(7)\\
0.003  & 0.986(5) & 0.9978(5) & 0.92(1) \\
0.01 & 0.987(0) & 0.9979(2) & 0.92(2) \\
0.03 & 0.986(5)& 0.9980(1)& 0.92(8)\\
0.1 & 0.986(5) & 0.9978(9)& 0.92(7)  \\
0.3 & 0.986(6) & 0.9981(8)& 0.92(7)  \\ \hline
 & 0.9866 & 0.99793 & 0.9247
\end{tabular}
\label{Tab:numerical_SPAM_1q}
\end{table}

\subsubsection{Two-qubit systems}

\begin{figure*}[tbh!]
    \centering
    \includegraphics[width=\textwidth]{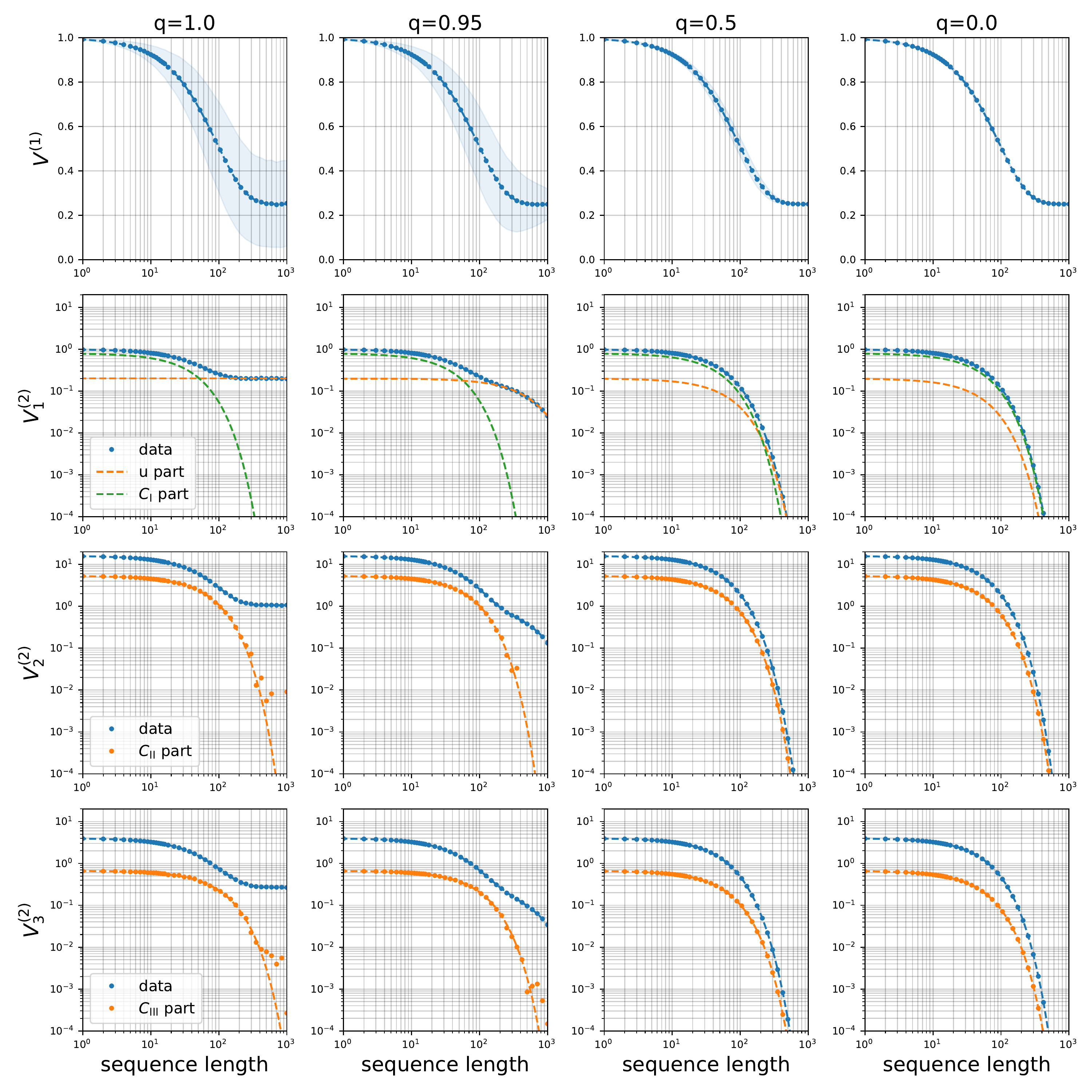}
    \caption{A step-by-step fitting process of $2$-RB for two-qubit systems are shown. The first line of the matrix are about the standard RB ($1$-RB). The second, third, and fourth lines show $2$-RB with two, three, and four exponentially decaying functions, respectively. The initial and measurement operators for $2$-RB are chosen according to Tab.\,\ref{Tab:coeff}. See the main text for detail.}
    \label{fig:two_qubit_decay}
\end{figure*}

For two-qubit systems, we investigate the noise given by
\begin{align}
\mathcal{E}_2(\rho) &= q e^{i \theta (X\otimes X)}  \rho e^{- i\theta (X \otimes X)} \nonumber \\ 
&+ (1-q)((1-p) \rho + p (X \otimes X) \rho (X\otimes X) ),
\end{align}
where we set $\theta$ to $p = \sin^2 \theta$.
From Theorem~\ref{Thm:2RB}, $V^{(2)}$ in this case shall be in the form of
\begin{equation}
V^{(2)}(m, \cE_2| \Delta, O_{\rm meas}) = A_0 u(\cE_2)^m + \sum_{\lambda={\rm I}, {\rm II}, {\rm III}} A_{\lambda} C_{\lambda}(\cE_2)^m, \label{Eq:1623}
\end{equation}
and $C_{\lambda}(\cE_2)$ satisfy
\begin{multline}
84 C_{\rm I}(\cE) + 20C_{\rm II}(\cE) + 15 C_{\rm III}(\cE) \\
=
\frac{225}{2} f(\cE)^2 + \frac{15}{2} h(\cE) - u(\cE). \label{Eq:19}
\end{multline}
With this setting, theoretical values are derived as follows.
\begin{align}
&f(\cE_2) = 1-\frac{16}{15}p,\\
&u(\mathcal{E}_2) = 1 - \frac{32}{15}p(1-p)(1-q^2),\\
&C_{\rm I}(\mathcal{E}_2) = 1 - \frac{4}{105}p \bigl(56-31p+14(1-p)q^2\bigr),\\
&C_{\rm II}(\mathcal{E}_2) = 1 - \frac{4}{15}p \bigl(8-5p-2(1-p)q^2 \bigr),\\
&C_{\rm III}(\mathcal{E}_2) = 1 - \frac{16}{15}p \bigl(2-p-(1-p)q^2 \bigr),
\end{align}
from which the theoretical value of the self-adjointness parameter $h(\cE)$ can be computed from Eq.\,\eqref{Eq:19}.

A step-by-step fitting process of two-qubit systems is shown in Fig.\,\ref{fig:two_qubit_decay} for $p=0.01$. The numbers of measurement and samplings of unitary sequences are both set to $10^4$.
The fidelity parameters $f(\mathcal{E}_2)$ can be obtained from $V^{(1)}(m, \mathcal{E}_1 | \ket{00}\bra{00}, \ket{00}\bra{00})$, which are plotted in the top line of the figure. In the figure, dashed lines are theoretical values, and the shaded area represents a standard deviation of each data. When $q$ is near to unity, standard deviations are large even when the sequence length increased. This is because the final quantum state is nearly pure state when $q \sim 1$, and probability distributions fluctuate randomly according to chosen random sequences. On the other hand, when $q \sim 0$, the final quantum state quickly converges to the maximally mixed state, and thus probability distribution becomes independent of the chosen random sequences.

We then fit four values $u, C_{\rm I}, C_{\rm II}, C_{\rm III}$ step by step where initial and measurement operators $\Delta$ and $O_{\rm meas}$ are chosen according to Tab.\,\ref{Tab:coeff}. The fitting results for several $q$ are shown in Tab.\,\ref{Tab:two_qubit_data}.
\begin{table}[t!]
\centering
\begin{tabular}{c||l|l|l|l}
$q$ & $u$ & $C_{\rm I}$ & $C_{\rm II}$ & $C_{\rm III}$ \\ \hline \hline
$0.00$ & $0.9792(7)$ & $0.9787(2)$ & $0.978869(7)$ & $0.97896(1)$ \\ 
       & $0.97888$   & $0.97879$   & $0.9788000$   & $0.978773$ \\ \hline
$0.50$ & $0.98416(9)$ & $0.97744(3)$ & $0.98015(1)$ & $0.98150(2)$ \\ 
       & $0.984160$   & $0.977465$   & $0.980120$   & $0.981413$ \\ \hline
$0.95$ & $0.997942(9)$ & $0.97401(3)$ & $0.98354(3)$ & $0.98830(5)$ \\ 
       & $0.9979410$   & $0.974020$   & $0.983565$   & $0.988304$ \\ \hline
$1.00$ & $0.999994(6)$ & $0.97349(6)$ & $0.98414(5)$ & $0.98930(9)$ \\ 
       & $1.0000000$   & $0.973505$   & $0.984080$   & $0.989333$
\end{tabular}
\caption{Fitting results for $p=0.01$ in the two-qubit 2-RB. In each cell, fitting results are written in the first line, and theoretical values are written in the second line.}
\label{Tab:two_qubit_data}
\end{table}

\begin{table}[t!]
\centering
\begin{tabular}{c||l|l|l}
$q$ & $F$ & $u$ & $H$ \\ \hline \hline
$0.00$ & $0.9920004(5)$ & $0.9792(7)$ & $0.999(2)$ \\ 
       & $0.99200000$   & $0.97888$   & $1.0000$    \\ \hline
$0.50$ & $0.991998(3)$ & $0.98416(9)$ & $0.9902(3)$ \\ 
       & $0.9920000$   & $0.984160$   & $0.99010$    \\ \hline
$0.95$ & $0.99203(1)$ & $0.997942(9)$ & $0.9630(5)$ \\ 
       & $0.992000$   & $0.9979408$   & $0.96426$    \\ \hline
$1.00$ & $0.99198(1)$ & $0.999994(6)$ & $0.9610(8)$ \\ 
       & $0.992000$   & $1.0000000$   & $0.96040$    
\end{tabular}
\caption{Processed data for $p=0.01$ in the two-qubit 2-RB. In each cell, fitting results are written in the first line, and theoretical values are written in the second line.}
\label{Tab:two_qubit_data_process}
\end{table}

First, we obtain $u$ and $C_{\rm I}$ from $V^{(2)}_1 := V^{(2)}(m, \mathcal{E}_2 | ZZ, \ket{00}\bra{00})$. The obtained results are shown in the second line of Fig.\,\ref{fig:two_qubit_decay}. The sampled data points and fitting results are shown as blue points and dashed lines, respectively. These lines are linear combinations of two exponentially decaying function. Exponential decays with coefficient $u$ and $C_{\rm I}$ are shown as orange and green lines, respectively. While we can clearly see two exponential decays for coherent noise, i.e., in the case of $q \sim 1$, an exponential decay of $C_{\rm I}$ part becomes dominant when noise becomes probabilistic. Even in this case, the fitting results are still reliable as shown in Tab.\,\ref{Tab:two_qubit_data}.

Then, we estimate $C_{\rm II}$ from $V^{(2)}_2 := V^{(2)}(m, \mathcal{E}_2 | ZZ, ZZ)$ and $C_{\rm III}$ from $V^{(2)}_3 := V^{(2)}(m, \mathcal{E}_2 | \rho_-, \rho_-)$, where these initial and measurement operators are chosen from the third and fourth columns of Tab.\,\ref{Tab:coeff}. The obtained results are shown in the third and fourth lines of Fig.\,\ref{fig:two_qubit_decay}.
In each figure, numerical data is plotted as blue circles and fitting results are shown as dashed lines. In each fitting process,  only a single exponentially decaying term is unknown in advance. We showed unfitted exponential decay as orange circles and fitting results as dashed lines. Although accuracy of orange data becomes not reliable when its value becomes much smaller than the others, we can fit $C_{\rm II}$ and $C_{\rm III}$ reliably.

We calculate the averaged fidelity $F(\cE)$, unitarity $u(\cE)$, and self-adjointness $H(\cE)$ from the fitting results. The processed values are shown in Tab.\,\ref{Tab:two_qubit_data_process}. We evaluate relative errors for all the plots with the same method as the single-qubit $2$-RB, and confirm that the relative errors are below $4\%$ for all the cases of $p=0.01, 0.02, 0.1, 0.2$ except the case when the theoretical value is exactly zero. While the relative errors of the self-adjointness become a few tens of percent in the case of $p=0.4$, we can say that reliable values can be obtained when the fidelities of operations are sufficiently high.
Note that, although standard deviations of the fitting results become large when $u$ is almost equal to $C_{\rm I}$, we confirmed that the fitted results are close to theoretical values even in such cases. Thus, we conclude that $2$-RB works also for two-qubit systems.

\subsection{Details of the experiments}\label{SS:ExperimentDetail}
In this section, we provide the details of the experiments in Sec.\,\ref{SS:Experiments}.
A superconducting qubit can be regarded as a sort of the LC resonant circuit, where a Josephson junction is an effective inductance, and has the Hamiltonian equivalent to that of the one-dimensional free particle trapped in anharmonic potential.
\begin{table}[t!]
\centering
\caption{The parameter fields of the qubits.}
\begin{tabular}{c||c|c|c|c|c}
    & $\omega_q/2\pi$   & $\alpha/2\pi$         & $T_1$                 & $T_2$ echo            \\ \hline \hline
Q1  & $9.077~{\rm GHz}$ & $-328.9~{\rm MHz}$    & $9.724~{\rm \mu s}$   & $13.670~{\rm \mu s}$  \\
Q2  & $8.927~{\rm GHz}$ & $-419.9~{\rm MHz}$    & $12.634~{\rm \mu s}$   & $15.763~{\rm \mu s}$
\end{tabular}
\label{Tab:Experiment_parameter_field}
\end{table}
The parameter fields of the qubits are summarized in Tab.\,\ref{Tab:Experiment_parameter_field}.

All unitary gates required in the $2$-RB for a single-qubit case were implemented by two $R_X(\pi/2)$ gates and three $R_Z$ gates with an arbitrary rotation angle, which are implemented by the shaped microwave pulse (Half-DRAG) with the length $11.70~{\rm ns}$~\cite{lucero2010reduced} and the Virtual-$Z$ gates~\cite{mckay2017efficient}, respectively.
The pre-measured averaged gate fidelity of the single-qubit Clifford gate is $0.991$.

The single-shot qubit readout was performed via the impedance-matched Josephson parametric amplifier~\cite{doi:10.1063/1.4886408}, and the assignment fidelity of the readout was $0.943$.

\begin{figure}[tb!]
    \centering
    \includegraphics[width=.4\textwidth]{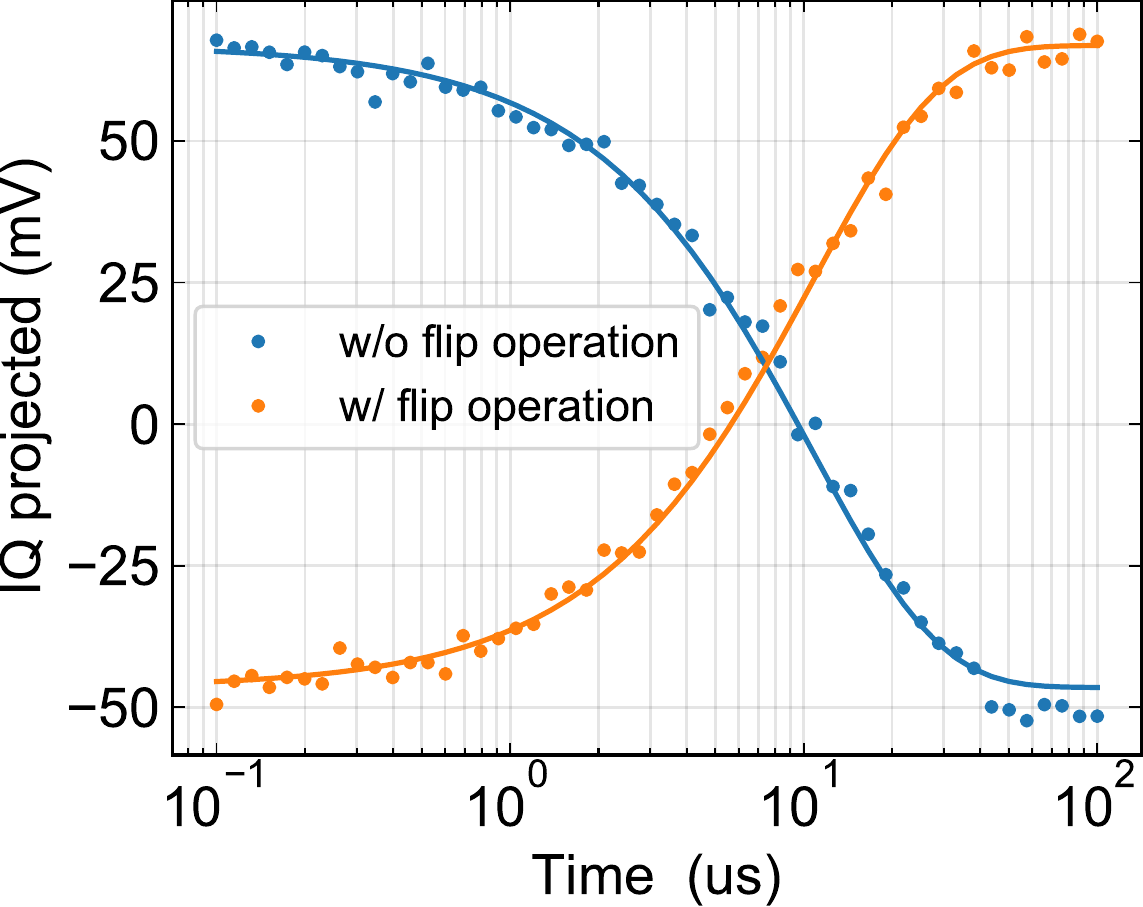}
    \caption{
    T1 decay experiment.
    The horizontal axis represents the delay time, the vertical axis represents the projected value of the IQ readout signal.
    The blue and orange dots represents the experimental results of the T1 decay experiments with and without flip operation just before measurement, respectively.
    The lines are fitting curves.
    }
    \label{Fig:experiment_t1}
\end{figure}

\begin{figure}[tb!]
    \centering
    \includegraphics[width=.4\textwidth]{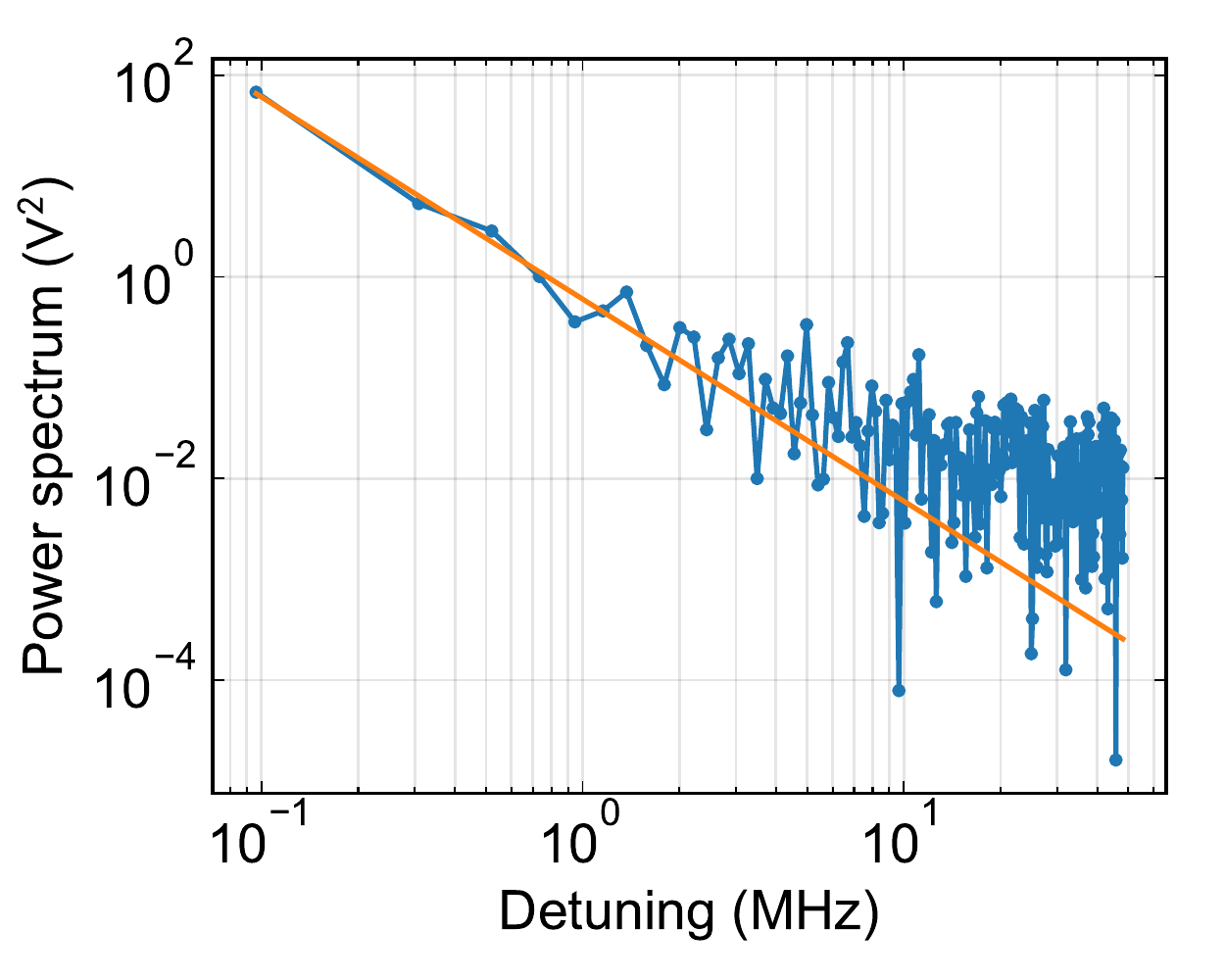}
    \caption{
    Ramsey oscillation experiment.
    The horizontal axis represents the detuning from the qubit eigenfrequency, and  the vertical axis represents the power spectrum of the Ramsey oscillation.
    The points connected by the blue line represent the experimental data, and the orange line represents the fitting curve.
    }
    \label{Fig:experiment_ramsey}
\end{figure}

As supplemental experiments, T1 decay and Ramsey oscillation were observed to clarify the background noise source of Q1.
The experimental results of the T1 decay is shown in Fig.\,\ref{Fig:experiment_t1}, where the horizontal and the vertical axes represent the delay time and the projected value of the IQ readout signal, respectively.
The blue and orange dots represent the experimental results with and without a flip operation just before measurement, respectively. The lines provide the fitting curves. As seen from the result, the T1 decay of Q1 follows exponential behavior, which is consistent with the expected behavior in isolated qubits.

The experimental results of the Ramsey oscillation is also given in Fig.\,\ref{Fig:experiment_ramsey}, where the horizontal and vertical axes represent the detuning from the qubit eigenfrequency and the power spectrum of the Ramsey oscillation, respectively.
The points connected by a blue line represent the experimental data, and the orange line provides a fitting curve.
In the fitting, we did not take the data in the small power spectrum region ($<1~\mathrm{V}^2$) into account. This is because the noise floor derived from the white noise is dominant there.
As seen from the result, the power spectrum of the Ramsey oscillation has no peaks other than the qubit eigenfrequency.
From the fitting curve, it was found that the detuning from the qubit eigenfrequency $\Delta$~($\mathrm{MHz}$) and the power spectrum of the Ramsey oscillation $PS(\Delta)$~($\mathrm{V^2}$) are related as
\begin{align}
PS(\Delta) \propto \Delta^{-2.004}.
\end{align}
This is consistent with the expected behavior when the transmon qubits are isolated well.

From the results of these supplemental experiments, we conclude that Q1 is not in the strong coupling regime with any noise source, which implies that the background noise of Q1 is time-independent. Thus, the requirements for $2$-RB are met in our experimental system.

\section{Summary and Discussions} \label{S:SD}

In this paper, we have provided an explicit constructions of \emph{exact} unitary $t$-designs for any $t$. In particular, quantum circuits for exact unitary $t$-designs on $N$ qubits have been provided for the first time. Our construction is inductive with respect to the number of qubits. Hence, all constructions obtained in this paper are inefficient when the number of qubits is large, implying that it is of practical use only when the size of the system is small.

As an application of exact unitary $2t$-designs on a small system, we have proposed the $t$-RB, which enables us to experimentally estimate higher-order properties of the noise on a quantum system.
Since the unitary designs are used in multiple times in a single run of the protocol, it is important for the design to be exact.
After providing a general scheme of the $t$-RB, we have studied the $2$-RB in detail.
It was shown that the $2$-RB reveals the self-adjointness of the noise, a new characterization of the noise that we argue to play an important role in QEC especially when decoders are based on applications of Pauli operators. Our results have been demonstrated numerically, which shows that the $2$-RB is experimentally tractable. We have then experimentally implemented the single-qubit $2$-RB on the superconducting qubit system. From the experimental results, we found that the characteristics of the background noise of a qubit changes depending on the presence of the interaction with the adjacent qubits.

Our results open a number of future problems. Regarding the implementation of $t$-designs, it is important to improve the efficiency. Despite that the inefficiency in our construction is likely to be intrinsic due to an inductive nature of the construction, the representation-theoretic method provides a way to searching more efficient ones. More specifically, the key in the construction is the relation between the representation of the whole unitary group and that of a certain subgroup of the unitary group. This indicates that finding the construction of exact unitary designs may be reduced to the problem of searching for a subgroup whose representation has a good relation to that of the whole unitary group.

It is also important to further develop the theory of the $t$-RB protocol. In this paper, we have analyzed only the $2$-RB in detail. It then turns out that self-adjointness of the noise can be revealed. It is of great interest to concretely investigate what characterization of the noise can be generally obtained from the $t$-RB. In the context of QEC, it is also important to comprehensively analyze quantitative relations between the self-adjointness and the feasibility of QEC.
Another promising future problem is to use exact higher-designs in the other RB-type protocols.
The $t$-RB is a straightforward generalization of the standard RB. However, there are numerous variant protocols~\cite{HHFFW2019}, most of which, if not all, are based on the Clifford group that is an exact unitary $2$-design. By extending such protocols to those with higher-designs, the noise on the system can be characterized in more detail.

\section{Acknowledgements}
The authors are grateful to H. Yamasaki and A. Darmawan for helpful discussions.
Y. Nakata is supported by JST, PRESTO (No. JPMJPR1865). 
Y. Suzuki is supported by JST, PRESTO (No.\,JPMJPR1916).
Y. Suzuki, K. Heya, Z. Yan, K. Zuo, S. Tamate., Y. Tabuchi, and Y. Nakamura are supported by JST ERATO (No. JPMJER1601) and by MEXT Q-LEAP (No. JPMXS0118068682).

\bibliographystyle{unsrt}

\bibliography{bib2.bib}

\appendix

\section{Decomposing $R_X(\theta_{\lambda})$ into two-qubit gates with arbitrary accuracy} \label{App:RX}

We here explicitly show how to decompose $R_X(\boldsymbol{\theta}_{\lambda})$ defined by
\begin{equation}
R_X(\boldsymbol{\theta}_{\lambda})= \sum_{\boldsymbol{j} \in \{0, 1\}^N}  e^{i \theta_{\lambda}^{(\boldsymbol{j})} X} \otimes  \ketbra{\boldsymbol{j}}{\boldsymbol{j}}
\end{equation}
into two-qubit gates with arbitrary precision under the assumption that ancillary qubits and an oracle that approximately compute $\theta^{(\boldsymbol{j})}_{\lambda}$ from $\boldsymbol{j}$ can be used.

More precisely, let $\tilde{\phi}^{(\boldsymbol{j})}_{\lambda}$ be a $m$-digit binary representation of $\theta^{(\boldsymbol{j})}_{\lambda}/(2\pi) \in [0,1)$ with an accuracy $2^{-m}$, and suppose that an oracle $Q$ works as $\ket{\boldsymbol{j}}\ket{\boldsymbol{x}} \mapsto \ket{\boldsymbol{j}}\ket{\boldsymbol{x} \oplus \tilde{\phi}^{(\boldsymbol{j})}_{\lambda}}$ for an arbitrary $m$-qubit computational basis $\ket{\boldsymbol{x}}$.
From the oracle $Q$ and an $m$-qubit working register, we can construct an $(N+1)$-qubit unitary gate $\tilde{R}_X(\boldsymbol{\theta}_{\lambda})$ which approximates $R_X(\boldsymbol{\theta}_{\lambda})$ as
\begin{equation}
    \tilde{R}_X(\boldsymbol{\theta}_{\lambda}) R_X(\boldsymbol{\theta}_{\lambda})^{\dagger} = \sum_{\boldsymbol{j}} e^{i 2\pi \epsilon_{\boldsymbol{j}} X} \otimes \ket{\boldsymbol{j}} \bra{\boldsymbol{j}},
\end{equation}
where $\epsilon_{\boldsymbol{j}} < 2^{-m}$ for all $\boldsymbol{j}$ with two queries to the oracle $Q$ and with two-qubit quantum gates whose number grows polynomially to $m$. 

This is done as follows.
Suppose that an initial state is $\ket{\psi_0} := \sum_{\boldsymbol{j}} \ket{\boldsymbol{j}}\ket{0}^{\otimes m}\ket{\psi_{\boldsymbol{j}}}$ without loss of generality. Use the oracle $Q$ to obtain $\sum_{\boldsymbol{j}} \ket{\boldsymbol{j}}\ket{\tilde{\phi}^{(\boldsymbol{j})}_{\lambda}}\ket{\psi_{\boldsymbol{j}}}$. Apply $N$ two-qubit gates $\Lambda_k = \ket{0}\bra{0} \otimes I + \ket{1}\bra{1} \otimes \exp(i2\pi 2^{-(k+1)} X)$ for $0 \le k < N$ to the quantum state where the first part of the tensor product in $\Lambda_k$ acts on the $k$-th qubit of the second register and the latter part acts on the last register, then we obtain $\sum_{\boldsymbol{j}} \ket{\boldsymbol{j}}\ket{\tilde{\phi}^{(\boldsymbol{j})}_{\lambda}} e^{i2\pi \tilde{\phi}^{(\boldsymbol{j})}_{\lambda} X}\ket{\psi_{\boldsymbol{j}}}$. 
We finally undo the second register with an oracle access to obtain $\sum_{\boldsymbol{j}} \ket{\boldsymbol{j}}\ket{0}^{\otimes m} e^{i2\pi \tilde{\phi}^{(\boldsymbol{j})}_{\lambda}X} \ket{\psi_{\boldsymbol{j}}} = 
(\sum_{\boldsymbol{j}} \ket{\boldsymbol{j}}\bra{\boldsymbol{j}} \otimes I^{\otimes m} \otimes  e^{i2\pi \tilde{\phi}^{(\boldsymbol{j})}_{\lambda}X} )\ket{\psi_0}$.

Since $2\pi \tilde{\phi}^{(\boldsymbol{j})}_{\lambda} - \theta^{(\boldsymbol{j})}_{\lambda} < 2^{-m}$, this process approximates $R_X(\boldsymbol{\theta}_{\lambda})$ within arbitrary accuracy by using sufficiently many number of ancillary qubits $m$.

\section{A Clifford-based $4$-design on two qubits} \label{App:4DesClifford}

Denoting by $R_W(\theta) = \exp[i \theta W]$ ($W = X, Y, Z$) a single-qubit rotation around the $W$-axis, we define a single-qubit rotation $R_1(\theta_1, \theta_2, \theta_3)$ by 
\begin{equation}
    R_1(\theta_1, \theta_2, \theta_3)
    =
    R_Z(\theta_1) R_Y(\theta_2) R_Z(\theta_3).
\end{equation}
We also define a two-qubit rotation $R_2(\varphi_X, \varphi_Y, \varphi_Z)$ by
\begin{equation}
    R_2(\varphi_X, \varphi_Y, \varphi_Z) = \exp \biggl[ -i \sum_{W=X,Y,Z } \varphi_W W \otimes W \biggr].
\end{equation}
We also let $U_c$ be a fixed two-qubit unitary given by
\begin{multline}
U_c = \bigl( R_1(\theta_1 , \theta_2, \theta_3) \otimes R_1(\theta_1' , \theta_2', \theta_3') \bigr)\\
R_2(\varphi_X, \varphi_Y, \varphi_Z)\\
\bigl( R_1(\theta_4 , \theta_5, \theta_6) \otimes R_1(\theta_4', \theta_5', \theta_6') \bigr),    
\end{multline}
where
\begin{align}
    &(\theta_1, \theta_2, \theta_3) = (1.50097, 5.69898, 2.53181)\\ 
    &(\theta_1', \theta_2', \theta_3') = (1.25383, 0.01700, 6.21127)\\    
    &(\varphi_X, \varphi_Y, \varphi_Z) = (0.376407, 0.368786, 3.69014)\\
    &(\theta_4, \theta_5, \theta_6) = (4.66335, 3.04854, 1.45524)\\ 
    &(\theta_4', \theta_5', \theta_6') = (0.337423, 3.38137, 3.82503).
\end{align}
Then, ${\sf C}(4) U_c {\sf C}(4)$ is an exact unitary $4$-design on $2$ qubits, up to the numerical precision.

These numbers are obtained by numerically searching a zero of a function that is related to the ${\sf C}(2)$-invariant functions.
See Subsec.~9.3 of Ref.\,\cite{BNOZ2020} for more details.

\section{Irreducible representations} \label{App:Irreps}

We here provide the irreps of a unitary group ${\sf U}(d)$ on the vector space
\begin{equation}
\cK := {\rm span}\{ \kett{\sigma_{\vec{n}_1  \otimes \vec{n}_2}} : \vec{n}_1, \vec{n}_2 \in \{0,1,2,3 \}^{q} \},
\end{equation}
under the action of $L_{\cV^{\otimes 2}}$ with $\cV(\rho) := V \rho V^{\dagger}$ $(V \in {\sf U}(d))$. 

To this end, we index the irreps of unitary group ${\sf U}(d)$ by non-increasing integer sequences: $(\lambda_1, \lambda_2, \dots, \lambda_d)$, where $\lambda_i \in \mathbb{Z}$, and $\lambda_i \geq \lambda_j$ ($i \geq j$). The above representation contains the irreps indexed by $\lambda$, where $\lambda^+ \leq 2$, with $\lambda^+$ being the sum of positive integers in $\lambda$, and $\sum_i \lambda_i = 0$.

For a given index $\lambda$, the dimension of the corresponding representation space is given by the Weyl's dimension formula:
\begin{equation}
\frac{\prod_{1 \leq i \leq j \leq d}(\lambda_i - \lambda_j + j - i)}{\prod_{k=1}^{d-1} k!}.
\end{equation}
The multiplicity  can be  obtained from the Littlewood-Richardson rule. Since a single-qubit case ($q=1$) is special, we below consider the single-qubit case and the multi-qubit case separately.

The dimension and the multiplicity of irreps are summarized in Tab.\,\ref{Tab:irrep1} for one-qubit systems, and Tab.\,\ref{Tab:irrepmulti} for multi-qubit systems.

\begin{table}[t!]
\centering
\caption{Irreducible representations for a single-qubit system. The definitions of the irreducible spaces are based on the irreps of the Clifford group given in Ref.\,\cite{HWW2018}, and are explained in the main text.}
\begin{tabular}{c||c|c|c}
Highest Weight & Dimension & Multiplicity & Irreducible spaces \\ \hline \hline
$(2,-2)$ & $5$ & $1$ & $\cK_{\rm I}$ \\ 
$(1,-1)$ & $3$ & $3$ & $\cK_{\rm l}, \cK_{\rm r}, \cK_{\rm \{A\}}$\\
$(0,0)$ & $1$ & $2$ & $\cK_0, \cK_{\rm id}$ \\
\end{tabular}
\label{Tab:irrep1}
\end{table}

\begin{table*}[t!]
\centering
\caption{Irreducible representations for a multi-qubit system, where $d=2^{q}$. The definitions of the irreducible spaces are based on the irreps of the Clifford group given in Ref.\,\cite{HWW2018}, and are explained in the main text.}
\begin{tabular}{c||c|c|c}
Highest Weight & Dimension & Multiplicity & Irreducible spaces \\ \hline \hline
$(2, 0, \dots ,0,-2)$ & $(d^2(d-1)(d+3))/4$ & $1$ & $\cK_{\rm I}$ \\ 
$(2,0, \dots, 0, -1, -1)$ & $((d^2-1)(d^2-4))/4$ & $1$ & $\cK_{\rm [A]}$\\
$(1,1,0, \dots, 0, -2)$ & $((d^2-1)(d^2-4))/4$ & $1$ & $\cK_{\{ {\rm adj}\}}^{\perp}$\\
$(1,1,0 \dots, 0, -1, -1)$ & $(d^2(d+1)(d-3))/4$ & $1$ & $\cK_{\rm II}$\\
$(1,0 \dots, 0, -1)$ & $d^2-1$ & $4$ & $\cK_{\rm l}, \cK_{\rm r}, \cK_{[{\rm adj}]}, \cK_{\{{\rm adj}\}}$\\
$(0, \dots, 0)$ & $1$ & $2$ & $\cK_0, \cK_{\rm id}$ \\
\end{tabular}
\label{Tab:irrepmulti}
\end{table*}

\subsection{Single-qubit systems}
To explicitly obtain all the irreps for one qubit, we start with the irreps of the Clifford group ${\sf C}(2)$, which are provided in Ref.\,\cite{HWW2018}. Since the Clifford group is a subgroup of the unitary group ${\sf U}(2)$, irreps of the unitary group are obtained by taking the union of some irreps of the Clifford group. Since the dimensions of the irreps, both of the Clifford and the unitary groups, are known, we can check which irreps of the Clifford group should be combined by dimension counting.

First, from Theorem 1 in Ref.\,\cite{HWW2018}, the irreducible decomposition of $\cK$ in terms of ${\sf C}(2)$ is given by
\begin{equation}
\cK_{\rm id}  \oplus \cK_0 \oplus \cK_1  \oplus \cK_{\rm r} \oplus \cK_{\rm l}\oplus \cK_{\rm \{S\}} \oplus \cK_{\rm \{A\}}.
\end{equation}
Here, the important subspaces in our analysis are 
\begin{align}
&\cK_0 := {\rm span} \{ \kett{\sigma_{1}^{\otimes 2} + \sigma_{2}^{\otimes 2} + \sigma_{3}^{\otimes 2}} \}, \\
&\cK_1 := {\rm span} \{ \kett{S_{1,2}}, \kett{S_{1,3}}, \kett{S_{2,3}} \}, \\
&\cK_{\rm \{S \}} := {\rm span} \{  \kett{\sigma_1^{\otimes 2} - 2 \sigma_2^{\otimes 2} + \sigma_3^{\otimes 2}}, \kett{\sigma_1^{\otimes 2} - \sigma_3^{\otimes 2}} \}.
\end{align}
See Ref.\,\cite{HWW2018} for the definitions of the other subspaces, where the vector space is denoted by $V$ instead of $\cK$. For instance, $\cK_{\rm id}$ in our notation corresponds to $V_{\rm id}$ in the notation of Ref.\,\cite{HWW2018}.

The dimension of each subspace is also given in Ref.\,\cite{HWW2018} as
\begin{align}
&\dim \cK_{\rm id} = \dim \cK_0 = 1,\\
&\dim \cK_1= 2,\\
&\dim \cK_{\rm r} =\dim \cK_{\rm l} =\dim \cK_{\rm \{S\}}=\dim \cK_{\rm \{A\}} = 3.
\end{align}
Comparing these with the dimensions of irreps of the unitary group, given in Tab.\,\ref{Tab:irrep1}, it is clear that $\cK_{\rm id}, \cK_0$ are also irreps of ${\sf U}(2)$.

We can also show that $\cK_I := \cK_{\rm \{ S\}} \oplus \cK_1$ is an irrep of ${\sf U}(2)$. To this end, we show that $\exists U \in {\sf U}(2)$, $\exists \kett{v} \in \cK_1$ such that $L_{\cU^{\otimes 2}} \kett{v}$ has a support on $\cK_{\rm \{ S\}}$. Together with the dimension counting, it immediately leads to that $\cK_{\rm I}$ is irreducible.

The statement is shown by construction. Let $T:= {\rm diag}(1, e^{i \pi/4})$ be a diagonal unitary matrix in ${\sf U}(2)$. Using the simple relation that
\begin{align}
&T X T^{\dagger} = \frac{1}{\sqrt{2}}( X+Y),\\
&T Z T^{\dagger} =Z,
\end{align}
it is straightforward to show that
\begin{equation}
L_{\cT^{\otimes 2}} \kett{\sigma_1^{\otimes 2} - \sigma_3^{\otimes 2}}
\propto
-\kett{\sigma_1^{\otimes 2} - 2\sigma_2^{\otimes 2} + \sigma_3^{\otimes 2} }
+
3 \kett{S_{1,2}}.
\end{equation}
Since $\kett{\sigma_1^{\otimes 2} - \sigma_3^{\otimes 2}} \in \cK_1$ and $\kett{S_{1,2}} \in \cK_{\rm \{S\}}$, we obtain the desired statement.

The fact that $\cK_{\rm I}$, $\cK_0$, and $\cK_{\rm id}$ are irreducible with respect to ${\sf U}(2)$ implies that so are the rest, i.e., $\cK_{\rm l}$, $\cK_{\rm r}$, and $\cK_{\rm \{A\}}$, due to the dimension condition.
We hence obtain that the irreducible decomposition of $\cK$:
\begin{equation}
\cK_{\rm id}  \oplus \cK_0  \oplus \cK_{\rm r} \oplus \cK_{\rm l}\oplus \cK_{\rm \{A\}}  \oplus \cK_{\rm I},
\end{equation}
where $\cK_{\rm id}$ and $\cK_0$ are equivalent representations, and $\cK_{\rm r}$, $\cK_{\rm l}$, and $\cK_{\rm \{A\}}$ are equivalent.

We finally mention the fact that the traceless symmetric ones are only $\cK_0$ and $\cK_{\rm I}$, which can be directly confirmed from their definitions. Thus, the traceless symmetric space $\cK_{TS}$ is decomposed into irreducible subspaces as
\begin{equation}
\cK_{TS} = \cK_0 \oplus \cK_{\rm I},
\end{equation}
which is multiplicity-free.

\subsection{Multi-qubit systems}
For $q \geq 2$, we also start with the irreps of the Clifford group ${\sf C}(d)$. From Theorem 1 in Ref.\,\cite{HWW2018}, the irreducible decomposition by ${\sf C}(d)$ is given by
\begin{multline}
\cK = \cK_{\rm id}  \oplus \cK_0 \oplus \cK_1  \oplus \cK_2  \oplus \cK_{\rm [adj]} \oplus \cK_{[1]} \oplus \cK_{[2]} \\ 
\oplus \cK_{\rm \{adj \}} \oplus \cK_{\{1\}} \oplus \cK_{\{2\}} \oplus \cK_{\rm \{A\}}\oplus \cK_{\rm \{adj \}}^{\perp}.
\end{multline}
Similarly to the single-qubit case, by comparing the dimension of each subspace with those in Tab.\,\ref{Tab:irrepmulti}, we obtain that $\cK_{\rm id}$ and $\cK_0$ are the trivial irreps, and that $\cK_{\rm l}, \cK_{\rm r}, \cK_{[{\rm adj}]}$, and $\cK_{\{{\rm adj}\}}$ are those corresponding to $(1, 0, \dots, 0, -1)$.
We can also observe that $\cK_{\rm [A]}$ and $\cK_{\rm \{adj\}}^{\perp}$ are also irrepds, corresponding to $(2, 0, \dots, 0, -1, -1)$ and $(1, 1, \dots, 0, -2)$, respectively.

We hence need to identify which of $\cK_1$, $\cK_2$, $\cK_{[1]}$, $\cK_{[2]}$, $\cK_{\{1\}}$, and $\cK_{\{2\}}$ consist of the irrep $\cK_{\rm I}$ with $(2, 0,\dots, 0, -2)$ and the irrep $\cK_{\rm II}$ with $(1,1,0, \dots, 0,-1,-1)$. This can be done again by dimension counting. From~\cite{HWW2018}, the dimension of each subspace is given by
\begin{align}
&\dim \cK_1 = \frac{d(d+1)}{2} - 1,\\
&\dim \cK_2= \frac{d(d-1)}{2} - 1,\\
&\dim \cK_{[1]} = (d^2-1)\biggl[ \frac{d(d+2)}{8}-1 \biggr],\\
&\dim \cK_{[2]} = (d^2-1)\biggl[ \frac{d(d-2)}{8}-1 \biggr],\\
&\dim \cK_{\{1\}} =\frac{d(d^2-1)(d+2)}{8},\\
&\dim \cK_{\{2\}} = \frac{d(d^2-1)(d-2)}{8}.
\end{align}
By a straightforward calculation, it can be shown that the only possible combination to satisfy the dimension condition that $\dim \cK_{\rm I} =(d^2(d-1)(d+3))/4$ and $\dim \cK_{\rm II} =(d^2(d+1)(d-3))/4$ is 
\begin{align}
\cK_{\rm I} := \cK_1 \oplus \cK_{[1]}  \oplus \cK_{\{1\}},\\
\cK_{\rm II} :=\cK_2 \oplus \cK_{[2]}  \oplus \cK_{\{2\}}.
\end{align}

We finally consider the irreducible decomposition of the traceless symmetric subspace $\cK_{TS}$. From the definitions of each subspace~\cite{HWW2018}, we can easily check that the irreducible decomposition of $\cK_{TS}$ by ${\sf U}(d)$ is given by
\begin{equation}
\cK_{TS} = \cK_0 \oplus \cK_{\rm I} \oplus \cK_{\rm II}  \oplus \cK_{\rm [adj]},
\end{equation}
which is multiplicity-free. In the main text, we have denoted $\cK_{\rm [adj]}$ by $\cK_{\rm III}$ for the simplicity of notation.

\section{Characterizations of a noise} \label{App:CharaNoise}

We here show the following properties of the self-adjointness $H(\cE)$ and the self-adjointness parameter $h(\cE)$.
\begin{enumerate}
\item In the Liouville representation, the self-adjointness parameter $h(\cE)$ is give by
\begin{align}
h(\cE) &= \frac{1}{d^2-1} \sum_{\vec{n} \neq \vec{0}} \braa{\sigma_{\vec{n}}} L_{\cE}^2 \kett{\sigma_{\vec{n}}}\label{Eq:B1}  \\
&=\frac{1}{d^2-1} \tr[ \tilde{L}_{\cE}^2].\label{Eq:B2} 
\end{align}
\item The self-adjointness parameter $h(\cE)$ satisfies
\begin{equation}
-\frac{1}{d^2-1} \leq h(\cE) \leq u(\cE), \label{Eq:B3}
\end{equation}
which immediately implies that $0 \leq H(\cE) \leq 1$.
\item $h(\cE) = u(\cE)$ if and only if $\tilde{L}_{\cE} = \tilde{L}_{\cE}^{\dagger}$. For a unital noise, $h(\cE) = u(\cE)$ if and only if the noise is self-adjoint.
\item $h(\cE) =-\frac{1}{d^2-1}$ if and only if $\tr[K_iK_j]  = 0$ for any $i, j$, where $\{ K_i \}$ are the Kraus operators of $\cE$.
\item The average gate fidelity $F(\cE)$ is bounded from above by $u(\cE)$ and $h(\cE)$:
 \begin{equation}F(\cE) \leq \frac{d-1}{d} \sqrt{\frac{h(\cE) + u(\cE)}{2}} + \frac{1}{d}.\label{Eq:B4}\end{equation}
\end{enumerate}

We first show Eq.\,\eqref{Eq:B1}. Recalling that $(L_{\cE} )_{\vec{0} \vec{n}}=0$ for any $\vec{n} \neq \vec{0}$ since $\cE$ is TP, we have
\begin{align}
&\sum_{\vec{n}, \vec{m} \neq \vec{0}} \tr\bigl[\sigma_{\vec{n}} \cE(\sigma_{\vec{m}})] \tr[\sigma_{\vec{m}} \cE(\sigma_{\vec{n}}) \bigr]\\
&=
\sum_{\vec{n}, \vec{m} \neq \vec{0}} \tr \bigl[\sigma_{\vec{n}}^{\otimes 2} (\cE \otimes \cE^{\dagger}) (\sigma_{\vec{m}}^{\otimes 2}) \bigr]\\
&=
\tr\bigl[ \bigl(\mathbb{F} - \sigma_{\vec{0}}^{\otimes 2} \bigr) (\cE \otimes \cE^{\dagger}) \bigl(\mathbb{F} - \sigma_{\vec{0}}^{\otimes 2} \bigr)\bigr]\\
&=
\tr\bigl[ \mathbb{F} (\cE \otimes \cE^{\dagger}) (\mathbb{F})\bigr] - 1, \label{Eq:brrr}
\end{align}
where $\mathbb{F} := \sum_{\vec{n}} \sigma_{\vec{n}}^{\otimes 2}$ is the swap operator on $\cK^{\otimes 2}$. Note that, in the last line, we have used the fact that $\cE$ is TP, also implying that $\cE^{\dagger}$ is unital.
We now use another expression of the swap operator, which is 
\begin{equation}
\mathbb{F}  = d\biggl[ (d+1) \int \varphi^{\otimes 2} d\varphi  - \sigma_{\vec{0}}^{\otimes 2} \biggr],
\end{equation}
which simply follows from the Schur's lemma.
Substituting this and using the relation that $\tr[ \mathbb{F} (A \otimes B)] = \tr[AB]$ for any operators $A$ and $B$, we obtain
\begin{multline}
\frac{1}{d^2-1} \sum_{\vec{n} \neq \vec{0}} \braa{\sigma_{\vec{n}}} L_{\cE}^2 \kett{\sigma_{\vec{n}}} \\
=\frac{1}{d-1}\biggl[ d \int \tr \bigl[ \cE(\varphi) \cE^{\dagger}(\varphi) \bigr] d\varphi  -1 \biggr].
\end{multline}

On the other hand, it is straightforward to show that 
\begin{align}
h(\cE) &:= \frac{d}{d-1}  \int \tr \bigl[ \cE'(\varphi) \cE'^{\dagger}(\varphi) \bigr] d\varphi, \\
&=\frac{1}{d-1}\biggl[ d \int \tr \bigl[ \cE(\varphi) \cE^{\dagger}(\varphi) \bigr] d\varphi  - 1 \biggr], \label{Eq:B12}
\end{align}
which follows from the definition of $\cE'$, i.e. $\cE'(\rho) := \cE(\rho - I/d)$. Hence, we have Eq.\,\eqref{Eq:B1}.
Note that Eq.\,\eqref{Eq:B2} follows simply from the definition of $\tilde{L}_{\cE}$.

We next show Eq.\,\eqref{Eq:B3}, and the properties 3 and 4. The lower bound of $h(\cE)$ in Eq.\,\eqref{Eq:B3} is obtained from Eqs.\,\eqref{Eq:B1} and ~\eqref{Eq:brrr}, which lead to
\begin{equation}
h(\cE) = \frac{1}{d^2-1} \bigl(\tr\bigl[ \mathbb{F} (\cE \otimes \cE^{\dagger}) (\mathbb{F})\bigr] - 1\bigr).
\end{equation}
Using the Kraus operators $\{K_i\}$ for $\cE$ and the swap trick, this can be rewritten as
\begin{equation}
h(\cE) = \frac{1}{d^2-1} \biggl(\sum_{i,j} | \tr[K_iK_j]|^2 -1\biggr),
\end{equation}
which is not smaller than $-1/(d^2-1)$. The equality holds if and only if $\tr[K_iK_j]=0$ for any $i,j$, which is the property 4. A simple instance of such a noise is a unitary noise that maps $\rho$ to $\sqrt{X} \rho \sqrt{X}^{\dagger}$.

The upper bound of $h(\cE)$ in Eq.\,\eqref{Eq:B3} is obtained from the relation that
\begin{multline}
\int d \psi \tr \biggl[ \bigl( \cE(\ketbra{\psi}{\psi} - I/d) - \cE^{\dagger}(\ketbra{\psi}{\psi}- I/d) \bigr)^2 \biggr]\\
=
\frac{2(d-1)}{d} \bigl( u(\cE) - h(\cE) \bigr),
\end{multline}
which can be checked by a direct calculation. Since the left-hand side is non-negative, we have $u(\cE) \geq h(\cE)$. About the property 3, it is obvious that $h(\cE) = u(\cE)$ if $\cE = \cE^{\dagger}$. The converse is shown as follows:
\begin{align}
&h(\cE) = u (\cE),\\
&\Leftrightarrow \tr[\tilde{L}_{\cE}^2] = \tr[\tilde{L}_{\cE}^{\dagger} \tilde{L}_{\cE}],\\
&\Leftrightarrow \sum_{\vec{n} > \vec{m}} \bigl( (L_{\cE})_{\vec{n}\vec{m}} -(L_{\cE})_{\vec{m}\vec{n}} \bigr)=0,\\
&\Leftrightarrow (L_{\cE})_{\vec{n}\vec{m}} = (L_{\cE})_{\vec{m}\vec{n}},\ \ \forall \vec{n},\vec{m},\\
&\Leftrightarrow L_{\cE} = L_{\cE}^{\dagger}.
\end{align}
Note that the last line follows from the fact that $L_{\cE}$ is a real matrix in the Pauli basis.

We finally show the relation between the average fidelity $F(\cE)$, the self-adjointness parameter $h(\cE)$, and the unitarity $u(\cE)$, i.e. Eq.\,\eqref{Eq:B4}.
To this end, we again use the fact that $L_{\cE}$ is a real matrix in the Pauli basis, leading to $\tr[ \tilde{L}_{\cE}] = \tr[ \tilde{L}_{\cE}^{\dagger}]$. We hence have
\begin{align}
\bigl( \tr[ \tilde{L}_{\cE}] \bigr)^2 &= \frac{1}{4} \bigl( \tr[ \tilde{L}_{\cE}] + \tr[ \tilde{L}_{\cE}^{\dagger}]  \bigr)^2,\\
&\leq \frac{1}{4} \bigl| \! \bigl| \tr[ \tilde{L}_{\cE}] + \tr[ \tilde{L}_{\cE}^{\dagger}]  \bigr| \! \bigr|_1^2,\\
&\leq \frac{d^2-1}{4} \bigl| \! \bigl| \tr[ \tilde{L}_{\cE}] + \tr[ \tilde{L}_{\cE}^{\dagger}]  \bigr| \! \bigr|_2^2,\\
&= \frac{d^2-1}{4} \tr \bigl[ \tilde{L}_{\cE}^2 + \bigl( \tilde{L}_{\cE}^{\dagger}\bigr)^2 + 2  \tilde{L}_{\cE}^{\dagger} \tilde{L}_{\cE} \bigr],\\
&= \frac{(d^2-1)^2}{2} \bigl( h(\cE) + u(\cE) \bigr),
\end{align}
where the second inequality follows from the relation that $|\!| M |\!|_1 \leq \sqrt{D} |\!| M |\!|_2$ for any $D \times D$ matrix $M$, and the last line from that $L_{\cE}$ is a real matrix, which implies that $\tr [\tilde{L}_{\cE}^2] = \tr[ (\tilde{L}_{\cE}^{\dagger})^2]$. Since $F(\cE) = \frac{\tr[L_{\cE}] + d}{d(d+1)}$ and $\tr[L_{\cE}] = \tr[\tilde{L}_{\cE}] + 1$, we obtain Eq.\,\eqref{Eq:B4}.

\end{document}